\documentclass[12pt]{article}
\pdfoutput=1
\usepackage[a4paper]{geometry}
\usepackage{jheppub, amsmath,amssymb,amsfonts,amsxtra, mathrsfs, makeidx,graphics,graphicx,amsthm,epsfig, ytableau,bm,longtable,float, color,tikz,mathtools,xfrac,footnote,rotating, lscape, makecell, environ,mathtools, empheq, colortbl, xcolor,enumerate, tikz-3dplot, slashed, pifont, multirow}

\newcommand\superequiv{\mathrel{\rlap{\raisebox{\fontdimen22\textfont2}{$=$}}\raisebox{-0.5\fontdimen22\textfont2}{$ = $}}}

\usetikzlibrary{positioning}
\usetikzlibrary{chains}
\usetikzlibrary{arrows, arrows.meta ,fit,decorations.pathreplacing}
\tikzstyle{every picture}+=[remember picture]
\tikzstyle{na} = [baseline]

\usetikzlibrary{arrows, decorations.markings, calc, fadings, decorations.pathreplacing, patterns, decorations.pathmorphing, positioning}
\tikzset{>={Latex[width=1.5mm,length=1.5mm]}}
\usetikzlibrary{graphs,graphs.standard,quotes}

\usepackage{mdframed}

\def\node#1#2{\overset{#1}{\underset{#2}{{\color{gray} \bullet}}}}

\def\node#1#2{\overset{#1}{\underset{#2}{\circ}}}

\tikzstyle{every picture}+=[remember picture]
\tikzstyle{na} = [baseline=-.5ex]

\usepackage{bbm}
\usepackage{subfigure}

\newcommand{\eg}{\textit{e.g.}}

\newcommand{\ie}{\textit{i.e.}}

\numberwithin{equation}{section}
\newcommand{\bes}[1]{\begin{equation} \begin{split} #1\end{split} \end{equation}}

\newcommand{\nn}{\nonumber}

\newcommand{\be}{\begin{equation}} \newcommand{\ee}{\end{equation}}
\newcommand{\bea}{\begin{equation} \begin{aligned}} \newcommand{\eea}{\end{aligned} \end{equation}}

\def\tilde{\widetilde}

\def\hat{\widehat}

\def\rt2{\sqrt{2}}

\def\Tr{\mathop{\rm Tr}}

\def\CD{{\cal D}}

\def\CN{{\cal N}}

\def\CS{{\cal S}}

\def\1{{\ds 1}}

\newcommand{\fm}{\mathfrak{m}}

\def\repa{\raise4pt\hbox{$\square$}\mkern-14mu\raise-4pt\hbox{$\square$}}
\def\repab{\overline{\raise4pt\hbox{$\square$}\mkern-14mu\raise-4pt\hbox{$\square$}\mkern-1mu}}

\def\smileface{\ensuremath{\hbox{\large$\bigcirc$}\mkern-15mu\raise-1pt\hbox{\scriptsize$\smallsmile$}%
\mkern-10mu\raise4pt\hbox{..}\mkern4mu}}
\def\frownface{\ensuremath{\hbox{\large$\bigcirc$}\mkern-15mu\raise-1pt\hbox{\scriptsize$\smallfrown$}%
\mkern-10mu\raise4pt\hbox{..}\mkern4mu}}

\newcommand{\ba}{\begin{array}}
\newcommand{\ea}{\end{array}}
\newcommand{\bi}{\begin{itemize}}
\newcommand{\ei}{\end{itemize}}
\def\vec#1{\bm{#1}}
\def\bea#1\eea{\allowdisplaybreaks \begin{align}#1\end{align}}
 \newcommand{\ben}{\begin{enumerate}}
\newcommand{\een}{\end{enumerate}}
\newcommand{\bean}{\begin{eqnarray*}}
\newcommand{\eean}{\end{eqnarray*}}
\newcommand{\eref}[1]{(\ref{#1})}

\newcommand{\comment}[1]{}

\definecolor{light-gray}{gray}{0.5}

\newcommand{\blue}{\color{blue}}
\newcommand{\gray}{\color{light-gray}}
\newcommand{\red}{\color{red}}

\def\aup#1 {\overset{#1}{\uparrow} \, \overset{\tilde{#1}}{\downarrow}}

\tikzset{snake it/.style={decorate, decoration={snake, amplitude=.4mm, segment length=2mm,
                       post length=0mm,pre length=0mm}}}
                       
\newcommand{\GCD}{\mathrm{GCD}}

\def\Gp{\Phi}

\tikzset{->-/.style={decoration={
  markings,
  mark=at position #1 with {\arrow{>}}},postaction={decorate}}}    
\tikzset{-<-/.style={decoration={
  markings,
  mark=at position #1 with {\arrow{<}}},postaction={decorate}}}          
  
\title{New aspects of Argyres--Douglas theories and their dimensional reduction}
\author[a]{Simone Giacomelli,}
\author[b,d]{Noppadol Mekareeya,}
\author[b,c]{and Matteo Sacchi}
\affiliation[a]{Mathematical Institute, University of Oxford, \\ Woodstock Road, Oxford, OX2 6GG, United Kingdom}
\affiliation[b]{INFN, sezione di Milano-Bicocca, \\Piazza della Scienza 3, I-20126 Milano, Italy}
\affiliation[c]{Dipartimento di Fisica, Universit\`a di Milano-Bicocca, \\ Piazza della Scienza 3, I-20126 Milano, Italy}
\affiliation[d]{Department of Physics, Faculty of Science, \\
Chulalongkorn University, Phayathai Road, \\
Pathumwan, Bangkok 10330, Thailand}
\emailAdd{simone.giacomelli@maths.ox.ac.uk}
\emailAdd{n.mekareeya@gmail.com}
\emailAdd{m.sacchi13@campus.unimib.it}
\abstract{Argyres-Douglas (AD) theories constitute an infinite class of superconformal field theories in four dimensions with a number of interesting properties.  We study several new aspects of AD theories engineered in $A$-type class $\mathcal{S}$ with one irregular puncture of Type I or Type II and also a regular puncture. These include conformal manifolds, structures of the Higgs branch, as well as the three dimensional gauge theories coming from the reduction on a circle.  We find that the latter admits a description in terms of a linear quiver with unitary and special unitary gauge groups, along with a number of twisted hypermultiplets.   The origin of these twisted hypermultiplets is explained from the four dimensional perspective. We also propose the three dimensional mirror theories for such linear quivers.  These provide explicit descriptions of the magnetic quivers of all the AD theories in question in terms of quiver diagrams with unitary gauge groups, together with a collection of free hypermultiplets. A number of quiver gauge theories presented in this paper are new and have not been studied elsewhere in the literature.}
\begin{document}
\maketitle

\section{Introduction} 

Superconformal theories with eight supercharges in four dimension represent an interesting laboratory for the exploration of strongly-coupled dynamics in field theory. Since the discovery of the Seiberg-Witten solution \cite{Seiberg:1994rs, Seiberg:1994aj} many four-dimensional interacting superconformal theories (SCFTs) have been identified at singular points of the Coulomb branch of $\mathcal{N}=2$ gauge theories. These are characterized by the presence of massless dyons which are relatively nonlocal, therefore making the theory intrinsically interacting. These models are usually referred to as Argyres-Douglas theories \cite{Argyres:1995jj}. The distinctive feature of these theories is the presence of Coulomb branch operators of fractional dimension and it is common to name all the models with this property Argyres-Douglas (AD) theories.

At present we still do not have a classification of AD theories and over the past twenty years a very vast landscape of such models has been discovered (see \eg~ \cite{Argyres:1995xn, Eguchi:1996vu, Eguchi:1996ds, Shapere:1999xr, Cecotti:2010fi, Bonelli:2011aa, Xie:2012hs, Wang:2015mra}). Moreover, over the years many techniques have been developed to study these elusive theories, in particular geometric methods based on stringy realizations of AD theories. The most commonly used ones are the class $\mathcal{S}$ construction \cite{Gaiotto:2009hg} and the geometric engineering in Type IIB via compactification on singular Calabi-Yau threefolds \cite{Shapere:1999xr, Cecotti:2010fi}. Both methods play a key role in the present paper and provide a detailed description of the Coulomb branch of these theories.  

The focus of this work is the analysis of the Higgs branch of AD theories, about which considerably less is known. In particular our focus is the construction of the 3d theories coming from the  reduction of 4d $\CN=2$ AD SCFTs on a circle and their 3d mirror theories \cite{Intriligator:1996ex}.  The latter are three-dimensional $\CN=4$ lagrangian theories whose Coulomb branch is identical to the Higgs branch of the 4d SCFT in question.  In a more contemporary terminology, such a mirror theory can be regarded as a ``magnetic quiver'' \cite{Ferlito:2017xdq, Cabrera:2018jxt, Cabrera:2019izd, Bourget:2019rtl} for the corresponding 4d $\CN=2$ AD theory. In this paper, we will focus on general AD theories engineered in A-type class $\mathcal{S}$ (the 6d SCFT is the worldvolume theory on a stack of M5 branes) with one irregular puncture of Type I or Type II \cite{Xie:2012hs, Xie:2013jc} and also a regular puncture (possibly trivial).    It should be noted that the magnetic quivers for a subclass of the theories we consider in this paper has so far been obtained from the class $\mathcal{S}$ description by exploiting known properties of the Hitchin moduli space \cite{boalch2008irregular, Xie:2012hs, Xie:2013jc, DelZotto:2014kka, Xie:2017vaf, Dey:2020hfe}, or from the analysis of $\mathcal{N}=1$ lagrangian theories which flow in the IR to AD theories \cite{Maruyoshi:2016tqk, Maruyoshi:2016aim, Agarwal:2016pjo, Benvenuti:2017lle, Benvenuti:2017kud, Agarwal:2017roi, Benvenuti:2017bpg}. See also \cite{Buican:2015ina, Buican:2015hsa, Buican:2015tda, Dedushenko:2019mnd} for an analysis based on chiral algebras and the superconformal index.

The new ingredient at the heart of our analysis is the recent observation of \cite{Simone:2020} that the above-mentioned class of AD theories becomes lagrangian upon dimensional reduction, being equivalent to {\it a linear quiver with unitary and special unitary gauge groups}. The derivation of this fact exploits the realization of these theories at special points in the parameter space of four-dimensional lagrangian theories. We then study the 3d mirror theories of such linear quivers.  The former can indeed be regarded as the magnetic quivers of the AD theories in question.   As we will see, such mirror theories in general involve {\it intricate quivers with unitary gauge groups plus a decoupled free sector}.  

The presence of the free sector was actually noticed in the special case of $A_1$ class $\mathcal{S}$ AD theories (usually called the $(A_1,A_n)$ and $(A_1,D_n)$ theories) \cite{Nanopoulos:2010bv}. This phenomenon has the following interpretation: Generically AD theories cannot be higgsed to free hypers and at a generic point of the Higgs branch the low-energy effective theory includes a strongly coupled ``non higgsable'' sector, which is a collection of multiple copies of other AD theories with no Higgs branch. Their number is related to the dimension of the Higgs branch, as will be derived in section \ref{sec:reviewAD} from the Type IIB geometric engineering of AD theories. These non higgsable theories without a Higgs branch become twisted hypermultiplets upon dimensional reduction and via mirror symmetry become a collection of free hypermultiplets.   We can provide an argument, based on the assumption that there are no nontrivial rank-zero theories in 3d with eight supercharges and nontrivial Higgs branch\footnote{In the theories discussed in \cite{Gang:2021hrd} both the Coulomb and Higgs branches are trivial. $\mathcal{N}=4$ models in 3d with this property can be interacting.}, suggesting that this statement is true in general. Indeed there are many examples of SCFTs in four (or higher dimension) with trivial Higgs branch. Upon compactification to 3d, the Coulomb branch becomes hyperK\"ahler and the Higgs branch remains trivial.  Applying mirror symmetry, which exchanges the Higgs and Coulomb branches, we obtain the 3d mirror of the non higgsable theory which has a trivial Coulomb branch.  However, this is free by the rank-zero assumption.  We indeed find that the number of free hypermultiplets in such a 3d mirror is always equal to the rank of the non higgsable sector, in perfect agreement with our expectations. Moreover, we manage to identify the non higgsable sectors in all theories in the class that we study in this paper.

The paper is organised as follows.  In Section \ref{sec:reviewAD}, we summarise various known properties of the AD theories corresponding to class $\mathcal{S}$ theories with one irregular puncture and those with one irregular and one regular full punctures.  In Section \ref{sec:newAD}, we discuss new aspects of the AD theories, including the $a$ and $c$ central charges for class $\mathcal{S}$ theories with regular and irregular punctures, conformal manifolds and properties of the Higgs branch.  We also propose a prescription for constructing the linear quivers, with unitary and special unitary gauge groups, arising from reduction of the AD theories to 3d.  We discuss the mirror theories for the $D_p(SU(N))$ theory, with $p\geq N$, and the $(A_{p-N-1},A_{N-1})=I_{p-N,N}$ theory in Section \ref{sec:DpSUNpgeqN}.  In Section \ref{sec:DpSUNplN}, the mirror theory for the $D_p(SU(N))$ theory, with $p \leq N$, is proposed. In Section \ref{sec:TypeIIp>Nm1}, we present the mirror theories for the $D_p^{N-1}(SU(N))$ theory, with $p\geq N-1$, and the $(I_{p-(N-1),N-1},S)$ theory.  The mirror theory for the $D_p^{N-1}(SU(N))$ theory, with $p\leq N-1$, is presented in Section \ref{sec:TypeIIp<Nm1}.  Many explicit examples are provided throughout the paper.  In Appendices \ref{app:aharony} and \ref{app:flipflip}, we review the Aharony duality \cite{Aharony:1997gp} for quivers and the flip-flip duality \cite{Aprile:2018oau} for $T(SU(N))$.  In Appendix \ref{app:deriv3dmirrD4SU10}, we present an explicit derivation the mirror theory for the $D_4(SU(10))$ theory utilising the flip-flip duality.  In Appendix \ref{app:SUenhanced}, we discuss global symmetry enhancement in linear quivers with special unitary gauge groups.  This appendix consists of new results, complementary to the unitary and symplectic cases discussed in \cite{Gaiotto:2008ak}.

\subsubsection*{Notations and conventions}
In the quiver diagram, we denote the unitary gauge group $U(k)$ by a circular node labelled by $k$, the flavour symmetry by a square node, and the bifundamental hypermultiplet between two groups by a line joining two nodes.  The special unitary gauge group is specified explicitly.  We also use the notation like
\bes{ \label{examplequiv}
[F_1]-(N_1)-SU(M)-(N_2)-[F_2]
}
Here $(N_1)$ and $(N_2)$ denote the unitary gauge groups $U(N_1)$ and $U(N_2)$ respectively; $SU(M)$ denotes the $SU(M)$ gauge group; $[F_1]$ and $[F_2]$ denote $F_1$ and $F_2$ flavours of fundamental hypermultiplets under the $U(N_1)$ and $U(N_2)$ gauge group respectively; and each line $-$ denotes the bifundamental hypermultiplets.  Whenever we would like to emphasise that the gauge group is of the unitary type, we write, for example $U(N_1)$, explicitly in the quiver. Moreover, quiver \eref{examplequiv} can also be written as
\bes{
[F_1]-(N_1)-[M] \,\, \longleftarrow SU(M) \longrightarrow \,\, [M]-(N_2)-[F_2]
}
where $\longleftarrow SU(M) \longrightarrow$ denotes the $SU(M)$ gauging.

\section{Review of Argyres-Douglas theories} \label{sec:reviewAD}

\subsection{Class $\mathcal{S}$ theories with one irregular puncture}
One can construct an infinite family of $\mathcal{N}=2$ SCFTs in M-theory by wrapping $N$ M5 branes on a sphere with one irregular puncture. The construction can be refined by adding a regular puncture and we will discuss both types of theories. There are two different types of irregular punctures, which employing the notation of \cite{Xie:2012hs, Xie:2013jc, Wang:2015mra} we denote as Type I and Type II (we will not discuss Type III in detail since their 3d mirror theory is known). For us it will be important to consider also the realization of these models as compactifications of Type IIB string theory on a threefold hypersurface singularity. 

The models described by a sphere with just one irregular puncture are associated with hypersurface singularities in $\mathbb{C}^4$ of the form 
\bes{ \label{sing} 
\text{{\bf Type I:}}\;\; &tw+x^N+z^k+\sum_{i,j}u_{ij}z^ix^j=0; \\ 
\text{{\bf Type II:}}\;\; &tw+x^N+xz^k+\sum_{i,j}u_{ij}z^ix^j=0.}
The theories of Type I are usually denoted as $(A_{N-1},A_{k-1})$ \cite{Cecotti:2010fi} or $I_{N,k}$ \cite{Xie:2013jc}.\footnote{Another nomenclature \cite{Wang:2015mra} for the $(A_{k-1},A_{N-1}) = I_{k,N}$ theory is $J^N[k]$ with $J=A_{N-1}=SU(N)$. In this notation, the Type II theory in \eref{sing} is denoted by $J^{N-1}[k]$ with $J=A_{N-1}$.}  We use both notations interchangeably in the following part of the paper. From (\ref{sing}) we can see that $(A_{N-1},A_{k-1})$ and $(A_{k-1},A_{N-1})$ are equivalent theories. In the above expression $u_{ij}$ denote deformation parameters, whose scaling dimension is determined by requiring homogeneity of the geometry (\ref{sing}) and demanding that the holomorphic threeform $\Omega_3$ has dimension one.
\be\label{O3} \Omega_3=\frac{dwdtdxdz}{dW},\ee 
where $W(w,t,x,z)=0$ is the hypersurface singularity in (\ref{sing}). The allowed deformations in (\ref{sing}) are those with $i,j\geq0$ such that $u_{ij}$ has positive scaling dimension. Furthermore, we can set to zero (modulo change of coordinates) all deformations with $j=N-1$ and also those with $i=k-1$ for Type I theories (for Type II theories the term with $i=k-1$ and $j=0$ is allowed). The parameters with dimension one describe mass deformations, whose number is equal to the rank of the global symmetry of the theory  (see Appendix B.6 of \cite{Giacomelli:2017ckh})
\be 
\mathrm{rk}(G_F)=N-1-b+\GCD(k,b)~, 
\ee
and those with dimension larger than one correspond to expectation values of Coulomb branch operators. From the geometric data we can compute the central charges $a$ and $c$ using the Shapere-Tachikawa relation \cite{Shapere:2008zf}
\be\label{2a-c} 8a-4c=\sum_i(2D(O_i)-1),\ee 
where the sum runs over CB operators and $D(O_i)$ denotes their scaling dimension, determined using the procedure described before. The rank of the theory (dimension of the Coulomb branch) is given by the expression 
\be r=\frac{(kN-b)(N-1)}{2b}-\frac{\mathrm{rk}(G_F)}{2},\ee 
where 
\be
b= \begin{cases} N & \quad\text{for Type I theories}~,\\
 N-1 &\quad \text{for Type II theories}~. 
 \end{cases}
\ee 
The $c$ central charge is given by the expression \cite{Xie:2016evu}
\be\label{ccentral} c=\frac{(kN-b)(N-1)}{12b(k+b)}(kN+k+b)-\frac{\mathrm{rk}(G_F)}{12}.\ee
We would also like to point out that Type II theories have also a class $\mathcal{S}$ realization involving a Type I irregular puncture and a regular simple puncture on the sphere. This works as follows: The Type II theory described by the geometry in (\ref{sing}) can also be realized as the $A_{k-1}$ $\mathcal{N}=(2,0)$ theory on the sphere with one regular simple puncture, corresponding to the partition $S=(k-1,1)$, and an irregular puncture of Type I \cite{Xie:2013jc}. The irregular puncture is the same which engineers the $I_{k, N-1} = (A_{k-1},A_{N-2})$ theory. Such a Type II theory is denoted as $(I_{k,N-1},S)$ \cite{Xie:2013jc}.

We would also like to review some basic properties about the conformal manifolds of these Argyres-Douglas theories. Exactly marginal deformations are described by $u_{ij}$ parameters in (\ref{sing}) with vanishing scaling dimension. It is easy to see that the number of such parameters (i.e. the dimension of the conformal manifold) for Type I theories is $\GCD(N,k)-1$ unless $k$ is a multiple of $N$ (or $N$ is a multiple of $k$), in which case the dimension of the conformal manifold is $\GCD(N,k)-2$. If we impose the condition $N=k$ the number of marginal deformations is further reduced to $N-3$. Since the addition of a regular puncture does not affect the dimension of the conformal manifold, the same analysis can be applied to theories of Type II with $\GCD(N,k)$ replaced by $\GCD(N-1,k)$. 

\subsection{Class $\mathcal{S}$ theories with one irregular and one regular full punctures}
If we add a full regular puncture, we find models with (at least) $SU(N)$ global symmetry described in Type IIB by hypersurface singularities in $\mathbb{C}^3\times\mathbb{C}^*$ of the form \cite{Cecotti:2012jx, Cecotti:2013lda, Wang:2015mra}
\bes{\label{def1}
\text{{\bf Type I:}}\;\; &wt+x^N+z^p+\sum_{i,j} u_{ij}z^ix^j=0; \\ 
\text{{\bf Type II:}}\;\; &wt+ x^N+xz^p+\sum_{i,j} u_{ij}z^ix^j=0. }
The $\mathbb{C}^*$ variable is $z$ and parametrizes the sphere on which the M5 branes are wrapped. The irregular puncture is located at $z=\infty$ and the regular puncture is at $z=0$. Now the holomorphic three-form is 
\be\label{OO3} \Omega_3=\frac{dwdtdxdz}{zdW}.\ee 
The scaling dimension of deformation parameters can be determined as before  by requiring $\Omega_3$ to have dimension one. The range of $i$ and $j$ is similar to the previous case: $i,j\geq 0$ and all terms with $j=N-1$ are dropped. This time however we have just $i\leq p$. The sum is again restricted to deformations $u_{ij}$ with positive dimension. In this class of theories the set of mass deformations includes all parameters $u_{ij}$ of dimension one, as in the previous case, and also all $u_{0j}$ parameters which represent the mass casimirs of the $SU(N)$ global symmetry. Coulomb branch operators are the remaining $u_{ij}$ parameters with scaling dimension larger than one.
From (\ref{def1}) and (\ref{OO3}) we find the CB spectrum 
\be D(u_{ij})=N-j-\frac{b}{p}i,\ee 
where again $b=N$ for Type I theories and $b=N-1$ for Type II theories. The rank of the theory $r$ (dimension of the Coulomb branch) and the rank of the global symmetry \cite[Appendix B]{Giacomelli:2017ckh} now are respectively  
\bes{\label{ranks} r&=\frac{pN(N-1)}{2b}-\frac{\mathrm{rk}(G_F)}{2}~; \\
\mathrm{rk}(G_F)&=2N-2+\GCD(p,b)-b~.} 
The models (\ref{def1}) are usually called $D_p^b(SU(N))$ \cite{Cecotti:2013lda, Wang:2015mra} and we adopt this notation. Furthermore, for $b=N$ (Type I theories) we will use the notation $D_p(SU(N))$, and when $p>N$ this theory will also be referred to as $(I_{p-N, N}, (1^N))$, where $(1^N)$ stands for the regular full puncture.
The central charges $a$ and $c$ are found from (\ref{2a-c}) and the following expression for $c$ (see \cite[(6.17)]{Cecotti:2013lda} and \cite[(2.15)]{Giacomelli:2017ckh}):
\be\label{ccentral2} c=\frac{(pN-b)(N-1)}{12b}(N+1)-\frac{N-1+\GCD(p,b)-b}{12}.\ee 
From (\ref{2a-c}) and (\ref{ccentral2}) one can show that for $D_p(SU(N))$ theories \cite[(6.84)]{Cecotti:2013lda}
\be\label{ccformula} 24(c-a)=\frac{1}{2}(N^2+\GCD(N,p)-2)-\frac{3N}{p}\sum_{j=1}^{N-1}\left\{\frac{jp}{N}\right\}\left(1-\left\{\frac{jp}{N}\right\}\right),\ee 
where $\{\;\}$ denotes the fractional part. The $SU(N)$ flavor central charge is given by the expression \cite{Xie:2016evu}
\be\label{fcc} k_{SU(N)}=2N-\frac{2b}{p}.\ee

From the Type IIB description discussed before it is easy to extract the SW curve of the $\mathcal{N}=2$ SCFTs. This is found simply by dropping the term $wt$ from (\ref{sing}) and (\ref{def1}). The SW differential is $\lambda_{SW}=xdz$ in the former case and $\lambda_{SW}=xdz/z$ in the latter. It's easy to check that the SW differential has always dimension 1 with the assignment discussed above, as it should be. 

If we close the regular puncture by turning on a principal nilpotent vev for the $SU(N)$ moment map, we find the family of theories (\ref{sing}), where $k=p-b$. This RG flow is allowed only if $p>b$, although the theories (\ref{def1}) exist for any $p>0$. The interpretation of this fact is that for $p<b$ one cannot close the regular puncture completely due to the chiral ring relations on the Higgs branch. We will have more to say about this point below. 

\section{New aspects of Argyres-Douglas theories and their dimensional reductions} \label{sec:newAD}
In this section, we discuss novel aspects of the AD theories and their dimensional reductions to 3d.  We first provide formulae for the $a$ and $c$ central charges of class $\CS$ theories with regular and irregular punctures.  The conformal manifolds of such AD theories are then investigated from the Type IIB geometry perspective.  We then discuss various properties of the Higgs branch, as well as non higgsable sectors.  Subsequently we provide the prescription for constructing linear quivers (with unitary and special unitary gauge groups) coming from dimensional reductions of the AD theories to 3d.  Some explicit examples are provided at the end of this section to illustrate the general principle discussed earlier.

\subsection{Central charges of class $\CS$ theories with regular and irregular punctures}
From the above discussion we see that we can interpret the difference between the $a$ and $c$ central charges of $D_p(SU(N))=(I_{p-N,N},(1^N))$ theory and those of $I_{p-N,N}=(A_{N-1},A_{p-N-1})$ as the contribution of the regular full puncture $(1^N)$.  For Riemann surfaces with regular punctures only, it is known that the contribution of the full puncture is \cite{Chacaltana:2012zy}
\be\label{acpunctures} a= \frac{N^3}{6}-\frac{5N^2}{48}-\frac{N}{16}~;\quad c= \frac{N^3}{6}-\frac{N^2}{12}-\frac{N}{12}.\ee  
However, on spheres with irregular punctures this formula does not hold as can be easily verified by comparing the central charges of 
$D_p(SU(N))$ and $(A_{N-1},A_{p-N-1})$. The explanantion of this discrepancy lies in the value of the $SU(N)$ flavor central charges, as we will now explain. 

In the RG flow from $D_p(SU(N))$ to $(A_{N-1},A_{p-N-1})$ the $U(1)_R$ symmetry is unbroken and descends to the infrared $U(1)_R$ symmetry. The $SU(2)_R$ symmetry is instead broken by the vev for the moment map but this can be fixed: As is well known, any nilpotent vev for a symmetry group $G$ is characterized by the embedding of a $SU(2)_{\rho}$ group in $G$. Both $SU(2)_R$ and $SU(2)_{\rho}$ are broken, but their (anti) diagonal combination is preserved along the RG flow and becomes the $SU(2)_R^{IR}$ symmetry in the infrared and therefore we can compute the $a$ and $c$ central charges via 't Hooft anomaly matching. We only need to subtract the contribution from the Goldstone multiplets.  

In the case of the principal nilpotent vev for $SU(N)$ the $SU(2)$ group is embedded in such a way that the fundamental representation of $SU(N)$ is identified with the $N$-dimensional irreducible representation of $SU(2)$. From this we can determine how the various components of the moment map (which transforms in the adjoint representation of $SU(N)$) transform under $SU(2)$. It turns out that all the components of the moment map apart from the lowest weight states of each $SU(2)$ irreducible representation become Goldstone modes \cite[Section 2.4]{Tachikawa:2015bga} and we should subtract their contribution to the central charges. We can notice that this computation only depends on the global symmetry of the theory and is therefore identical for Argyres-Douglas theories and class $\mathcal{S}$ models with regular punctures only. The anomaly matching equations are 
\begin{align}
\label{match1} \Tr U(1)_R&=48(a_{UV}-c_{UV})= 48(a_{IR}-c_{IR}) + \text{``G.M. contr''}, \\
 \Tr U(1)_R(SU(2)_R^{IR})^2&=4a_{IR}-2c_{IR}+ \text{``G.M. contr''} \nn \\
&= 4a_{UV}-2c_{UV}-I_{\rho}\frac{k_{SU(N)}}{2}~, \label{match2}
\end{align}
where ``G.M. contr'' is the contribution of Goldstone multiplets and in the second equation we used the fact that $SU(2)_R^{IR}=SU(2)_R^{UV}-SU(2)_{\rho}$. We have also used the relation $ \Tr U(1)_R(SU(2)_{\rho})^2=I_{\rho} \Tr U(1)_RSU(N)^2$ where $I_{\rho}$ denotes the embedding index\footnote{If under an embedding $H$ in $G$ of Lie algebras, a representation $\vec{r}$ of $G$ decomposes into $\oplus_i \vec{r}_i$ under $H$, then the embedding index of $H$ in $G$ is $I_{H \hookrightarrow G} = \frac{\sum_i T(\vec{r}_i)}{T(\vec{r})}$, where $T(\vec r)$ denotes the Dynkin index of the representation $\vec r$.  For the principal embedding, the representation $\vec N$ of $SU(N)$, whose Dynkin index is $1/2$, decomposes into the $N$-dimensional representation of $SU(2)$, whose Dynkin index is $\frac{1}{12}(N-1)N(N+1)$, and so the corresponding embedding index is $\frac{N(N^2-1)}{6}$.} of $SU(2)_{\rho}$ in $SU(N)$. In the case of the principal nilpotent vev 
\be I_{\rho}=\frac{N(N^2-1)}{6}.\ee 
From (\ref{match1}) we conclude that the contribution from the regular puncture to $a-c$ is the same as in (\ref{acpunctures}), whereas (\ref{match2}) tells us that the contribution to $2a-c$ depends on the value of $k_{SU(N)}$, which is equal to $2N$ in the conventional case with regular punctures only and is $2N-2\frac{b}{p}$ for $D_p^b(SU(N))$ theories. We therefore conclude that (\ref{acpunctures}) does not hold for $D_p^b(SU(N))$ theories and is replaced by 
\be\label{acpunctNew} a= \frac{N^3}{6}-\frac{5N^2}{48}-\frac{N}{16}-\frac{N(N^2-1)b}{12p};\quad c= \frac{N^3}{6}-\frac{N^2}{12}-\frac{N}{12}-\frac{N(N^2-1)b}{12p}.\ee 
We therefore find that the contribution to $a$ and $c$ is the same as in (\ref{acpunctures}), modulo a shift proportional to the scaling dimension of the coordinate on the sphere ($D(z)=\frac{b}{p}$) times a coefficient which depends on the choice of 6d theory only. In (\ref{acpunctNew}) we give the result for the full puncture but the derivation holds in general. The central charges of the theory $(I_{N,k},Y)$ with $Y$ a generic $SU(N)$ puncture are equal to those of $(A_{N-1},A_{k-1})$ plus 
\be\label{acgeneral} a=a_{Y}+\frac{6I_{\rho_Y}-N(N^2-1)}{12}\frac{N}{N+k};\quad c=c_Y+\frac{6I_{\rho_Y}-N(N^2-1)}{12}\frac{N}{N+k},\ee 
where $a_Y$ and $c_Y$ are the standard contributions from the puncture $Y$ (see \cite[Section 3.3]{Chacaltana:2012zy}) and $I_{\rho_Y}$ is the embedding index of $SU(2)$ in $SU(N)$ associated with the nilpotent vev leading to the puncture $Y$ starting from the full puncture. For example, when $Y$ is trivial $I_{\rho_Y}=(N^3-N)/6$ and when $Y$ is the full puncture $I_{\rho_Y}=0$.

\subsection{Conformal manifolds from Type IIB}

In this subsection we discuss the conformal manifold of Argyres-Douglas theories from the perspective of geometric engineering in Type IIB string theory. It is well known that whenever the spectrum includes a Coulomb branch operator of dimension two there is a corresponding $\mathcal{N}=2$-preserving marginal deformation and therefore a nontrivial conformal manifold. In all known cases $\mathcal{N}=2$  marginal deformations are associated with gauge couplings and there is a ``weak-coupling" cusp in which the SCFT is described by vectormultiplets weakly gauging isolated SCFTs. This is true in particular for the models we are discussing in the present paper. One natural question is therefore to provide explictly a weak-coupling like description for a given theory.  For example we have the relation \be D_4(SU(6))\quad = \quad \left[ D_2(SU(9))\longleftarrow SU(3) \longrightarrow D_2(SU(3)) \right],\ee 
where $\longleftarrow SU(3) \longrightarrow$ denotes the $SU(3)$ gauging, so at some point of the one-dimensional conformal manifold there is a weakly-gauged $SU(3)$ vector multiplet. 

Here we would like to point out that the Type IIB description of Argyres-Douglas SCFTs is a convenient framework to approach this question\footnote{See \cite{Buican:2016arp} for a chiral algebra analysis of $\mathcal{N}=2$ conformal manifolds.}. The reason is that it is easy to describe geometrically in Type IIB the gauging (provided the gauge group is simply-laced) of two SCFTs whenever both theories admit such a description. Let us illustrate this point by considering the gauging of two $D_p(SU(N))$ theories, which is the case of interest for us, although the statement is indeed more general. Consider $D_{p_1}(SU(N))$ and $D_{p_2}(SU(N))$, which are described in Type IIB by the hypersurface singularities 
\be wt+x^N+z^{p_1}+u_{ij}z^ix^j=0;\quad  wt+x^N+z^{p_2}+v_{ij}z^ix^j=0.\ee 
If we want to gauge the diagonal $SU(N)$, the resulting geometry is (see \cite{Tachikawa:2011yr})
\be\label{iibgeom} \frac{1}{z^{p_1}}+u_{ij}\frac{x^j}{z^i}+wt+P_N(x)+z^{p_2}+v_{ij}z^ix^j=0;\quad \Omega_3=\frac{dwdtdxdz}{zdW},\ee
where $P_N=x^N+u_2x^{N-2}+\dots$ Roughly speaking, we have a $SU(N)$ gauge group, described by the corresponding ADE singularity $wt+x^N$, coupled to a matter sector localized at $z=0$ (in this case $D_{p_1}(SU(N))$) and a matter sector localized at $z=\infty$ (for us $D_{p_2}(SU(N))$). 

The procedure outlined above can be exploited to provide a weak coupling like description for all the theories described before with a nontrivial conformal manifold. Let us explain how it works. The simplest case is that of $D_p(SU(N))$ models, for which the holomorphic threeform is already in the canonical form  $\frac{dwdtdxdz}{zdW}$. For $N>2$ we have a nontrivial conformal manifold whenever $\GCD(p,N)>1$. Setting $n=N/\GCD(N,p)$ and $q=p/\GCD(N,p)$ we find that e.g. the term $x^{N-n}z^{q}$ describes a marginal deformation and the geometry can be written as follows: 
\be x^N+\dots+wt+P_{N-n}(x)z^{q}+\dots + z^p=0,\ee 
where $\dots$ stand for other deformation terms. Dividing everything by $z^{q}$ we get 
\be\label{dpiib} \frac{x^N}{z^{q}}+\dots+wt+P_{N-n}(x)+\dots+z^{p-q}=0;\quad \Omega_3=\frac{dwdtdxdz}{zdW}\ee 
where we have suitably rescaled $w$, $t$ and $W$. 

The expression (\ref{dpiib}) is similar to (\ref{iibgeom}) and therefore we conclude that we have a $SU(N-n)$ vector multiplet coupled to two matter sectors. The matter sector at $z=\infty$ is $D_{p-q}(SU(N-n))$. The matter sector at $z=0$ instead is not in general of the $D_p(SU(k))$ type and we call it $\mathcal{D}_{q}(N,N-n)$ as in \cite{Simone:2020}. We therefore have the following description for $D_p(SU(N))$
\be \label{descDpSUn} D_p(SU(N)) \quad = \quad \left[  \mathcal{D}_{q}(N,N-n)\longleftarrow SU(N-n)\longrightarrow D_{p-q}(SU(N-n)) \right]~.\ee 

More generally, we may introduce the $\mathcal{D}^{b}_{q}(N,N-n)$ theory \cite{Simone:2020} in such a way that  
\be D^b_p(SU(N)) \quad = \quad \left[  \mathcal{D}^{b}_{q}(N,N-n)\longleftarrow SU(N-n)\longrightarrow D^b_{p-q}(SU(N-n)) \right]\ee
holds for both $b=N$ (Type I) and $b=N-1$ (Type II).  In this generalisation, we just need to take $n$ to be $n=b/\GCD(b,p)$.  As usual, when $b=N$, we drop the superscript $b$ and simply write $ \mathcal{D}^{N}_{q}(N,N-n)$ as $\mathcal{D}_{q}(N,N-n)$.

The $\mathcal{D}^b_{q}(N,N-n)$ theory has $SU(N)\times U(1)\times SU(N-n)$ global symmetry, which enhances to $SU(2N-n)$ for $q=2$. The SW curves and differential are 
\be x^N+x^{N-n}z^q+\text{defs.}=0;\quad \lambda=x\frac{dz}{z}.\ee 
The allowed deformations ``defs.'' are the terms $u_{ij}x^jz^i$ with $i\leq q$ such that the scaling dimension of $u_{ij}$ is positive. All the terms with $i=0$ or $i=q$ describe mass parameters associated with the global symmetry of the theory.

It is easy to check that $\mathcal{D}^{b}_{q}(N,N-n)$ is an isolated SCFT, without a conformal manifold, whereas in general $D_{p-q}(SU(N-n))$ has marginal couplings, unless $\GCD(N,p)=2$. We find in total $\GCD(N,p)-1$ marginal deformations for $D_p(SU(N))$, with gauge groups (at a specific cusp in the conformal manifold) $SU(N-n), \, SU(N-2n), \, \ldots, \, SU(n)$.  The only exception is when $p$ is a multiple of $N$ since $SU(1)$ is trivial as a gauge group and it should be treated as giving one flavour to the gauge group attached to it. In this case the dimension of the conformal manifold is $N-2$. Analogously, we can easily prove that for $D_p^{N-1}(SU(N))$ theories we always have $\GCD(p,N-1)-1$ marginal couplings and, setting $n=(N-1)/\GCD(p,N-1)$, the corresponding gauge groups are $SU(N-n), \, SU(N-2n), \dots, \, SU(n+1)$.    

\subsubsection*{Theories with only irregular puntures}

The analysis is more tricky for theories with just the irregular puncture and we illustrate the procedure with a Type I example: the $(A_5,A_7)$ theory. The SW curve and differential are 
\be x^6+(u_{10}z+u_{21}xz^2+{\color{blue}u_{43}}x^3z^4)+\dots+z^8=0;\quad \lambda=xdz,\ee 
where we have highlited some specific deformations, which turn out to describe the Coulomb branch operators of a $SU(3)$ vector multiplet and the corresponding marginal coupling (the term in blue). With the redefinition $\tilde{x}=xz$ and dividing everything by $z$ we find \be \frac{\tilde{x}^6}{z^7}+P_3(\tilde{x})+\dots+z^7=0;\quad \lambda=\tilde{x}\frac{dz}{z},\ee 
which can be interpreted (as we have anticipated) as a $SU(3)$ vector multiplet coupled to $D_7(SU(3))$ and to a second SCFT localized at $z=0$. In order to understand what it is, we exploit the identity $(A_5,A_7)=(A_7,A_5)$ (so that the second sector is now localized at $z=\infty$) and rewrite the curve as 
\be x^8+(u_{01}x+u_{12}x^2z+{\color{blue}u_{34}}x^4z^3)+\dots+z^6=0;\quad \lambda=xdz.\ee 
Performing again the change of variables $\tilde{x}=xz$ and multiplying everything by $z$ we find 
\be \frac{\tilde{x}^8}{z^7}+\tilde{x}P_3(\tilde{x})+\dots+z^7=0;\quad \lambda=\tilde{x}\frac{dz}{z}.\ee 
The interpretation we propose is the following: The fact that we have a polynomial of the form $xP_3(x)$ means that the matter sector localized at $z=\infty$ has a $SU(4)$ global symmetry but only a $SU(3)$ subgroup is gauged. The second SCFT is therefore identified with $D_7(SU(4))$. Overall, from these manipulations we find the proposal 
\be \label{A5A7} (A_5,A_7) \quad = \quad \left[  D_7(SU(3)) \longleftarrow SU(3) \longrightarrow D_7(SU(4)) \right]~.\ee 
It is easy to check that the Coulomb branch spectra and $a$, $c$ central charges match. It's also easy to check that the $SU(3)$ beta function is zero using (\ref{fcc}). 

This analysis can be extended to generic $(A_{N-1},A_{k-1})$ models, although in general the two matter sectors are not both of $D_p(SU(N))$. We can illustrate this point by analyzing the $(A_3,A_9)=I_{4,10}$ model, whose curve and differential are 
\be x^4+(u_{30}+{\color{blue}u_{52}}x^2z^2)z^3+\dots +z^{10}=0;\quad \lambda=xdz.\ee 
With the usual change of variable $\tilde{x}=xz$ and dividing everything by $z^3$ we find 
\be \frac{\tilde{x}^4}{z^7}+P_2(\tilde{x})+\dots+z^7=0;\quad \lambda=\tilde{x}\frac{dz}{z},\ee 
from which we infer that there is a $SU(2)$ gauge group and the matter sector at $z=\infty$ is $D_7(SU(2))$. Considering now the relation $(A_3,A_9)=(A_9,A_3)$ we find that, after setting $\tilde{x}=xz$ and multiplying everything by $z^3$, we get the geometry 
\be\label{geom39} \frac{\tilde{x}^{10}}{z^7}+\tilde{x}^3P_2(\tilde{x})+\dots+z^7=0;\quad \lambda=\tilde{x}\frac{dz}{z}.\ee 
Again we are led to interpret the term $\tilde{x}^3P_2(\tilde{x})$ as evidence for a $SU(5)$ flavor symmetry, an $SU(2)$ subgroup of which is gauged. The matter sector at $z=\infty$ now looks like $D_7(SU(5))$, even though we can easily see that some of the Coulomb branch operators of this theory are missing in (\ref{geom39}). More precisely, we would recover the CB spectrum of $D_7(SU(5))$ if we had three operators of dimension $\frac{30}{7}, \frac{25}{7}$ and $\frac{23}{7}$ which are not present in (\ref{geom39}). We can now notice that by partially closing the $SU(5)$ full puncture, labelled by the partition $(1^5)$, to the puncture labelled by the partition $(3,1^2)$, the effect on the CB spectrum is precisely to remove the three unwanted operators\footnote{In order to see this, we recall from \cite{Chacaltana:2010ks} that the puncture $(1^5)$ has the pole structure $\{1,2,3,4\}$ and the puncture $(3,1^2)$ has the pole structure $\{1,2,2,2\}$ where $\{p_k \} = \{p_1, \ldots, p_5\}$ are the orders of the poles in the $k$-differential $\phi_k$ in the Seiberg-Witten differential $\lambda$ such that $\lambda^5 = \lambda^3 \phi_2+\lambda^2 \phi_3+\lambda \phi_4+\phi_5$. Partially closing the puncture $(1^5)$ to $(3,1^2)$, we see that the orders of the poles $(4,3)$ in $(\phi_4, \phi_3)$ become $(2,2)$ in $(\phi_4, \phi_3)$, whereas the orders of the poles in the other $k$-differentials do not change. Recall that the Type IIB geometry of $I_{5,2}$ is $x^5+z^2=0$ and that the Seiberg-Witten differential is $\lambda = x d z$ and so the scaling dimensions of $x$ and $z$ are $\frac{2}{7}$ and $\frac{5}{7}$ respectively.  We see that the three CB operators that have been removed correspond to the coefficients of $z^{-4}$, $z^{-3}$ and $x z^{-3}$; their scaling dimensions are respectively $\frac{10}{7}-4 \left( -\frac{5}{7} \right) = \frac{30}{7}$, $\frac{10}{7}-3 \left( -\frac{5}{7} \right) = \frac{25}{7}$ and $\frac{10}{7}-\frac{2}{7}-3 \left( -\frac{5}{7} \right) = \frac{23}{7}$.}. We therefore identify the second matter sector with $(I_{5,2},(3,1^2))$, where $D_7(SU(5))=(I_{5,2},(1^5))$. Notice that the partial closure of the puncture breaks $SU(5)$ to $SU(2)\times U(1)$ and it is this $SU(2)$ group which is being gauged. We therefore conclude that the $SU(5)$ global symmetry is not actually there. Overall, we find 
\be (A_3,A_9) \quad = \quad \left[ (I_{5,2},(3,1^2))\longleftarrow SU(2)\longrightarrow D_7(SU(2)) \right]~.\ee 
Notice that the gauging is conformal since $D_7(SU(2))$ contributes $12/7$, which is half of \eref{fcc}, to the $SU(2)$ beta function and the required further contribution of $16/7$ to cancel the beta function comes from $(I_{5,2},(3,1^2))$: The contribution from $D_7(SU(5))$ is $30/7$ and the $SU(2)$ preserved by the partial closure of the puncture has embedding index one in $SU(5)$. From $30/7$ we should then subtract the contribution of Goldstone multiplets. The Goldstone modes charged under $SU(2)$ are organized into 4 doublets (these are the non lowest components of 4 triplets of the $SU(2)$ whose embedding in $SU(5)$ defines the nilpotent orbit\footnote{Explicitly, the embedding of $SU(2)_t$ into $SU(5)$ associated with the partition $(3,1^2)$, whose symmetry is $U(1)_q \times SU(2)_x$, can be realised as follows.  The character of the fundamental representation of $SU(5)$ can be written as $q(t^2+1+t^{-2})+(x+ x^{-1})=q[2]_t + [1]_x$, where $t$ is the fugacity for the $SU(2)_t$ that is embedded into $SU(5)$ and $(q, x)$ are the fugacities for the $U(1)_q \times SU(2)_x$ symmetry of $(3,1^2)$.  Here $[h]_v$ denotes the character of $SU(2)$ representation $[h]$, with dimension $h+1$, written in the fugacity $v$.  The character of the adjoint of representation of $SU(5)$ can then be written as $[4]_t +[2]_t \left( 1+ q^{-1}[1]_x +q [1]_x \right) + (1+ [2]_x)$. The four triplets mentioned above corresponds to the representations $[2]_t$ that multiply $q^{-1}[1]_x +q [1]_x$ (indeed, the representation $q^{-1}[1]_x +q [1]_x$ is 4 dimensional). On the other hand, the representation $[4]_t$ and the other $[2]_t$ do not contribute to the anomaly matching \eref{match2}, since they do not multiply any $x$-dependent character and thus are invariant under $SU(2)_x$. The non-lowest weights in the aforementioned four $[2]_t$'s correspond to the Goldstone modes in question.}). Since their fermions have charge $-1$ under $U(1)_r$, their contribution to the beta function is $4\times 1/2 \times (-1)^2=2$ and subtracting this quantity from $30/7$ we find precisely $16/7$ as expected.

We can now approach the general case of $(A_{N-1},A_{k-1})$ theories. We assume without loss of generality that $k>N$ and set 
\be\label{defgcd} \GCD(N,k)=m;\quad N=mn;\quad k=m \kappa~,\ee 
where $n$ and $\kappa$ are coprime. The curve can be written as 
\be x^N+(u_{k-\kappa-n,0}+\dots+{\color{blue}u_{k-\kappa,n}}x^nz^n)z^{k-\kappa-n}+\dots+z^k=0,\ee 
and with the same manipulations as in the previous examples we find the duality 
\be\label{SdualAD} (A_{N-1},A_{k-1}) \quad = \quad \left[ (I_{N-n,k-\kappa},(k-\kappa-n,1^n))\longleftarrow SU(n)\longrightarrow D_{\kappa+n}(SU(n)) \right].\ee 
We can check as before that the gauging is conformal and the $D_{\kappa+n}(SU(n))$ sector is isolated. The rest of the conformal manifold for $m>2$ is in $ (I_{N-n,k-\kappa},(k-\kappa-n,1^n))$. This duality can also be exploited to explore the Higgs branch and we would like to explain this next. 

\subsection{Properties of the Higgs branch}
Let us now discuss some properties of the Higgs branch of Argyres-Douglas theories.  The main goal is to show that, at a generic point of the Higgs branch, the theory can be reduced to a set of hypermultiplets plus a collection of SCFTs with empty Higgs branch.  We shall refer to the latter as ``non higgsable theories'', where upon reduction to 3d, we expect that a non higgsable theory of rank $r$ becomes $r$ twisted hypermultiplets.

Starting from (\ref{SdualAD}) we perform the following move: We turn on a principal nilpotent vev for the $SU(n)$ moment map of $D_{\kappa+n}(SU(n))$, which has the effect of closing the puncture and reduce $D_{\kappa+n}(SU(n))$ to $(A_{n-1},A_{\kappa-1})$. The F-term coming from the $SU(n)$ gauging implies that the same vev should be given to the second sector, therefore reducing the regular puncture $(k-\kappa-n,1^n)$ to $(k-\kappa-n,n)$. This operation indeed fully breaks the $SU(n)$ gauge symmetry. Now we move on the Higgs branch of the theory by further closing the puncture to $(k-\kappa)$. 

From this discussion we therefore conclude that by partially moving on the Higgs branch of the $(A_{N-1},A_{k-1})$ theory we find a low-energy theory given by two decoupled sectors: one is $(A_{n-1},A_{\kappa-1})$ and the other $(A_{N-n-1},A_{k-\kappa-1})$. There is also a free hypermultiplet parametrizing the position on the Higgs branch. Notice that $(A_{n-1},A_{\kappa-1})$ has no mass parameters and trivial Higgs branch. As we will see later, upon dimensional reduction this theory becomes a collection of free twisted hypermultiplets. 

Now we can just proceed recursively and conclude that at a generic point of the Higgs branch the $(A_{N-1},A_{k-1})=(A_{mn-1},A_{m\kappa-1})$ theory reduces to $m-1$ hypermultiplets parametrizing the Higgs branch and $m$ copies of $(A_{n-1},A_{\kappa-1})$: \be\label{hbflow} (A_{N-1},A_{k-1})\longrightarrow \text{$m-1$ hypermultiplets}+ \text{$m$ copies of}\; (A_{n-1},A_{\kappa-1}).\ee 
We can check using (\ref{ccformula}) and (\ref{acpunctures}) that the quantity $c-a$  is the same in both descriptions, as expected for a Higgs branch flow. In proving this we exploit the identity 
\be \sum_{j=1}^{N-1}\left\{\frac{j(k+N)}{N}\right\}\left(1-\left\{\frac{j(k+N)}{N}\right\}\right)=m\sum_{j=1}^{n-1}\left\{\frac{j(\kappa+n)}{n}\right\}\left(1-\left\{\frac{j(\kappa+n)}{n}\right\}\right).\ee
Using formula \eref{ccformula} and subtracting the contribution $24(c-a)[\eref{acpunctNew}]$ of the full puncture, we then obtain
\begin{align}
~24(c-a)[(A_{mn-1},A_{m\kappa-1})] &= \frac{1}{2} (-2+m+mn) -\frac{3n}{\kappa+n} m S ~, \label{24cma1} \\
~24(c-a)[(A_{n-1},A_{\kappa-1})] &= \frac{1}{2} (-1+n) -\frac{3n}{\kappa+n} S~, \label{24cma2}
\end{align}
where $S$ is defined as
\bes{
S= \sum_{j=1}^{n-1}\left\{\frac{j(\kappa+n)}{n}\right\}\left(1-\left\{\frac{j(\kappa+n)}{n}\right\}\right)~.
}
Since $\eref{24cma1} = (m-1)+ m\eref{24cma2}$, we learn that the dimension of the Higgs branch is $m-1$ and at a generic point we have $m$ copies of the non higgsable theory $(A_{n-1},A_{\kappa-1})$, where $n$ and $\kappa$ are coprime. Using the guess that the dimensional reduction of the non higgsable theory is a collection of free twisted hypermultiplets, we expect that the 3d mirror includes a collection of $\frac{1}{2}m(n-1)(\kappa-1)$ free hypermultiplets. As we will see, the 3d mirrors proposed below are perfectly consistent with this expectation; see \eref{freehypTypeIp>N} with $\kappa = q-n$.

We can easily generalize to the Type II case. The curve now takes the form 
\be\label{thcurve} x^N+(u_{k-\kappa-n-1,0}+\dots+{\color{blue}u_{k-\kappa,n+1}}x^{n+1}z^{n+1})z^{k-\kappa-n-1}+\dots+xz^k=0,\ee
where \be\label{defgcd2} \GCD(N-1,k)=m;\quad N-1=mn;\quad k=m \kappa.\ee 
In this case the S-dual description is 
\be\label{SdualAD2}  (I_{N-n-1,k-\kappa},(k-\kappa-n-1,1^{n+1}))\longleftarrow SU(n+1)\longrightarrow D_{\kappa+n}^n(SU(n+1)).\ee 
Again we can easily check that the gauging is again conformal. The $D_{\kappa+n}^n(SU(n+1))$ sector is isolated but, contrary to the Type I case, it has a nontrivial one dimensional Higgs branch. The low-energy theory at a generic point of the Higgs branch includes a free hypermultiplet and the $I_{n,\kappa}$ theory. The sector $ (I_{N-n-1,k-\kappa},(k-\kappa-n-1,1^{n+1}))$ can be treated as before. We conclude that at a generic point of the Higgs branch of (\ref{thcurve}) the low-energy theory is given by $m$ hypermultiplets and $m$ copies of the $I_{n,\kappa}=(A_{n-1},A_{\kappa-1})$ theory:
\be\label{hbflow2} \text{$m$ hypermultiplets}+ \text{$m$ copies of}\; (A_{n-1},A_{\kappa-1}).\ee 
The dimension of the Higgs branch is therefore $m$ and upon dimensional reduction we expect to find $\frac{1}{2}m(n-1)(\kappa-1)$ free twisted hypermultiplets. This again agrees with the 3d mirrors we will find later; see \eref{freehypTypeIIp>Nm1} with $\kappa = q-n$.

\subsection{The lagrangian for $D_p^b(SU(N))$ theories in 3d} \label{sec:DpbSUn3d}

One of the crucial observations for us is that the dimensional reduction of $D_p^b(SU(N))$ theories is always a lagrangian theory. It is a linear quiver with unitary and special unitary nodes which can be determined as follows (see \cite{Simone:2020}).

We start from (\ref{def1}) and collect all terms with the same power of $z$. The number of colors $n_i$ of the $i$-th gauge group (associated with all the terms proportional to $z^i$) is the largest value of $j$ appearing in (\ref{def1}) (i.e. the exponent of the largest power of $x$)\footnote{In case the coefficient of the term with $j=N-1$ has positive dimension, the number of colours for the corresponding gauge node in 3d is $N-1$, even though the term can be removed with a change of coordinates as we have explained before.}. If the corresponding parameter is dimensionless (meaning the 4d SCFT has a nontrivial conformal manifold) the gauge group is special unitary, otherwise it is unitary. The first gauge group is always coupled to $N$ fundamentals and, in case there is a dimensionless parameter $u_{i1}$, the interpretation is that the $(i-1)$-th gauge group is coupled to a fundamental flavor. In particular, for $b=N-1$ the quiver always has $p-1$ gauge groups, $N$ flavors on one side and one flavor on the other. 

The idea for the derivation of this result is to exploit the decomposition  \eref{descDpSUn}: 
\be D_p(SU(N))\,\, =\,\, \left[ \mathcal{D}_{q}(N,N-n))\longleftarrow SU(N-n) \longrightarrow D_{p-q}(SU(N-n)) \right],\ee 
from which we infer that the dimensional reduction of $D_p(SU(N))$ is an $SU(N)$ gauging of the dimensional reduction of $D_{p-q}(SU(N-n))$ and $\mathcal{D}_{q}(N,N-n))$. The problem is therefore reduced to determining the dimensional reduction of $\mathcal{D}_{q}(N,N-n))$, from which we can then  obtain the linear quiver coming from 3d reduction of a general $D_p(SU(N))$ theory in a recursive way. 

The linear quiver coming from the 3d reduction of the $\mathcal{D}_{q}(N,N-n)$ theory turns out to be:
\bes{ \label{3dCD}
(\mathcal{D}_{q}(N,N-n))_{3d} : \quad [N] -U(m_1) - U(m_2)- \ldots -U(m_{q-1}) - [N-n]
}
where the number of gauge groups is $q-1$ and
\bes{
m_j = N-\lceil j n/q \rceil~.
} 
For $n=1$ and $q=2$ the theory $\mathcal{D}_{2}(N,N-1)\equiv D_2(SU(2N-1))$ can be realized as the effective low-energy theory at the origin of the Coulomb branch of $SU(N)$ SQCD with $2N-1$ flavors, for which we turn on a mass for all the flavors proportional to the SQCD scale of the theory. Exploiting now the fact that in 3d the Coulomb branch of  $SU(N)$ SQCD with $2N-1$ massive flavors (all with the same mass) is identical to that of $U(N-1)$ SQCD with $2N-1$ flavors (see \cite{Simone:2020}) we conclude that the dimensional reduction of $\mathcal{D}_{2}(N,N-1)$ is equivalent to $U(N-1)$ SQCD with $2N-1$ flavors, in agreement with (\ref{3dCD}) and with the known 3d mirror derived in \cite{Xie:2016uqq}. The general statement (\ref{3dCD}) follows by generalizing the above argument: The $\mathcal{D}_{q}(N,N-n)$ theory is the low-energy effective theory at a singular point in the moduli space of a linear quiver with $q-1$ SU gauge groups, which reduce to unitary groups upon dimensional reduction (see \cite{Simone:2020} for further details).

Let's ilustrate the algorithm with the example of $D_{6}(SU(4))$: The fully deformed curve is 
\be\begin{array}{l} x^4+u_{02}x^2+u_{01}x+u_{00}+z(u_{10}+u_{11}x+u_{12}x^2+{\color{red}u_{13}x^3})+z^2(u_{20}+u_{21}x+u_{22}x^2)+ \\
z^3(u_{30}+u_{31}x+{\color{blue}u_{32}}x^2)+z^4(u_{40}+u_{41}x)+z^6=0\end{array}\ee
The term in red is redundant and can be traded for $z^5u_{50}$, but we keep it for the purpose of specifying the lagrangian in 3d. The coefficient in blue is a marginal coupling. From our algorithm we therefore find 
\be\label{lagr3d} (D_{6}(SU(4)))_{3d}: \quad [4]-U(3)-U(2)-SU(2)-U(1)\ee
We will give several other examples below.

Notice that the linear quiver we have just described has by construction the same rank as the parent $D_p^b(SU(N))$ theory and we can also provide an explicit map between the Coulomb branch operators in 4d and 3d: Given a parameter $u_{ij}$ describing the vev of a CB operator in 4d, the corresponding operator in 3d is the Casimir invariant of degree $n_i-j$ of the $i$-th gauge group $\Tr\Phi_i^{n_i-j}$.

In some cases, a certain unitary gauge group in the linear quiver obtained by the above procedure may be underbalanced. In which case such a gauge group can be replaced by the one with lower rank using the duality \cite{Gaiotto:2008ak}
\bes{ \label{dualunderb}
&\text{$U(N)$ with $2N-1$ flavours} \\ 
&\longleftrightarrow \,\,  \text{$U(N-1)$ with $2N-1$ flavours + one twisted hypermultiplet}~. 
}
For example, using the above procedure, one obtains the 3d reduction of the $D_7(SU(4))$ theory as 
\bes{
(D_7(SU(4)))_{3d}:  \qquad [4]-U(3)-U(2)-U(2)-U(1)-U(1)~.
}
This quiver can be dualised and the result is
\bes{
(D_7(SU(4)))_{3d}:  \quad [4]-U(3)-U(2)-U(1) \,\, \text{+ 3 twisted hypermultiplets}~.
}
Indeed, this 3d thoery flows to the $T(SU(4)$ theory with 3 twisted hypermultiplets in the IR.

We present explicit descriptions of the linear quivers (after dualising and neglecting all possible twisted hypermultiplets) for the reduction to 3d of $D_p(SU(N))$ in \eref{linearquiv3dDpSUn} for $p\geq N$ and in \eref{linearquiv3dDpSUnp<n} for $p\leq N$.  The linear quivers for the reduction of the $D^{N-1}_p(SU(N))$ are depicted in \eref{linearquiv3dDpSUnTypeII} for $p\geq N-1$ and in \eref{linearquiv3dDpSUnp<nII} for $p\leq N-1$.

\subsection{Further examples}
\subsubsection*{Example: $D_9(SU(3))$}
Let us consider the $D_9(SU(3))$ theory.  This is a rank 7 theory, with the Coulomb branch spectrum $\left(\frac{4}{3},\frac{4}{3},\frac{5}{3},\frac{5}{3},2,\frac{7}{3},\frac{8}{3} \right)$.  We propose that this theory can be constructed as an $SU(2)$ gauging of the following theories
\bes{ \label{systemD9SU3}
D_9(SU(3)) \quad = \quad \left[ \CD^3_{3,2} \,\, \longleftarrow SU(2) \longrightarrow D_6(SU(2)) \right] \,\, 
}
where the $SU(2)$ gauge group arises because of the 4d marginal coupling which can be seen from the blue terms in the following Type IIB geometry:
\bes{
x^3 + z^9 +  {\blue c x^2 z^3 }=0~.
}
The $\CD^3_{3,2}$ theory is associated with the geometry $x^3 + x^2 z^3=0$, and the 3d reduction of this theory is given by \eref{3dCD}:
\bes{
[3]-2-2-[2]
}
On the other hand, the $D_6(SU(2))$ theory\footnote{Note that the $D_6(SU(2))$ theory can be identified with the $(A_1, D_6)$ theory whose Coulomb branch spectrum is $\left(\frac{4}{3},\frac{5}{3}\right)$.} is associated with the geometry $x^3 z^{-3} +z^6 +c x^2=0$, and the 3d reduction of this theory is
\bes{
[2]-1-1-[1]
}
Therefore the 3d reduction of \eref{systemD9SU3} is
\bes{ \label{SU2gaugingD9SU3}
(D_9(SU(3)))_{3d}: \quad  [3]-2-2-[2] \,\, \longleftarrow SU(2) \longrightarrow \,\, [2]-1-1-[1]
}
The Higgs branch of this theory is 5, in agreement with the Higgs branch of the $D_9(SU(3))$ theory, and the Coulomb branch of this theory is 7, in agreement with the rank of the $D_9(SU(3))$ theory.

The mirror of the left quiver in \eref{SU2gaugingD9SU3} can be written as
\bes{
\begin{array}{lll}
1-&2-&2-1\\
&\, | &\, | \\
&\! [2] &\! [1]
\end{array}
\qquad = \qquad 
\begin{tikzpicture}[baseline=0, font=\small]
\tikzstyle{every node}=[minimum size=0.5cm]
\node[draw, circle] (c1) at (-2,0) {$1$};
\node[draw, circle] (c2) at (-1,0) {$2$};
\node[draw, circle] (c3) at (1,0)  {$2$}; 
\node[draw, circle] (c4) at (2,0)  {$1$}; 
\node[draw, circle] (f1) at (0,1)  {$1$}; 
\draw[draw, solid] (c1)--(c2)--(c3)--(c4);
\draw[draw, solid] (f1)--(c2);
\draw[double, thick] (c3)--(f1);
\end{tikzpicture}
}
whereas the quiver on the right in \eref{SU2gaugingD9SU3} is self-mirror and can be written as
\bes{
[2]-1-1-[1] \qquad = \qquad 
\begin{tikzpicture}[baseline=0, font=\footnotesize]
\tikzstyle{every node}=[minimum size=0.2cm]
\node[draw, circle] (c1) at (0,1) {$1$};
\node[draw, circle] (c2) at (1,0) {$1$};
\node[draw, circle] (c3) at (-1,0) {$1$};
\draw (c3)--(c1);
\draw[double, thick] (c1)--(c2);
\draw (c2)--(c3);
\end{tikzpicture}
}
The $SU(2)$ gauging in \eref{SU2gaugingD9SU3} can be viewed in the mirror theory as discussed in \cite{Xie:2016uqq}.  The result is
\bes{
\begin{tikzpicture}[baseline=0, font=\small]
\tikzstyle{every node}=[minimum size=0.5cm]
\node[draw, circle] (c1) at (-2,0) {$1$};
\node[draw, circle] (c2) at (-1,0) {$2$};
\node[draw, circle, thick, red] (c3) at (1,0)  {$2$}; 
\node[draw, circle, thick, red] (c4) at (2,0)  {$1$}; 
\node[draw, circle] (f1) at (0,1)  {$1$}; 
\draw[draw, solid] (c1)--(c2)--(c3)--(c4);
\draw[draw, solid] (f1)--(c2);
\draw[double, thick] (c3)--(f1);
\end{tikzpicture}
\qquad + \qquad
\begin{tikzpicture}[baseline=0, font=\footnotesize]
\tikzstyle{every node}=[minimum size=0.2cm]
\node[draw, circle] (c1) at (0,1) {$1$};
\node[draw, circle] (c2) at (1,0) {$1$};
\node[draw, circle, thick, red] (c3) at (-1,0) {$1$};
\draw (c3)--(c1);
\draw[double, thick] (c1)--(c2);
\draw (c2)--(c3);
\end{tikzpicture}
}
where the red nodes are fused together.  As a result, the mirror theory associated with the $D_9(SU(3))$ is
\bes{ \label{mirrorD9SU3}
\begin{tikzpicture}[baseline=0, font=\footnotesize]
\tikzstyle{every node}=[minimum size=0.2cm]
\node[draw, circle] (c1) at (0,1) {$1$};
\node[draw, circle] (c2) at (1,0) {$1$};
\node[draw, circle] (c3) at (0,-1) {$1$};
\node[draw, circle] (c4) at (-1,0) {$2$};
\node[draw, circle] (t1) at (-3,0) {$1$};
\draw[double, thick] (c1)--(c2);
\draw[double, thick] (c2)--(c3);
\draw (c3)--(c4);
\draw (c4)--(c1);
\draw[double, thick] (c1)--(c3);
\draw (c2)--(c4);
\draw (c4)--(t1);
\end{tikzpicture}
}
This is in agreement with the discussion in \cite{Xie:2012hs, Xie:2013jc}.  The Coulomb branch of this theory is 5 quaternionic dimensional, in agreement with the Higgs branch of the $D_9(SU(3))$ theory.  The Coulomb branch symmetry is $SU(3) \times U(1)^2$, where the $SU(3)$ factor comes from the balanced nodes in the $1-2$ tail and the factor $U(1)^2$ comes from the three overbalanced $U(1)$ gauge nodes, with an overall $U(1)$ decoupled.  The Higgs branch of this theory is 7 quaternionic dimensional, in agreement with the rank of the $D_9(SU(3))$ theory.   The Higgs branch symmetry of \eref{mirrorD9SU3} is $SU(2)^3 \times U(1)^3$.

The $I_{6,3}=(A_5, A_2)$ theory can be obtain from the $D_{9}[SU(3)]$ theory by closing the full $SU(3)$ puncture $[1^3]$. The mirror theory associated with the $(A_5, A_2)$ theory can thus be obtained by decoupling the tail $1-2$ in \eref{mirrD14SU8}:
\bes{ \label{mirrorD9SU3}
(A_5, A_2)_{\text{3d mirr}}: \quad
\begin{tikzpicture}[baseline=0, font=\footnotesize]
\tikzstyle{every node}=[minimum size=0.2cm]
\node[draw, circle] (c1) at (0,1-0.5) {$1$};
\node[draw, circle] (c2) at (1,0-0.5) {$1$};
\node[draw, circle] (c3) at (-1,0-0.5) {$1$};
\draw[double, very thick] (c1)--(c2)--(c3)--(c1);
\end{tikzpicture}
}
This is in agreement with the result presented in \cite{Xie:2012hs, Xie:2013jc} (see also \cite{DelZotto:2014kka, Dey:2020hfe}). The Higgs branch of this theory is $6-2=4$ quaternionic dimensional, in agreement with the fact that the $(A_5, A_2)$ theory is a rank 4 theory\footnote{The Coulomb branch spectrum is $\left( \frac{4}{3},\frac{4}{3},\frac{5}{3},2 \right)$.}. The Higgs branch symmetry is $SU(2)^3 \times U(1)$. The Coulomb branch of this theory is 2 quaternionic dimensional; this is in agreement with the Higgs branch dimension $24(c-a) = 24 \left( \frac{1}{5}-\frac{7}{60} \right) = 2$ of the $(A_5, A_2)$ theory.  The Coulomb branch symmetry is $U(1)^2$. 

\subsubsection*{Example: $D_{14}(SU(8))$}
This theory has rank $45$ and the value of $24(c-a)$ is $209/7 \approx 29.8571$.  We propose that this theory can be constructed as an $SU(4)$ gauging of the following theories
\bes{ \label{systemD14SU8}
D_{14}(SU(8)) \quad = \quad \left[ \CD^7_{8,4} \,\, \longleftarrow SU(4) \longrightarrow D_7(SU(4)) \right] \,\, 
}
where the $SU(4)$ gauge group arises because of the 4d marginal coupling which can be seen from the blue terms in the following Type IIB geometry:
\bes{
x^8 + z^{14} +  {\blue c x^4 z^7 }=0~.
}
The $\CD^7_{8,4}$ theory is associated with the geometry $x^8+x^4 z^7=0$, and the 3d reduction is given by \eref{3dCD}:
\bes{
[8]-7-6-6-5-5-4-[4]
}
On the other hand, the $D_7(SU(4))$ theory is associated with the geometry $x^8z^{-7}+z^7+cx^4=0$, and the 3d reduction gives
\bes{
[4]-3-2-2-1-1
}
The reduction to $3d$ of \eref{systemD14SU8} is therefore
\bes{
[8]-7-6-6-5-5-4-[4] \,\, \longleftarrow SU(4) \,\, \longrightarrow [4]-3-2-2-1-1
}
Dualising the underbalanced unitary gauge nodes using \eref{dualunderb}, we obtain the following theory
\bes{ \label{gauge D7SU4andD784}
(D_{14}(SU(8)))_{3d}:  \quad &[8]-7-6-5-4-4-4-[4] \,\, \longleftarrow SU(4) \,\, \longrightarrow [4]-3-2-1 \\
&+\,\, \text{6 twisted hypers}~.
}
The Higgs branch of the quiver in the first line of \eref{gauge D7SU4andD784} is 29 quaternionic dimensional.  Note that the difference between this quantity and the value of $24(c-a)$ of the $D_{14}(SU(8))$ is due to the non higgsable SCFTs.  We shall discuss this later in \eref{deltaH814}.  The Coulomb branch of the quiver in the first line of \eref{gauge D7SU4andD784} is 39 quaternionic dimensional.  Adding to the latter 6, which is the contribution of the twisted hypermultiplets, we obtain 45, which is the rank of the $D_{14}(SU(8))$ theory.

The quiver on the left of the arrow in \eref{gauge D7SU4andD784} corresponds to the $T^{[6^4,1^8]}_{[5^4, 4^3]}(SU(32))$ theory, and its mirror theory is $T^{[5^4, 4^3]}_{[6^4,1^8]}(SU(32))$, whose quiver is
\bes{
\begin{array}{lll}
1-2-3-&4-&7-6-5-4-3-2-1\\
          &\,|   &\,|  \\
          &\! [3] &\! [4]
\end{array}
}
whereas the quiver on the right of the arrow in \eref{gauge D7SU4andD784} is the $T(SU(4))$ theory, which is self-mirror:
\bes{ \label{TSU4}
[4]-3-2-1
}

To obtain the mirror theory of the system \eref{gauge D7SU4andD784}, we fuse the tail $3-2-1$ of the two theories together.  If the gauge group in \eref{gauge D7SU4andD784} were $U(4)$ instead of $SU(4)$, we would obtain the theory
\bes{ \label{ifU4}
\begin{array}{lll}
&{\blue [4]}-&7-6-5-4-3-2-1\\
          &\,|   &\,|  \\
          &\! [3] &\! [4]          \\
\end{array}
}
together with $6$ free hypermultiplets.
However, since the gauge group in question is $SU(4)$, we gauge the $U(1)$ inside the flavour symmetry $U(4)$ indicated in blue in the above quiver.  We rewrite the $T(SU(4))$ theory \eref{TSU4} as $1 \superequiv 3-2-1$, where $\superequiv$ denotes four bifundamental hypermultiplets.  Upon fusing the quiver tail, we obtain the mirror theory associated with the $D_{14}[SU(8)]$ theory:
\bes{ \label{mirrD14SU8}
(D_{14}[SU(8)])_{\text{3d mirr}}: \quad 
&\begin{array}{ll}
[12]-1 \superequiv &\, 7-6-5-4-3-2-1 \\
 &\, | \\
 &\! [4] \\
& 
\end{array} \quad
\text{+ 6 hypers} \\
& \text{or equivalently} \\
&\begin{tikzpicture}[baseline=0, font=\small]
\tikzstyle{every node}=[minimum size=0.5cm]
\node[draw, circle] (c2) at (-1,0) {$1$};
\node[draw, circle] (c3) at (0,0) {$7$};
\node[draw, circle] (c4) at (1,0)  {$6$}; 
\node[draw=none] (c5) at (2,0)  {$\cdots$}; 
\node[draw, circle] (c6) at (3,0)  {$2$}; 
\node[draw, circle] (c7) at (4,0)  {$1$}; 
\node[draw, circle] (f1) at (-0.5,1)  {$1$}; 
\draw[draw, solid, blue, very thick] (c2)--(c3) node[midway,below] {$4$};
\draw[draw, solid] (c3)--(c4)--(c5)--(c6)--(c7);
\draw[draw, solid] (f1)--(c2);
\draw[very thick, blue] (c2)--(f1) node at (-1,0.6)  {\footnotesize $12$};
\draw[very thick, blue] (c3)--(f1) node at (0,0.6)  {\footnotesize $4$};
\end{tikzpicture} \qquad \quad \text{+ 6 hypers}
}
where the 12 hypermultiplets in $[3]-[4]$ of \eref{ifU4} becomes  $[12]-1$ in the above quiver.  

The Coulomb branch symmetry of this theory is $SU(8)\times U(1)$, where the factor $SU(8)$ comes from the tail $7-6-5-\ldots-1$ and the $U(1)$ factor comes from the overbalanced gauge node $1$ next to $[12]$.  This Coulomb branch symmetry matches with the global symmetry of the $D_{14}[SU(8)]$ theory in 4d, as it should be.  The Higgs branch symmetry of this quiver, neglecting the contribution of the 6 free hypermultiplets, is $SU(4)^2 \times SU(12) \times U(1)$.  This matches with the Coulomb branch symmetry of the quiver in the first line of \eref{gauge D7SU4andD784}, as explained in Appendix \ref{app:SUenhanced}.

The $I_{8, 6}= (A_7, A_5)$ theory can be obtain from the $D_{14}[SU(8)]$ theory by closing the full $SU(8)$ puncture $[1^8]$. The mirror theory associated with the $(A_7, A_5)$ theory can thus be obtained by decoupling the tail $7-6-5-4-3-2-1$ in \eref{mirrD14SU8}:
\bes{ \label{mirrA7A5}
(A_7, A_5)_{\text{3d mirr}}: \quad (1)-[12] \,\, \text{+ 6 hypers}~.
}
The Coulomb branch symmetry is $U(1)$ and the Higgs branch symmetry, neglecting the contribution of the six free hypermultiplets, is $SU(12)$.

\section{Mirror theories for $D_p(SU(N))$ and $I_{p-N,N}$ with $p \geq N$} \label{sec:DpSUNpgeqN}
Let us write
\bes{ \label{defabc}
\GCD(p,N)=m~, \quad N = m n, \quad p =m q~,
}
where $n$ and $q$ are coprime.  The mirror theory associated with the $D_p(SU(N))$ theory can be constructed as follows:
\ben
\item Start with a complete graph with $m$ vertices such that each vertex corresponds to a $U(1)$ gauge group and each edge has multiplicity 
\bes{ \label{multgraphDpSUNp>N} n(q-n)  = \frac{1}{m^2}N(p-N)~. }
The multiplicity of the edge is equal to the number of bifundamental hypermultiplet between the corresponding pair of gauge groups. 
\item Construct the tail $(N-1)-(N-2)-\cdots-2-1$, where each node with label $\ell$ is the gauge group $U(\ell)$ and $-$ denotes a bifundamental hypermultiplet.
\item Connect the $(N-1)$ node of tail to the complete graph such that the node $U(N-1)$ of the tail is connected to each $U(1)$ node in the complete graph by the edges, each with equal multiplicity $n= \frac{N}{m}$.  Note that this quiver has an overall $U(1)$ that needs to be decoupled.  \label{Step3}
\item The number of free hypermultiplets is equal to \be \label{freehypTypeIp>N} H_{\text{free}} = \frac{1}{2}m(n-1)(q-n-1)=\frac{(N-m)(p-N-m)}{2m}~.\ee \label{Step4}
\item The quiver obtained in Step \ref{Step3}, together with the free hypermultiplets in Step \ref{Step4}, defines a mirror theory for the circle reduction of the $D_p(SU(N))$ with $p\geq N$.
\een

The mirror theory for the $I_{p-N,N} = (A_{p-N-1}, A_{N-1})$ theory can be obtained from the above quiver by decoupling the tail $(N-1)-(N-2)-\cdots-2-1$.  Explicitly, it consists of a complete graph with $m$ $U(1)$ nodes such that each edge has multiplicity $\frac{1}{m^2}N(p-N)$, together with $H_{\text{free}}$ free hypermultiplets.

\subsubsection*{The number of free hypermultiplets}
The number of free hypermultiplets can be computed as follows. The rank of the $D_p(SU(N))$ theory is
\bes{
\mathrm{rank}(D_p(SU(N))) = \frac{1}{2} p (N-1) -\frac{1}{2} (N-2+m)~.
}
whereas the quaternionic dimension of the Higgs branch of the quiver is
\bes{
&\frac{1}{2} m(m-1)(n q-n^2) + n m(N-1) -\frac{1}{2}N(N-1) -m +1 
\\&= (1-m)+ \frac{1}{2m} N(N-p) +\frac{1}{2}N(p-1)~.
}
The difference between the two quantities is $H_{\text{free}}=\frac{(N-m)(p-N-m)}{2m}$, which is the number of free hypermultiplets as stated above.

\subsection{Examples}\label{secmir41}
When $\GCD(p,N)=1$, the mirror theory for the $D_p(SU(N))$ theory is 
\bes{ \label{gcd1}
(D_p(SU(N)))^{\GCD(p,N)=1}_{\text{3d mirr}}: \qquad &1 \,\, {\begin{tikzpicture}[baseline=0] \draw[blue, thick] (0,0.1) --(0.5,0.1) node[midway,above] {\blue \scriptsize $N$}; \end{tikzpicture}} \,\,(N-1)-(N-2) - \cdots-2 -1 \\
& \text{+ $\frac{(N-1)(p-N-1)}{2}$ hypers}
}
where the blue line with the label $N$ denotes $N$ bifundamental hypermultiplets.  The quiver depicted in the first line can actually be identified with that of the $T(SU(N))$ theory, which is self-mirror, and the Higgs and Coulomb branch symmetries are both $SU(N)$.  Upon decoupling an overall $U(1)$ from the leftmost $U(1)$ gauge node, we can rewrite it as 
\bes{  [N]-(N-1)-(N-2)-\cdots-2-1}
Upon decoupling the tail $(N-1)-(N-2)-\cdots-2-1$, we obtain the mirror theory for the $I_{p-N, N} = (A_{p-N-1}, A_{N-1})$ as follows:
\bes{
(I_{p-N, \, N})^{\GCD(p,N)=1}_{\text{3d mirr}}:  \qquad  \text{$\frac{1}{2}(N-1)(p-N-1)$ free hypermultiplets}
}

When $\GCD(p,N)=2$, we have the following mirror theories:
\bes{
(D_p(SU(N)))^{\GCD(p,N)=2}_{\text{3d mirr}}: \qquad\,\,
&\scalebox{0.9}{
\begin{tikzpicture}[baseline=0, font=\small]
\tikzstyle{every node}=[minimum size=0.5cm]
\node[draw, circle] (c2) at (-2,0) {$1$};
\node[draw, circle] (c3) at (0,0) {} node[draw=none] at (0,-0.5) {\footnotesize $N-1$};
\node[draw, circle] (c4) at (1,0) {} node[draw=none] at (1,-0.5) {\footnotesize $N-2$}; 
\node[draw=none] (c5) at (2,0)  {$\cdots$}; 
\node[draw, circle] (c6) at (3,0)  {$2$}; 
\node[draw, circle] (c7) at (4,0)  {$1$}; 
\node[draw, circle] (f1) at (-1,1)  {$1$}; 
\draw[draw, solid, red, very thick] (c2)--(c3) node[midway,below] {$n$};
\draw[draw, solid] (c3)--(c4)--(c5)--(c6)--(c7);
\draw[draw, solid] (f1)--(c2);
\draw[very thick, blue] (c2)--(f1) node at (-2.2,0.6)  {\footnotesize $n(q-n)$};
\draw[very thick, red] (c3)--(f1) node at (-0.2,0.6)  {\footnotesize $n$};
\end{tikzpicture}} \\
& \text{+ $\frac{1}{4}(N-2)(p-N-2)$ hypers} \\
(I_{p-N, \, N})^{\GCD(p,N)=2}_{\text{3d mirr}}:  \qquad  &1-[n(q-n)] \\ & \text{+ $\frac{1}{4}(N-2)(p-N-2)$ hypers}
}
where $n$ and $q$ are defined as in \eref{defabc}.

When $\GCD(p,N)=3$, we have the following mirror theories:
\bes{ \label{DpSUngcd3}
(D_p(SU(N)))^{\GCD(p,N)=3}_{\text{3d mirr}}:  \quad
&\begin{tikzpicture}[baseline=0, font=\small]
\tikzstyle{every node}=[minimum size=0.5cm]
\node[draw, circle] (c2) at (-2,0) {$1$};
\node[draw, circle] (c3) at (0,0) {} node[draw=none] at (0, -0.5) {\scriptsize $N-1$};
\node[draw, circle] (c4) at (1,0)   {} node[draw=none] at (1, -0.5) {\scriptsize $N-2$}; 
\node[draw=none] (c5) at (2,0)  {$\cdots$}; 
\node[draw, circle] (c6) at (3,0)  {$2$}; 
\node[draw, circle] (c7) at (4,0)  {$1$}; 
\node[draw, circle] (f1) at (-1,1.2)  {$1$}; 
\node[draw, circle] (f2) at (-1,-1.2)  {$1$}; 
\draw[draw, solid, red, very thick] (c2)--(c3); 
\draw[draw, solid] (c3)--(c4)--(c5)--(c6)--(c7);
\draw[draw, solid] (f1)--(c2);
\draw[very thick,blue] (c2)--(f1) node at (-2.2,0.7)  {\footnotesize $n(q-n)$};
\draw[very thick, red] (c3)--(f1) node[midway, above] {\footnotesize $n$};
\draw[very thick, blue] (c2)--(f2) node at (-2.2,-0.6)  {\footnotesize $n(q-n)$};
\draw[very thick, red] (c3)--(f2) node[midway, below] {\footnotesize $n$};
\draw[very thick, blue] (f1)--(f2); 
\end{tikzpicture} \\
& \text{+ $\frac{1}{6}(N-3)(p-N-3)$ hypers}\\
(I_{p-N, \, N})^{\GCD(p,N)=3}_{\text{3d mirr}}: \qquad
&\begin{tikzpicture}[baseline=0, font=\footnotesize]
\tikzstyle{every node}=[minimum size=0.2cm]
\node[draw, circle] (c1) at (0,1-0.5) {$1$};
\node[draw, circle] (c2) at (1,-0.5) {$1$};
\node[draw, circle] (c3) at (-1,-0.5) {$1$};
\draw[very thick, blue] (c1)--(c2) node[midway, right] {$n(q-n)$};
\draw[very thick, blue] (c1)--(c3)  node[midway, left] {$n(q-n)$};
\draw[very thick, blue] (c2)--(c3) node[midway, below] {$n(q-n)$};
\end{tikzpicture}
\\  &\text{+ $\frac{1}{6}(N-3)(p-N-3)$ hypers}
}

When $\GCD(p,N)=4$, we have the following mirror theories:
\bes{
(D_p(SU(N)))^{\GCD(p,N)=4}_{\text{3d mirr}}:  \quad
&\scalebox{0.8}{
\begin{tikzpicture}[baseline=0]
\def\n{5}
\node[circle,minimum size=3 cm] (b) {};
\foreach\x in{1,...,4}{
  \node[minimum size=0.75cm,draw,circle] (n-\x) at (b.{360/\n*\x}){1};
}
\foreach\x in{5}{
  \node[minimum size=0.75cm,draw,circle] (n-\x) at (b.{360/\n*\x}) {} node[draw=none] at (1.6,-0.5) {\scriptsize $N-1$};
}
\foreach\x in{1,...,4}{
  \foreach\y in{1,...,4}{
    \ifnum\x=\y\relax\else
      \draw (n-\x) edge[very thick, blue] (n-\y);
    \fi
  }
 \foreach\y in{5}{
    \ifnum\x=\y\relax\else
      \draw (n-\x) edge[very thick, red] (n-\y);
    \fi
  } 
\node[draw=none] at (-2,0) {\blue $n(q-n)$};  
\node[draw=none] at (1.2,1) {\red $n$};  
\node[draw,circle, minimum size=0.75cm,] (6a) at (3,0) {} node[draw=none] at (3,-0.5) {\scriptsize $N-2$};
\node[draw=none] (4a) at (4.5,0) {$\cdots$};
\node[draw,circle] (2a) at (6,0) {2};
\node[draw,circle] (1a) at (7.5,0) {1};
\draw (n-5)--(6a)--(4a)--(2a)--(1a);
}
\end{tikzpicture}}\\
&\text{+ $\frac{1}{8}(N-4)(p-N-4)$ hypers} \\
(I_{p-N, \, N})^{\GCD(p,N)=4}_{\text{3d mirr}}: \qquad
&\scalebox{0.7}{
\begin{tikzpicture}[baseline=0]
\def\n{4}
\node[draw=none] at (-1.3,1) {\blue $n(q-n)$};
\node[circle,minimum size=3 cm] (b) {};
\foreach\x in{1,...,\n}{
  \node[minimum size=0.75cm,draw,circle] (n-\x) at (b.{360/\n*\x}){1};
}
\foreach\x in{1,...,\n}{
  \foreach\y in{1,...,\n}{
    \ifnum\x=\y\relax\else
      \draw (n-\x) edge[very thick, blue] (n-\y);
    \fi
  }
} 
\end{tikzpicture}} \\ 
& \text{+ $\frac{1}{8}(N-4)(p-N-4)$ hypers}
}

Finally, let us consider the case of $\GCD(p,N)=N$, \ie~ $p=N \fm$ for some integer $\fm \geq 1$.  In this case, $m=N$, $n=1$ and $q=\fm$.  The 3d reduction of the corresponding $D_{N \fm} (SU(N))$ is
\bes{ \label{3dredpmultN}
&[N]-(N-1)^{\fm-1}-SU(N-1)-(N-2)^{\fm-1}-SU(N-2)-\cdots \\
&-(2)^{\fm-1}-SU(2)- (1)^{\fm-1}-[1]
}
where $[1]$ can be identified with the group $SU(1)$ and $(x)^k$ denotes the chain $(x)-(x)-\cdots-(x)$ such that $(x)$ appears $k$ times.  The mirror theory consists of 
\bi
\item a complete graph with $N$ $U(1)$ nodes such that each edge has multiplicity $\fm-1$, 
\item the tail $(N-1)-\cdots-(2)-(1)$, and 
\item the edges (each with multiplicity $1$) that connect the $U(N-1)$ node of the tail to each $U(1)$ node in the complete graph.  
\ei
As before, we can decouple the tail and obtain the mirror theory for $I_{N(\fm-1), N}$, which is a complete graph with $N$ $U(1)$ nodes where each edge has multiplicity $\fm-1$.  There is no free hypermultiplet for these theories.

\subsection{Properties of the Higgs and Coulomb branches}
\subsubsection*{The Coulomb branch of the mirror theory}
The Coulomb branch symmetry of the mirror theory for $D_p(SU(N))$ is, in general, $SU(N) \times U(1)^{m-1}$, where the $SU(N)$ factor comes from the tail and the $U(1)^{m-1}$ comes from the complete graph.  Note that this can get enhanced further when certain nodes in the complete graph are balanced.  In any case, the rank of the Coulomb branch symmetry is $(N-1)+(m-1)$.  This agrees with the rank of the flavour symmetry of the $D_p(SU(N))$ theory \cite{Cecotti:2013lda}. The Coulomb branch of the mirror theory is 
\bes{ \label{CouldimmirrDpSUNp>N}
\frac{1}{2} nm (nm-1)+(m-1) =  \frac{1}{2}N(N-1)+(m-1)
}
quaternionic dimensional.  It is worth comparing this with the quantity $24(c-a)$ of the $D_p(SU(N))$ theory which is given by \eref{ccformula} (see \cite[(6.84)]{Cecotti:2013lda}):
\bes{ \label{24cma4d}
&\text{$24(c-a)[D_{p}(SU(N))]$} \\
&= \frac{1}{2} (N^2+m -2) - \frac{3N}{p} \left[\sum_{j=1}^{N-1}  \left \{ \frac{j p}{N} \right \}   \left(1-  \left \{  \frac{j p}{N} \right \} \right) \right]~.
}
The difference between $24(c-a)$ and \eref{CouldimmirrDpSUNp>N} is
\bes{\label{deltaHNp}
\delta_H(N, p)= \frac{1}{2} (N-m) - \frac{3N}{p} \left[\sum_{j=1}^{N-1}  \left \{ \frac{j p}{N} \right \}   \left(1-  \left \{  \frac{j p}{N} \right \} \right) \right]~,
}
where $m= \GCD(N,p)$ and $p \geq N$.  Recall that the quantity $\delta_H(N,p)$ is the value of $24(c-a)$ of the SCFTs without Higgs branch (``non higgsable'' SCFTs) that are components of the $D_p(SU(N))$ theory where upon reduction to 3d such SCFTs become a collection of twisted hypermultiplets.  The general discussion of the non higgsable SCFTs can be found around \eref{hbflow}.

As an explicit example, let us try to identify the ``non higgsable'' component of $D_{14}(SU(8))$.  We first recall the identification $D_{14}(SU(8)) = (I_{6,8}, [1]^8)$, \ie~ the $D_{14}(SU(8))$ theory can be realised as a class $\mathcal{S}$ theory associated with a sphere with a full puncture $(1)^8$ and an irregular puncture that engineers the theory $I_{6,8}=(A_5,A_7)$.   From \eref{A5A7}, the latter can be realised as 
\bes{
I_{6,8}: \,\,  \left[ D_7(SU(3)) = (I_{4,3}, [1^3])\right]\,\, \longleftarrow SU(3) \longrightarrow\,\, \left[D_7(SU(4)) = (I_{3,4} ,[1^4])\right]~.
}
Closing the puncture $[1]^3$ in the above system, we see that the $F$-terms lead to the system consisting of the $I_{4,3}$ theory and the $(I_{3,4},[3,1])$ theory, where $[3,1]$ is the simple puncture.  Moving along the Higgs branch, we can further close the simple puncture and obtain a system of $I_{3,4}$ and $I_{4,3}$.  Since $I_{4,3} = I_{3,4}$, this is a system of two copies of the $I_{3,4}$ SCFT.   Let us denote this system by $(I_{3,4})^{\otimes 2}$.  We claim that $(I_{3,4})^{\otimes 2}$ is the component with empty Higgs branch of $I_{6,8}$ and of $D_{14}(SU(8))$.  This claim can be tested as follows.  The $a$ and $c$ central charges of $I_{3,4}$ are (see \eg~ \cite[Table 1]{Xie:2013jc})
\bes{
~(a, c)[I_{3,4}] = \left(\frac{75}{56},\frac{19}{14} \right)~, \qquad 24(c-a)[I_{3,4}] = \frac{3}{7}~.
}
We observe that 
\bes{ \label{deltaH814}
24(c-a)[(I_{3,4})^{\otimes 2}] = \frac{6}{7} \overset{\eref{deltaHNp}}{=} \delta_H (8,14)~.
}
This provides a nontrivial test of our claim.

Let us tabulate a few values of $(N,p)$, where $\delta_H(N, p)$ is non-zero, and identify the ``non higgsable'' SCFTs.
\be
\begin{tabular}{|c|c|c|c|}
\hline
$(N,p)$ & $\delta_H$ & non higgsable SCFT & $H_{\text{free}} $\\
\hline
$(2,5)$ & $1/5$ & $I_{2,3} = (A_1, A_2)$ & 1 \\
$(2,7)$ & $2/7$ & $I_{2,5}=(A_1,A_4)$ & 2 \\
$(2,9)$ & $1/3$ & $I_{2,7}=(A_1,A_6)$  & 3\\
$(3,5)$ & $1/5$ & $I_{2,3} = (A_1, A_2)$ & 1\\
$(3,7)$ & $3/7$ & $I_{3,4} = (A_2, A_3)$ &3 \\
$(3,8)$ & $1/2$ & $I_{3,5} = (A_2,A_4)$ & 4\\
$(3,10)$ & $3/5$ & $I_{3,7}=(A_2,A_6)$ &6 \\
\hline
\end{tabular}
\ee
This is in agreement with the discussion around \eref{hbflow}.  Note that $H_{\text{free}} $ is the rank of the corresponding non higgsable SCFT.

\subsubsection*{The Higgs branch of the mirror theory}
The Higgs branch dimension of the mirror quiver theory is
\bes{ \label{HiggsdimmirrDpSUNp>N}
= \frac{1}{2} \left[ N(p-1) -\frac{N(p-N)}{m}-2m+2 \right]~.
}
After adding to it the contribution $H_{\text{free}}$ of the free hypermultiplets, this agrees with the rank of the corresponding 4d $D_p(SU(N))$ theory given by \eref{ranks}.
The Higgs branch symmetry (neglecting the contribution of the free hypermultiplets) is, in general, $U(1)^{m} \times SU(n)^m \times SU(nq-n^2)^{\frac{1}{2}m(m-1)}$ (assuming that $m \geq 3$).\footnote{It should be noted that for $m=1$ we do not have any $U(1)$ factor, see the discussion around \eref{gcd1}, and for $m=2$ we have precisely one $U(1)$ factor.  The Coulomb branch symmetry of the linear quiver \eref{linearquiv3dDpSUn} is explained in Appendix \ref{app:SUenhanced}.  These symmetries can also be seen from the Higgs branch symmetry of the mirror theory. }
It is interesting to compare with this with the Coulomb branch symmetry of the linear quiver coming from the 3d reduction of the $D_p(SU(N))$ theory (after dualising and neglecting all possible twisted hypermultiplets)\footnote{It can be checked that the quaternionic Coulomb branch dimension of this linear quiver is \eref{HiggsdimmirrDpSUNp>N}; adding to it $H_{\text{free}}$, which is the number of twisted hypermultiplets obtained upon compactification, we obtain the rank of the $D_p(SU(N))$ theory.  Moreover, the quaternionic Higgs branch dimension of this linear quiver is equal to \eref{CouldimmirrDpSUNp>N}; adding to it the values of $24(c-a)$ of the non higgsable theories, we obtain the value of $24(c-a)$ of the $D_p(SU(N))$ theory.}:
\be \label{linearquiv3dDpSUn}
\scalebox{0.9}{$
\begin{split} 
&[N]-(N-1)-(N-2)-\cdots-(N-n+1)-(N-n)^{q-n}-SU(N-n)\\ 
&-(N-n-1)-(N-n-2)-\cdots-(N-2n+1)-(N-2n)^{q-n}-SU(N-2n)- \cdots \\
&-SU(2n)-(2n-1)- \cdots-(n+1)-(n)^{q-n} -SU(n) - (n-1) -(n-2) - \cdots 2-1
\end{split}$}
\ee
where there are in total $(m-1)$ special unitary gauge groups\footnote{As remarked earlier, when $p$ is a multiple of $N$, which means $m=N$ and $n=1$, there is an $SU(1)$ group present in the linear quiver and this should be treated as one flavour under the gauge group next to it; see \eref{3dredpmultN}.  In this case, the total number of non-trivial special unitary gauge groups is $m-2=N-2$.} $SU(N-jn)$ with $1\leq j \leq m-1$ and the notation $(x)^{q-n}$ indicates that the chain $(x)-\cdots-(x)$ where $(x)$ appears $q-n$ times.  An example for $N=8$ and $p=14$, where $n=4$, $q=7$ and $m=2$, is depicted in the first line of \eref{gauge D7SU4andD784}.  The Coulomb branch symmetry of \eref{linearquiv3dDpSUn} is discussed in more detail in Appendix \ref{app:SUenhanced}.  We briefly summarise the discussion there as follows.  Observe that the leftmost $(N-jn)$ node in each chain $(N-j n)^{q-n}$, with $1\leq j \leq m-1$, is overbalanced.  Together with the $SU(N-n)$ node, these give rise to the $U(1)^{m}$ factor in the Coulomb branch symmetry of \eref{linearquiv3dDpSUn}.  Furthermore, the sequences of the balanced unitary gauge nodes $(N-\ell n-1)-(N- \ell n -2)- \cdots -(N- \ell n - (n-1))$, with $0\leq \ell \leq m-2$, give rise to an $SU(n)^{m-1}$ factor. Moreover, the chain
\bes{
[n]-(n)^{q-n-1}-SU(n) - (n-1)-\cdots - (n-2)- \cdots (2)-(1)
}
gives rise to the Coulomb branch symmetry $SU(n) \times SU(n(q-n))$.  This is a typical way we have a factor of $SU(n(q-n))$.  In general, we have the factors $SU(n(q-n))^{\frac{1}{2}m(m-1)}$, where each $SU(n(q-n))$ arises from the operators with $R$-charge $1$ associated with the following pair of chains:
\bes{
&(N-j n)^{q-n-1} {\gray -SU(N-j n)} ~, \\
&(N- \ell n -1) -(N-\ell n-2)- \cdots - (N-\ell n -1) - (N- \ell n)
}
with $1 \leq j \leq \ell \leq m-1$.  Note that there are $\frac{1}{2}(m-1)[(m-1)+1] = \frac{1}{2}m(m-1)$ such pairs.  In summary, we have $U(1)^{m} \times SU(n)^m \times SU(nq-n^2)^{\frac{1}{2}m(m-1)}$ for the Coulomb branch symmetry of \eref{linearquiv3dDpSUn}; this is in agreement with the Higgs branch symmetry of the mirror theory.

\subsection{Argyres-Douglas quivers from susy enhancement RG flows}
\label{subsec:susyRGflow}

It is known (see \cite{Giacomelli:2017ckh}) that $D_p(SU(N))$ theories flow in the infrared to $(A_{N-1},A_{p-1})$ models upon a relevant deformation of the Maruyoshi-Song type \cite{Maruyoshi:2016tqk, Maruyoshi:2016aim}. This RG flow involves introducing a chiral multiplet $M$ transforming in the adjoint of $SU(N)$, couple it to the $SU(N)$ moment map of $D_p(SU(N))$ (flipping of the moment map) and turning on a principal nilpotent vev for $M$ which breaks $SU(N)$ completely. This deformation breaks supersymmetry to $\mathcal{N}=1$ but supersymmetry enhances back to $\mathcal{N}=2$ in the infrared. For $p>N$ we can therefore flow to $(A_{N-1},A_{p-N-1})$ by closing the regular puncture, or to $(A_{N-1},A_{p-1})$ by implementing the RG flow described above. As we will now see, this imposes strong constraints on the structure of the 3d mirror which is remarkably solved by our construction. This provides a nontrivial check of our method. 

In order to analyze the effect of the RG flow on the 3d mirror theory, we make use of the flip-flip duality for $T(SU(N))$ of \cite{Aprile:2018oau}. In the appendix we describe a field theoretic derivation of it, based on recursively applying Aharony-duality \cite{Aharony:1997gp}, which has been proposed in \cite{Hwang:2020wpd}. The duality states that $T(SU(N))$ is equivalent to the same theory with both $SU(N)_C$ and $SU(N)_H$ moment maps flipped. Under this duality the moment maps $\mu_{C,H}$ are mapped to the corresponding flipping fields $M_{C,H}$. It will be more convenient for us to use a slightly modified version of this duality, described as follows: If we flip the $SU(N)_C$ moment map of $T(SU(N))$, under the flip-flip duality this deformation is mapped to a flipping for $M_C$, which then becomes massive and can be integrated out. We therefore obtain the following variant of the duality: On one side the $SU(N)_C$ moment map is flipped while on the other only the $SU(N)_H$ moment map is flipped. Each flipping field is mapped to the corresponding moment map in the other duality frame. Schematically we have
\be\label{flipflip}
\begin{aligned}
T(SU(N))\quad\mathcal{W}=\Tr(\mu_C M_C)\quad &\longleftrightarrow \quad T(SU(N))\quad\mathcal{W}=\Tr(\mu_H M_H)\\
M_C\quad &\longleftrightarrow \quad \mu_C\\
 \mu_H \quad &\longleftrightarrow \quad M_H\\
\end{aligned}
\ee
where we have included the superpotential deformation describing the flipping. 

Let us come to the main point of this section, namely how to implement the RG flow in the 3d mirror. A feature of these RG flows is the decoupling of chiral multiplets hitting the unitarity bound. It is known that for $p>N$ only the Coulomb branch operators of $D_p(SU(N))$ can decouple, and precisely $N-1$ of them do so. These can be described as follows: denoting the deformation terms as before $u_{ij}z^ix^j$, for each $j$ the operator which decouples is identified by the largest value of $i$ such that $Ni/p<N-j-1$. There are precisely $N-1$ of them. These are the CB operators with smallest scaling dimension among those labelled by a given $j$.

The statement of the previous section can be derived as follows. In the RG flow analysis we can parametrize the trial R-symmetry in such a way that all CB operators $u_{ij}$ have trial R-symmetry $D_{UV}(u_{ij})(1+\epsilon)$. Furthermore, we can show (see \cite{Giacomelli:2018ziv} and also appendix A of \cite{Giacomelli:2017ckh}) that the trial $a$ central charge is maximized at $\epsilon=-\frac{p+3N}{3p+3N}$ and therefore the dimension of CB operators in the IR is 
\be D_{IR}(u_{ij})=D_{UV}(u_{ij})\frac{p}{p+N}.\ee 
We therefore conclude that only CB operators whose dimension in the UV is in the range $1<D_{UV}(u_{ij})\leq 1+N/p$ can hit the unitarity bound and decouple. Since the dimension in the UV of $u_{ij}$ is $D_{UV}(u_{ij})=N-j-Ni/p$ and the dimension of $z$ is $N/p$, we immediately see that for each value of $j$ there is exactly one value of $i$ such that $u_{ij}$ decouples in the infrared. The same discussion can be applied to $D_p^{N-1}(SU(N))$ theories with $p>N-1$ as well, simply by replacing $N$ with $b=N-1$.

In order to make the RG flow and dimensional reduction commute, we flip the operators which decouple in order to set them to zero in the chiral ring \cite{Benvenuti:2017lle}. The question is then the mapping of these operators to the Lagrangian theory in 3d. Let us see how it works. From our algorithm to build the 3d Lagrangian theory, we see that the various $u_{ij}$ CB operators are mapped to Casimir invariants of the various $U(n_i)$ vector multiplets and in particular those which decouple will be mapped to operators of the form $\Tr\Phi_i$. Once all the ugly nodes in 3d have been dualized, the structure of the linear quiver in 3d for $p>N$ is as described in \eref{linearquiv3dDpSUn}; schematically, this can be summarised as follows:
\bes{
[N]&-(N-1)-(N-2)-\cdots-(N-n)-SU(N-n)\\ 
&-(N-n-1)-(N-n-2)-\cdots-(N-2n)-SU(N-2n)- \cdots~.
}
In other words, we start with a $U(N-1)$ with $N$ flavors, followed by a $U(N-2)$ and so on. In short we have a decreasing ramp which terminates with a $U(N-n)$ gauge group where $n=N/\GCD(N,p)$. This is followed by a sequence of $U(N-n)$ groups which terminates with a $SU(N-n)$ group. Then the ramp restarts all the way to $U(N-2n)$ and so on.   We can identify the flipped operators with the linear Casimirs of the unitary nodes along the ramps. These have R-charge 1 and are the Cartan components of the moment map of the CB global symmetry arising from the ramps. This is a product of $U(n)$ factors. 

In the mirror theory the flipped operators are mapped to R-charge one Higgs branch operators, which are necessarily quadratic in the bifundamentals. In the mirror theory the $U(n)$ factors arise from the various $U(N-1)\times U(1)$ bifundamentals, which have all multiplicity $n$ (denoted by red lines in Section \ref{secmir41}). We therefore conclude that in the 3d mirror we should flip the cartan components of the $SU(N)$ meson, namely the cartan components for the $SU(N)$ HB moment map $\mu_H$ of the $T(SU(N))$ tail. 

Next we add the flipping field $M$ for the global symmetry of $D_p(SU(N))$ and turn on the nilpotent vev. Let's study this step-by-step. Upon dimensional reduction and mirror symmetry $M$ is mapped to a flipping field $M_C$ for the CB moment map of the $T(SU(N))$ tail. In order to see the effect of the nilpotent vev, it is convenient to use the flip-flip duality (\ref{flipflip}) and focus on a dual frame in which the HB moment map is flipped, which is easier to understand. In this duality frame the nilpotent vev is given directly to the CB moment map of the tail. The effect is just to remove the tail altogether. We are therefore left with the abelian quiver with $\GCD(N,p)$ $U(1)$ nodes only. Overall, apart from removing the tail we are just adding the (dual of the) flipping field $M$. 

The $N-1$ singlets we have introduced before to make RG flow and dimensional reduction commute, are now coupled to the diagonal components of $M$ (because of (\ref{flipflip}) and because before the dualization they were flipping the diagonal components of the HB moment map). Taking this fact into account, we see that the field $M$ decomposes as $n^2$ bifundamentals between each pair of $U(1)$ nodes and also $n^2-n$ chiral adjoints for each $U(1)$ node. The latter are just $\GCD(N,p)\frac{n^2-n}{2}$ gauge invariant hypermultiplets which decouple and become free. 

Overall, the net effect is to remove the $T(SU(N))$ tail, add $n^2$ bifundamentals and produce $\GCD(N,p)\frac{n^2-n}{2}$ extra free hypers. Since closing the puncture simply amounts in the 3d mirror to remove the $T(SU(N))$ tail, we conclude that the difference between the 3d mirror of $(A_{N-1},A_{p-N-1})$ and that of $(A_{N-1},A_{p-1})$ is the addition of $n^2$ bifundamentals between each pair of nodes and the addition of $\GCD(N,p)\frac{n^2-n}{2}$ free hypers, in perfect agreement with our algorithm!

To illustrate the far reaching consequences of this analysis, let's consider the $D_p(SU(N))$ theory with $p=N+\GCD(N,p)=N+m$, where as before $m=\GCD(N,p)$. Upon closing the regular puncture, we find in the IR the theory $(A_{N-1},A_{m-1})$ which is in the subclass for which the mirror is already known. This is given by the complete graph with $m$ $U(1)$ nodes such that each edge has multiplicity $n=N/m$. Then from the analysis of the susy enhancing RG flow we conclude that the 3d mirror of   $(A_{N-1},A_{N+m-1})$ is given by the same quiver but with edge multiplicity $n^2+n$ and $\GCD(N,p)\frac{n^2-n}{2}$ free hypermultiplets. Analogously, this fixes the 3d mirror of $(A_{N-1},A_{2N+m-1})$ and so on. We are therefore given a set of consistency conditions which constrain significantly the structure of the 3d mirror and they are automatically satisfied by our algorithm. The same consistency check can be performed also for Type II punctures, with identical conclusions. This represents a highly non trivial consistency check of our procedure.

\section{Mirror theory for $D_p(SU(N))$ with $p \leq N$} \label{sec:DpSUNplN}
Let us define the following shorthand notations:
\bes{
\begin{array}{lll}
x =\lfloor N/p \rfloor~,  & \qquad M= N-(x+1)~, &  \qquad m = \GCD(p,N) ~,\\
n= N/m~, & \qquad q= p/m~. &
\end{array}
}
The mirror theory for the $D_p(SU(N))$ theory with $p \leq N$ can be constructed as follows:
\ben
\item Start with a complete graph with $m$ $U(1)$ nodes where each edge has multiplicity 
\bes{
m_G = m_A m_B = (q(1+x)-n)(n-q x)= \left[ \frac{p(1+x)-N}{m} \right]\left[ \frac{N- px}{m} \right]
}
where $m_A = q(1+x)-n$ and $m_B = n-q x$.  Note that this formula can be obtained from \eref{multgraphDpSUNp>N} by taking $n \rightarrow n-qx$.
\item Construct two types of tails:
\bes{ \nn
&\text{Tail A}: \quad (p-1)-2(p-1)-\cdots-(p-1)(x-2)-(p-1)(x-1)-(p-1)x \\
&\text{Tail B}: \quad M-(M-1)-(M-2)-\ldots-2-1
}
\item Attach the node $(p-1)x$ of tail A to each $U(1)$ node in the complete graph with edges with equal multiplicity 
\bes{
m_A = q(1+x)-n = \frac{p-(N-p x)}{m}~.
}
\item Attach the node $M$ of tail B to each $U(1)$ node in the complete graph with edges with equal multiplicity 
\bes{
m_B = n-q x = \frac{N - p x}{m}~.
}
\item Connect the node $(p-1)x$ of tail A to the node $M$ of tail B with an edge with multiplicity $1$.
\item The quiver constructed above has an overall $U(1)$ that needs to be decoupled.
\item There are 
\bes{
H_{\text{free}} &= \frac{1}{2}m (n-qx-1)(q(1+x)-n -1)  \\
&= \frac{ (N-px-m)(p(1+x)-N -m) }{2m}
}
free hypermultiplets.  Note that this formula can be obtained from \eref{freehypTypeIp>N} by taking $n\rightarrow n-qx$.
\item The above quiver, together with $H_{\text{free}}$ free hypermultiplets, is the mirror theory for the $D_p(SU(N))$ theory with $p \leq N$.
\een

\subsubsection*{The Coulomb branch of the mirror theory}
In general, the Coulomb branch symmetry of this mirror theory is $SU(N) \times U(1)^{m-1}$. The quaternionic dimension of the Coulomb branch is 
\bes{ \label{CplN}
\frac{1}{2}N(N-1)+(m-1)+ \frac{1}{2} p x (x+1) - N x~.  
}
The difference between the quantity $24(c-a)$ of  the $D_p(SU(N))$ theory and this quantity is
\bes{
\delta_H(N, p)&=\eref{24cma4d} - \eref{CplN} \\
&=  \frac{1}{2} (N-m) -\frac{1}{2} p x (x+1) + Nx \\
& \qquad - \frac{3N}{p} \left[\sum_{j=1}^{N-1}  \left \{ \frac{j p}{N} \right \}   \left(1-  \left \{  \frac{j p}{N} \right \} \right) \right]~.
}
As discussed earlier, the quantity $\delta_H(N,p)$ is the value of $24(c-a)$ of the ``non higgsable'' SCFTs along the Higgs branch of the $D_p(SU(N))$ theory where upon reduction to 3d such SCFTs become a collection of twisted hypermultiplets.

Let us tabulate a few values of $(N,p)$, where $\delta_H(N, p)$ is non-zero, and identify the ``non higgsable'' SCFTs.
\be
\begin{tabular}{|c|c|c|c|}
\hline
$(N,p)$ & $\delta_H$ & non higgsable SCFT & $H_{\text{free}} $ \\
\hline
$(7,5)$ & $1/5$ & $I_{2,3} = (A_1, A_2)$ & 1 \\
$(8,5)$ & $1/5$ & $I_{2,3} = (A_1, A_2)$ & 1 \\
$(9,7)$ & $2/7$ & $I_{2,5}=(A_1,A_4)$ & 2 \\
$(10,7)$ & $3/7$ & $I_{3,4} = (A_2, A_3)$ & 3 \\
$(11,7)$ & $3/7$ & $I_{3,4} = (A_2, A_3)$ & 3\\
$(11,8)$ & $1/2$ &  $I_{3,5} = (A_2,A_4)$ & 4 \\
$(11,9)$ & $1/3$ & $I_{2,7}=(A_1,A_6)$ & 3\\
\hline
\end{tabular}
\ee
Note that $H_{\text{free}} $ is equal to the rank of the corresponding non higgsable SCFT.

\subsubsection*{The Higgs branch of the mirror theory}
It can be checked that the quaternionic Higgs branch dimension of the quiver theory constructed using the above prescription is
\bes{ \label{HplN}
 \frac{1}{2} \left[ \frac{(N-p x) (N-p (x+1))}{m}+N (p-1)-2 m+2\right]~.  
}
Adding to it $H_{\text{free}}$, this is equal to the rank of the 4d $D_{p}(SU(N))$ theory given by \eref{ranks}. In general the Higgs branch symmetry (neglecting the contribution of the free hypermultiplets) is $U(1)^{m+1} \times SU(m_A)^m\times SU(m_B)^m \times SU(m_G)^{\frac{1}{2}m(m-1)}$, where we emphasize that $m_G= m_A m_B$.  It is interesting to compare this with the Coulomb branch symmetry of the linear quiver coming from the 3d reduction of the $D_p(SU(N))$ theory (after dualising and neglecting all possible twisted hypermultiplets)\footnote{It can be checked that the quaternionic Coulomb branch dimension of this linear quiver is \eref{HplN}.  After adding to it $H_{\text{free}}$, which is the number of twisted hypermultiplets obtained upon compactification, is equal to the rank of the $D_p(SU(N))$ theory.  Moreover, the quaternionic Higgs branch dimension of this linear quiver is \eref{CplN}; after adding to it the values of $24(c-a)$ of the non higgsable theories, this is equal to $24(c-a)$ of the $D_p(SU(N))$ theory.} (which is empty when $m_A=1$):
\be \label{linearquiv3dDpSUnp<n}
\scalebox{0.9}{$
\begin{split} 
&[N]-(N-(x+1))-\cdots -(N-m_B(x+1)) -C[N-n]-SU(N-n)\\ 
&-(N-n-(x+1))-\cdots -(N-n-m_B(x+1)) -C[N-2n]-SU(N-2n)- \cdots \\
&-SU(2n)-(2n-(x+1))-\cdots -(2n-m_B(x+1))-C[n]-SU(n)\\
&  - (n-(x+1))  - \cdots -(n-m_B(x+1))-C[0]
\end{split}$}
\ee
where we use the shorthand notation for the following chain:
\bes{
C[Y] = (Y+(m_A-1) x )- (Y+(m_A-2) x ) -\ldots -(Y+2x)-(Y+x)~.
}
Note that there are in total $(m-1)$ special unitary gauge groups and that each ``ramp'' has length $m_B +(m_A-1)=q-1$.  As argued in Appendix \ref{app:SUenhanced}, the Coulomb branch symmetry of \eref{linearquiv3dDpSUnp<n} is in general $U(1)^{m+1} \times SU(m_A)^m\times SU(m_B)^m \times SU(m_A m_B)^{\frac{1}{2}m(m-1)}$.

\subsection{Examples}

\label{subsec:DpSUNplN:Examples}

Let us first consider an example of $D_{4}(SU(6))$ theory whose reduction to 3d gives
\bes{ \label{D4SU6linear}
(D_4(SU(6))_{3d}: \quad [6]-4-SU(3)-1
}
This can be realised as gluing $[6]-4-[3]$ and $[3]-1$ via an $SU(3)$ gauge group.  In terms of the mirror theories, the gluing we are interested in is
\bes{ \label{D4SU6gluing}
\begin{tikzpicture}[baseline=0, font=\scriptsize]
\tikzstyle{every node}=[minimum size=0.5cm]
\node[draw, circle] (a1) at (0,0) {1};
\node[draw, circle] (a2) at (1,0) {2};
\node[draw, circle] (a3) at (2,0) {3};
\node[draw, circle] (a4) at (3,0) {4};
\node[draw, circle] (a6) at (4,0) {4};
\node[draw, circle] (a7) at (5,0) {3};
\node[draw, circle] (a8) at (6,0) {2};
\node[draw, circle] (a9) at (7,0) {1};
\node[draw, circle] (b) at (3.5,-1) {1};
\draw[draw, solid] (a1)--(a2);
\draw[draw, solid] (a2)--(a3);
\draw[draw, solid] (a3)--(a4);
\draw[draw, solid] (a4)--(a6);
\draw[draw, solid] (a6)--(a7);
\draw[draw, solid] (a7)--(a8);
\draw[draw, solid] (a8)--(a9);
\draw[draw, solid] (a4)--(b);
\draw[draw, solid] (a6)--(b);
\end{tikzpicture}
\qquad + \qquad
\begin{tikzpicture}[baseline=0, font=\scriptsize]
\tikzstyle{every node}=[minimum size=0.5cm]
\node[draw, circle] (a1) at (0,0) {1};
\node[draw, circle] (a2) at (2,0) {1};
\node[draw, circle] (b) at (1,-1) {1};
\draw[draw, solid] (a1)--(a2);
\draw[draw, solid] (b)--(a1);
\draw[draw, solid] (b)--(a2);
\end{tikzpicture}
}
Specifically, we would like to gauge the diagonal combination of the $SU(3)$ topological symmetry coming from the $3-2-1$ tail of the left quiver and the one of the right quiver, which after decoupling an overall $U(1)$ corresponds to the standard linear abelian quiver $[1]-1-1-[1]\equiv T_{[2,1]}(SU(3))$.
This gluing is different from those that we have dealt with so far, which always involved two identical tails of the form of $T(SU(N))$ theories. In order to understand what is the result of this operation we proceed as follows.

First, we perform the gluing replacing the left quiver in \eqref{D4SU6gluing} with a copy of the $T(SU(3))$ theory, which we represent as $1-2\equiv 3$
\bes{
\begin{tikzpicture}[baseline=0, font=\scriptsize]
\tikzstyle{every node}=[minimum size=0.5cm]
\node[draw, circle] (a1) at (0,0) {1};
\node[draw, circle] (a2) at (1,0) {2};
\node[draw, circle] (a3) at (2,0) {3};
\node[draw, circle] (a4) at (3,0) {4};
\node[draw, circle] (a6) at (4,0) {4};
\node[draw, circle] (a7) at (5,0) {3};
\node[draw, circle] (a8) at (6,0) {2};
\node[draw, circle] (a9) at (7,0) {1};
\node[draw, circle] (b) at (3.5,-1) {1};
\draw[draw, solid] (a1)--(a2);
\draw[draw, solid] (a2)--(a3);
\draw[draw, solid] (a3)--(a4);
\draw[draw, solid] (a4)--(a6);
\draw[draw, solid] (a6)--(a7);
\draw[draw, solid] (a7)--(a8);
\draw[draw, solid] (a8)--(a9);
\draw[draw, solid] (a4)--(b);
\draw[draw, solid] (a6)--(b);
\end{tikzpicture}
\qquad + \qquad
\begin{tikzpicture}[baseline=0, font=\scriptsize]
\tikzstyle{every node}=[minimum size=0.5cm]
\node[draw, circle] (a1) at (0,0) {1};
\node[draw, circle] (a2) at (1,0) {2};
\node[draw, circle] (a3) at (2,0) {1};
\draw[draw, solid] (a1)--(a2);
\draw[very thick, blue] (a2)--(a3) node at (1.5,0.3)  {\footnotesize $3$};
\end{tikzpicture}
}
This gluing is standard and gives
\bes{\label{D4SU6beforevev}
\begin{tikzpicture}[baseline=0, font=\scriptsize]
\tikzstyle{every node}=[minimum size=0.5cm]
\node[draw, circle] (a1) at (0,0) {1};
\node[draw, circle] (a2) at (1,0) {2};
\node[draw, circle] (a3) at (2,0) {3};
\node[draw, circle] (a4) at (3,0) {4};
\node[draw, circle] (a6) at (4,0) {4};
\node[draw, circle] (a7) at (5,0) {1};
\node[draw, circle] (b) at (3.5,-1) {1};
\draw[draw, solid] (a1)--(a2);
\draw[draw, solid] (a2)--(a3);
\draw[draw, solid] (a3)--(a4);
\draw[draw, solid] (a4)--(a6);
\draw[very thick, blue] (a6)--(a7) node at (4.5,0.3)  {\footnotesize $3$};
\draw[draw, solid] (a4)--(b);
\draw[draw, solid] (a6)--(b);
\end{tikzpicture}
}
Now we give the nilpotent vev $[2,1]$, which makes $T(SU(3))$ reduce to $T_{[2,1]}(SU(3))$, to the three flavors at the right end of the tail depicted in blue\footnote{Remember that gluing two copies of $T(SU(N)$ by gauging a diagonal combination of one $SU(N)$ global symmetry from each theory, we obtain a singular theory that behaves as a delta-function theory that identifies the two remaining $SU(N)$ symmetries (see \cite{Benvenuti:2011ga} for a proof at the level of the $S^3$ partition function). Because of this delta, if we give a nilpotent vev that breaks one of the two $SU(N)$ symmetries to a subgroup, then also the other $SU(N)$ receives the same nilpotent deformation.}. We claim that the result of such a vev is the following quiver, which corresponds to the mirror theory for the 3d reduction \eqref{D4SU6linear} of $D_{4}(SU(6))$:
\bes{ \label{D4SU6mirr}
(D_4(SU(6))_{\text{3d mirr}}: \quad
&\begin{tikzpicture}[baseline=0, font=\scriptsize]
\tikzstyle{every node}=[minimum size=0.5cm]
\node[draw, circle] (a1) at (0,0) {1};
\node[draw, circle] (a2) at (1,0) {2};
\node[draw, circle] (a3) at (2,0) {3};
\node[draw, circle] (a4) at (3,0) {4};
\node[draw, circle] (a6) at (5,0) {3};
\node[draw, circle] (u) at (4,1) {1};
\node[draw, circle] (b) at (4,-1) {1};
\draw[draw, solid] (a1)--(a2);
\draw[draw, solid] (a2)--(a3);
\draw[draw, solid] (a3)--(a4);
\draw[draw, solid] (a4)--(a6);
\draw[draw, solid] (a4)--(u);
\draw[draw, solid] (a4)--(b);
\draw[draw, solid] (b)--(a6);
\draw[draw, solid] (b)--(u);
\draw[draw, solid] (a6)--(u);
\end{tikzpicture}}
Every node on the central line is balanced and each $U(1)$ on top and bottom is overbalanced.  Decoupling an overall $U(1)$, we have the Coulomb branch symmetry $SU(6) \times U(1)$, in agreement with the global symmetry of $D_{4}(SU(6))$.    In this particular example, the Coulomb branch symmetry of \eref{D4SU6linear} is $U(1)^3$ (see Appendix \ref{app:SUenhanced}), which is in agreement with the Higgs branch symmetry of \eref{D4SU6mirr}.

The fact that the result of the vev is precisely the quiver in \eqref{D4SU10mirr} can be understood exploiting the flip-flip that we already used in Subsection \ref{subsec:susyRGflow}. Notice indeed that, upon decoupling an overall $U(1)$, the quiver in \eqref{D4SU6beforevev} can be understood as a $T^\sigma_\rho(SU(N)$ theory. The flip-flip duality for $T(SU(N)$ of \cite{Aprile:2018oau} was extended in \cite{Hwang:2020wpd} to the entire class of $T^\sigma_\rho(SU(N)$ theories and was used in that reference to study exactly vevs of the type we are interested in.

The strategy of \cite{Hwang:2020wpd} for analysing nilpotent vevs of linear quiver gauge theories with unitary groups was to study the deformation in the flip-flip dual frame, where it translates into a massive deformation. In our specific case, the massive deformation is $\mathcal{J}_{[2,1]}M$, where $\mathcal{J}_{[2,1]}$ is the Jordan matrix associated to the partition $[2,1]$, while $M$ is the meson matrix constructed with the 3 flavors at the right end of the quiver in \eqref{D4SU6beforevev}. This deformation is now easier to study, since we can just integrate out the massive fields, so to get the following quiver theory (see equation (2.28) of \cite{Hwang:2020wpd}):
\bes{\label{D4SU6flipflip}
\begin{tikzpicture}[baseline=0, font=\scriptsize]
\tikzstyle{every node}=[minimum size=0.5cm]
\node[draw, circle] (a1) at (0,0) {1};
\node[draw, circle] (a2) at (1,0) {2};
\node[draw, circle] (a3) at (2,0) {3};
\node[draw, circle] (a4) at (3,0) {4};
\node[draw, circle] (a6) at (4,0) {4};
\node[draw, circle] (a7) at (5,0) {1};
\node[draw, circle] (b) at (3.5,-1) {1};
\draw[draw, solid] (a1)--(a2);
\draw[draw, solid] (a2)--(a3);
\draw[draw, solid] (a3)--(a4);
\draw[draw, solid] (a4)--(a6);
\draw[black,solid] (a6) edge [out=30,in=150,loop,looseness=1]  (a7);\draw[red,solid] (a6) edge [out=-30,in=-150,loop,looseness=1]  (a7);
\draw[draw, solid] (a4)--(b);
\draw[draw, solid] (a6)--(b);
\end{tikzpicture}
}
We stress the fact that this representation of the theory is schematic at this stage.
In the quiver, the red line stands for the fact that the two chirals contained in the associated hypermultiplet don't interact with the adjoint chiral through the standard $\mathcal{N}=4$ superpotential, contrary to all the other fundamental chirals represented with the standard black color. Calling $Q$, $\tilde{Q}$ the fundamental chirals in question and $\Phi$ the adjoint chiral at the rightmost $U(4)$ gauge group, their interaction is indeed $\tilde{Q}\Phi^2 Q$. Moreover, applying flip-flip duality we also produce some gauge singlet chiral fields. For these two reasons, in this frame only $\mathcal{N}=2$ supersymmetry is expected. Nevertheless, since the nilpotent vev is a deformation that preserves the full supersymmetry of the original 3d $\mathcal{N}=4$ quiver, we expect supersymmetry to get enhanced back to $\mathcal{N}=4$ in the IR. For this reason, in the following steps we will ignore all possible gauge singlets: since the final result has to be explicitly $\mathcal{N}=4$ supersymmetric, they should all disappear after the manipulations we are going to perform.

The idea of \cite{Hwang:2020wpd} was that, after having studied the effect of the deformation in the flip-flip dual frame where it is more tractable, we can then go back to the original frame by sequentially applying Aharony duality \cite{Aharony:1997gp} along the quiver. In Appendix \ref{app:aharony} we review some interesting facts about Aharony duality applied to linear quivers and in Appendix \ref{app:flipflip} we explain how to apply this strategy to derive the flip-flip duality of $T(SU(N))$. The steps we are going to perform here are very similar (for more details about the following manipulations we refer the reader to \cite{Hwang:2020wpd}).

We start applying Aharony duality at the leftmost $U(1)$ gauge node, since the adjoint chiral there is just a singlet. The rank of this gauge node remains the same, but after the dualization the adjoint chiral of the adjacent $U(2)$ node becomes massive. This allows us to apply the duality again on this second node. Also in this case the rank of this node remains the same after the dualization and we remove the adjoint chiral at the adjacent $U(3)$ gauge node. It is clear that we can in this way iterate the procedure. Notice also that we don't produce any link between the $U(1)$ node on the left and the $U(3)$ node on the right at this step, since the corresponding field is actually massive. We keep applying Aharony duality along the whole quiver, until we reach the right $U(4)$ gauge node. Now the rank of this node gets lowered to $4+2+1-4=3$ after the dualization. Moreover, at this step we do produce some links, since the field denoted in red in \eqref{D4SU6flipflip} is involved in a different superpotential interaction from the previous ones. Hence, we get to the quiver
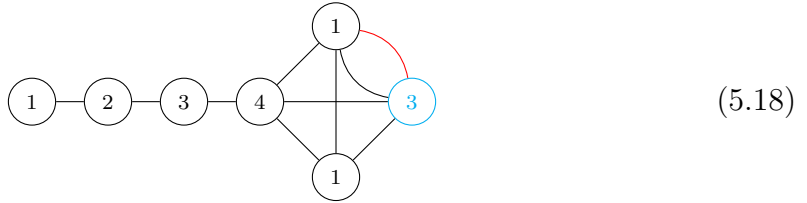
\bes{
\begin{tikzpicture}[baseline=0, font=\scriptsize]
\tikzstyle{every node}=[minimum size=0.5cm]
\node[draw, circle] (a1) at (0,0) {1};
\node[draw, circle] (a2) at (1,0) {2};
\node[draw, circle] (a3) at (2,0) {3};
\node[draw, circle] (a4) at (3,0) {4};
\node[draw, circle, cyan] (a6) at (5,0) {3};
\node[draw, circle] (u) at (4,1) {1};
\node[draw, circle] (b) at (4,-1) {1};
\draw[draw, solid] (a1)--(a2);
\draw[draw, solid] (a2)--(a3);
\draw[draw, solid] (a3)--(a4);
\draw[draw, solid] (a4)--(a6);
\draw[draw, solid] (a4)--(u);
\draw[draw, solid] (a4)--(b);
\draw[draw, solid] (b)--(a6);
\draw[draw, solid] (b)--(u);
\draw[black,solid] (a6) edge [out=170,in=280,loop,looseness=1]  (u);\draw[red,solid] (a6) edge [out=100,in=350,loop,looseness=1]  (u);
\end{tikzpicture}
}
It is important to keep in ming that this quiver is again only schematic. In particular, single lines should now be regarded as pairs of chirals, because of the $\mathcal{N}=2$ superpotential involving the red line. Moreover, we are not depicting explicitly all of the matter content of the theory, for example in the quiver it is not manifest the fact that at this stage we don't have the adjoint chiral at the rightmost $U(3)$ node anymore\footnote{As said before we are also neglecting gauge singlets. Some of them were present in \eqref{D4SU6flipflip} after the application of flip-flip duality, while others have been produced after the applications of Aharony duality. At the end of the procedure all these singlets should give mass to each other, since we expect to recover manifest $\mathcal{N}=4$ supersymmetry.} (to highlight this fact we coloured the node without the adjoint chiral in light blue). In order to solve this problem, we have to apply again Aharony duality along the quiver starting from the leftmost $U(1)$ node, but stopping this time at the second to last $U(4)$ node. In particular, after the last dualization the field associated to the red line becomes massive and we get the quiver
 \bes
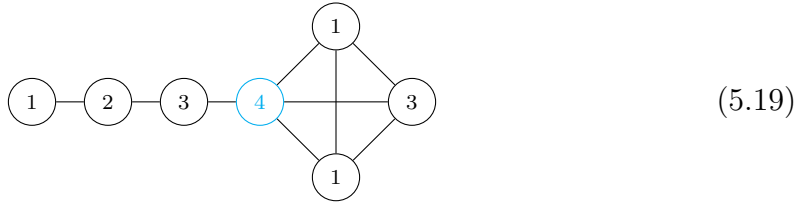
{
\begin{tikzpicture}[baseline=0, font=\scriptsize]
\tikzstyle{every node}=[minimum size=0.5cm]
\node[draw, circle] (a1) at (0,0) {1};
\node[draw, circle] (a2) at (1,0) {2};
\node[draw, circle] (a3) at (2,0) {3};
\node[draw, circle, cyan] (a4) at (3,0) {4};
\node[draw, circle] (a6) at (5,0) {3};
\node[draw, circle] (u) at (4,1) {1};
\node[draw, circle] (b) at (4,-1) {1};
\draw[draw, solid] (a1)--(a2);
\draw[draw, solid] (a2)--(a3);
\draw[draw, solid] (a3)--(a4);
\draw[draw, solid] (a4)--(a6);
\draw[draw, solid] (a4)--(u);
\draw[draw, solid] (a4)--(b);
\draw[draw, solid] (b)--(a6);
\draw[draw, solid] (b)--(u);
\draw[draw, solid] (a6)--(u);
\end{tikzpicture}
}
This is almost the desired theory \eqref{D4SU6mirr}, but now we lack the adjoint chiral at the second to last $U(4)$ node. The solution is to keep iterating the application of Aharony duality along the tail, starting from the leftmost $U(1)$ node and stopping each time one node before, so to correctly reconstruct all the adjoint chirals at each gauge node and recover the manifestly $\mathcal{N}=4$ theory \eqref{D4SU6mirr}.

Next we consider the $D_{4}(SU(10))$ theory whose reduction to 3d gives
\bes{ \label{D4SU10linear}
(D_4(SU(10))_{3d}: \quad [10]-7-SU(5)-2
}
This can be realised as gluing $[10]-7-[5]$ and $[5]-2$ via an $SU(5)$ gauge group.  In terms of the mirror theories, this can be written as
\bes{ \label{D4SU10gluing}
\begin{tikzpicture}[baseline=0, font=\scriptsize]
\tikzstyle{every node}=[minimum size=0.5cm]
\node[draw, circle] (a1) at (0,0) {1};
\node[draw, circle] (a2) at (1,0) {2};
\node[draw=none] (a3) at (2,0) {$\cdots$};
\node[draw, circle] (a6) at (3,0) {6};
\node[draw, circle] (a7) at (4,0) {7};
\node[draw, circle] (a8) at (5,0) {7};
\node[draw, circle] (a9) at (6,0) {6};
\node[draw=none] (a12) at (7,0) {$\cdots$};
\node[draw, circle] (a13) at (8,0) {2};
\node[draw, circle] (a14) at (9,0) {1};
\node[draw, circle] (b) at (4.5,-1) {1};
\draw[draw, solid] (a1)--(a2);
\draw[draw, solid] (a2)--(a3);
\draw[draw, solid] (a3)--(a6);
\draw[draw, solid] (a6)--(a7);
\draw[draw, solid] (a7)--(a8);
\draw[draw, solid] (a8)--(a9);
\draw[draw, solid] (a9)--(a12);
\draw[draw, solid] (a12)--(a13);
\draw[draw, solid] (a13)--(a14);
\draw[draw, solid] (a7)--(b);
\draw[draw, solid] (a8)--(b);
\end{tikzpicture}
\quad + \quad
\begin{tikzpicture}[baseline=0, font=\scriptsize]
\tikzstyle{every node}=[minimum size=0.5cm]
\node[draw, circle] (a1) at (1,0) {1};
\node[draw, circle] (a2) at (2,0) {2};
\node[draw, circle] (a3) at (3,0) {2};
\node[draw, circle] (a4) at (4,0) {1};
\node[draw, circle] (b) at (2.5,-1) {1};
\draw[draw, solid] (a1)--(a2);
\draw[draw, solid] (a2)--(a3);
\draw[draw, solid] (a3)--(a4);
\draw[draw, solid] (a2)--(b);
\draw[draw, solid] (a3)--(b);
\end{tikzpicture}
}
Similarly to the above example, we obtain the following mirror theory (see Appendix \ref{app:deriv3dmirrD4SU10} for more details on this gluing)
\bes{ \label{D4SU10mirr}
(D_4(SU(10))_{\text{3d mirr}}: \quad
&\begin{tikzpicture}[baseline=0, font=\scriptsize]
\tikzstyle{every node}=[minimum size=0.5cm]
\node[draw, circle] (c3) at (-2,0) {3}; \node[below= 0.1cm of c3] {};
\node[draw, circle] (c4) at (-1,0) {6}; \node[below= 0.13cm of c4] {};
\node[draw, circle] (c6) at (1,0)  {7}; \node[below= 0.1cm of c6] {};
\node[draw, circle] (c7) at (2,0) {6}; \node[below= 0.1cm of c7] {};
\node[draw, circle] (c8) at (3,0) {5}; \node[below= 0.1cm of c8] {};
\node[draw=none] (c9) at (4,0) {$\cdots$}; \node[below= 0.1cm of c9] {};
\node[draw, circle] (c10) at (5,0)  {2}; \node[below= 0.1cm of c10] {};
\node[draw, circle] (c11) at (6,0)  {1}; \node[below= 0.1cm of c11] {};
\node[draw, circle] (f1) at (0,1)  {$1$}; \node[below= 0.1cm of f1] {};
\node[draw, circle] (f2) at (0,-1)  {$1$}; \node[below= 0.1cm of f1] {};
\draw[draw, solid] (c3)--(c4)--(c6)--(c7)--(c8)--(c9)--(c10)--(c11);
\draw[draw, solid] (c4)--(f1) node[midway,above, xshift=0.1cm] {} ;
\draw[draw, solid] (c4)--(f2) node[midway,above, xshift=0.1cm] {} ;
\draw[draw, solid]  (f1)--(c6) node[midway,above, xshift=-0.1cm] {};
\draw[draw, solid]  (f2)--(c6) node[midway,above, xshift=-0.1cm] {};
\draw[draw, solid]  (f1)--(f2) node[midway,above, xshift=-0.1cm] {};
\end{tikzpicture}}
where this is in agreement with the result obtained from the steps described earlier with $m_B=m_A=m_G=1$ and $H_{\text{free}}=0$.  Every node on the central line is balanced and each $U(1)$ on top and bottom is overbalanced.  Decoupling an overall $U(1)$, we have the Coulomb branch symmetry $SU(10) \times U(1)$, in agreement with the global symmetry of $D_{4}(SU(10))$.    In this particular example, the Coulomb branch symmetry of \eref{D4SU10linear} is $U(1)^3$ (see Appendix \ref{app:SUenhanced}), which is in agreement with the Higgs branch symmetry of \eref{D4SU10mirr}.

Next we consider $D_{9}(SU(12))$, where we have $m=3$, $x=1$, $m_A=2$, $m_B=1$ and $m_G=2$.  The reduction to 3d gives
\bes{
(D_{9}(SU(12)))_{3d}:  \quad [12]-10-9-SU(8)-6-5-SU(4)-2-1
}
Following the above procedure, we obtain the mirror theory
\be
\tdplotsetmaincoords{70}{110}
\begin{tikzpicture}[tdplot_main_coords,baseline=0,font=\scriptsize]
\tikzstyle{every node}=[minimum size=0.5cm]
\node[draw, circle] (c4) at (0,-1,0) {8}; \node[below= 0.13cm of c4] {};
\node[draw, circle] (c6) at (0,3,0)  {10}; \node[below= 0.1cm of c6] {};
\node[draw, circle] (c7) at (0,4,0) {9}; \node[below= 0.1cm of c7] {};
\node[draw, circle] (c8) at (0,5,0) {8}; \node[below= 0.1cm of c8] {};
\node[draw=none] (c9) at (0,6,0) {$\ldots$}; \node[below= 0.1cm of c9] {};
\node[draw, circle] (c10) at (0,7,0)  {2}; \node[below= 0.1cm of c10] {};
\node[draw, circle] (c11) at (0,8,0)  {1}; \node[below= 0.1cm of c11] {};
\node[draw, circle] (f1) at (0,1,2)  {$1$}; \node[below= 0.1cm of f1] {};
\node[draw, circle] (f2) at (0,1,-2)  {$1$}; \node[below= 0.1cm of f1] {};
\node[draw, circle] (f3) at (-4.5,1,0)  {$1$}; \node[below= 0.1cm of f1] {};
\draw[draw, solid] (c4)--(c6)--(c7)--(c8)--(c9)--(c10)--(c11);
\draw[draw, solid, double, thick] (c4)--(f1) node[midway,above, xshift=0.1cm] {} ;
\draw[draw, solid, double, thick] (c4)--(f2) node[midway,above, xshift=0.1cm] {} ;
\draw[draw, solid, double, thick] (c4)--(f3) node[midway,above, xshift=0.1cm] {} ;
\draw[draw, solid]  (f1)--(c6) node[midway,above, xshift=-0.1cm] {};
\draw[draw, solid]  (f2)--(c6) node[midway,above, xshift=-0.1cm] {};
\draw[draw, solid]  (f3)--(c6) node[midway,above, xshift=-0.1cm] {};
\draw[draw, solid,double, thick]  (f1)--(f2)--(f3)--(f1) node[midway,above, xshift=-0.1cm] {};
\end{tikzpicture}
\ee
with $H_{\text{free}} =0$. The Higgs branch symmetry of this quiver is $U(1)^4 \times SU(2)^6$.  The Coulomb branch symmetry of this quiver is $SU(12) \times U(1)^2$.

Next let us consider the case of $N=p\mathfrak{m}$ with a positive integer $\mathfrak{m}$, the $D_p(SU(p\fm))$ theory and the 3d theory upon reduction admit the following quiver description:
\bes{ \label{linearSU}
(D_p(SU(p\fm)))_{3d}: \quad [p \fm] - SU((p-1)\fm) - SU((p-2)\fm) -\cdots -SU(\fm)
}
In this case we have $x=\fm$, $m=p$, $M=(p-1)\fm-1$, $m_B=0$, $m_A =1$, $m_G=0$ and $H_{\text{free}} =0$, and so we have the mirror theory
\bes{\label{mirrpdividesN}
&\begin{tikzpicture}[baseline=0, font=\scriptsize]
\tikzstyle{every node}=[minimum size=0.5cm]
\node[draw,circle] (cm1) at (-6,0) {}; \node[below= 0.1cm of cm1] {\!\!\!\!\!\! $p-1$};
\node[draw,circle] (c0) at (-5,0) {}; \node[below= 0.1cm of c0] {\,\,\, $2(p-1)$};
\node[draw=none] (c1) at (-4,0) {$\cdots$}; \node[below= 0.1cm of c1] {};
\node[draw, circle] (c2) at (-3,0) {}; \node[below= 0.1cm of c2] {\!\!\!\!\!\! $\substack{(p-1)\\ (\fm-2)}$};
\node[draw, circle] (c3) at (-2,0) {}; \node[below= 0.1cm of c3] {$\substack{(p-1) \\(\fm-1)}$};
\node[draw, circle] (c4) at (-1,0) {}; \node[below= 0.13cm of c4] {\,\,\,\, $(p-1)\fm$};
\node[draw, circle] (c6) at (1,0)  {}; \node[below= 0.1cm of c6] {$\substack{(p-1)\fm \\-1}$};
\node[draw, circle] (c7) at (2,0) {}; \node[below= 0.1cm of c7] {$\substack{(p-1)\fm \\-2}$};
\node[draw, circle] (c8) at (3,0) {}; \node[below= 0.1cm of c8] {$\substack{(p-1)\fm \\-3}$};
\node[draw=none] (c9) at (4,0) {$\cdots$}; \node[below= 0.1cm of c9] {};
\node[draw, circle] (c10) at (5,0)  {}; \node[below= 0.1cm of c10] {$2$};
\node[draw, circle] (c11) at (6,0)  {}; \node[below= 0.1cm of c11] {$1$};
\node[draw,circle] (b1) at (-2,1) {$1$}; \node[above= 0.1cm of b1] {};
\node[draw,circle] (b2) at (-1,1) {$1$}; \node[above= 0.1cm of b2] {};
\node[draw=none] (b3) at (0,1) {$\ldots$}; \node[above= 0.1cm of b3] {};
\node[draw,circle] (b4) at (1,1) {$1$}; \node[above= 0.1cm of b3] {};
\draw[draw, solid] (cm1)--(c0)--(c1)--(c2)--(c3)--(c4)--(c6)--(c7)--(c8)--(c9)--(c10)--(c11);
\draw[draw, solid] (c4)--(b1);
\draw[draw, solid] (c4)--(b2);
\draw[draw, solid] (c4)--(b4);
\draw [decorate,decoration={brace,amplitude=10pt}] (-2.5,1.3) -- (1.5,1.3) node [black,midway, yshift=0.6cm] {\footnotesize $p$};
\end{tikzpicture}
}
This is as expected for the mirror theory for \eref{linearSU} (see \eg~ \cite{Hanany:2018vph, Bourget:2019rtl, talkZhong}).  For $p=2$, this is simply the mirror theory of the $SU(\fm)$ gauge theory with $2\fm$ flavours \cite{Hanany:1996ie}.

Finally, let us consider the case of $\GCD(p,N)=m=1$.  The 3d theory upon reduction can be identified with $H_{\text{free}}$ twisted hypermultiplets and the $T_{\vec \rho}(SU(N))$ theory, such that
\bes{
\vec{\rho} = \left((x+1)^{N-px}, x^{p(x+1)-N} \right)~,} 
whose quiver description is 
\bes{
&[N] - (N-(x+1))- (N-2(x+1)) - \cdots- (x(p(x+1)-N))- \\
&- (x(p(x+1)-N-1))- \cdots - (x)~.
}
Note that every gauge node is balanced, except $(x(p(x+1)-N))$, which is overbalanced.
The corresponding mirror theory is then $T^{\vec \rho} (SU(N))$ theory whose quiver description is
\bes{ 
&\begin{tikzpicture}[baseline=0, font=\scriptsize]
\tikzstyle{every node}=[minimum size=0.5cm]
\node[draw,circle] (cm1) at (-6,0) {}; \node[below= 0.1cm of cm1] {\!\!\!\!\!\! $p-1$};
\node[draw,circle] (c0) at (-5,0) {}; \node[below= 0.1cm of c0] {\,\,\, $2(p-1)$};
\node[draw=none] (c1) at (-4,0) {$\cdots$}; \node[below= 0.1cm of c1] {};
\node[draw, circle] (c2) at (-3,0) {}; \node[below= 0.1cm of c2] {\!\!\!\!\!\! $\substack{(p-1)\\ (x-2)}$};
\node[draw, circle] (c3) at (-2,0) {}; \node[below= 0.1cm of c3] {$\substack{(p-1) \\(x-1)}$};
\node[draw, circle] (c4) at (-1,0) {}; \node[below= 0.13cm of c4] {$ (p-1)x$};
\node[draw, circle] (c6) at (1,0)  {}; \node[below= 0.1cm of c6] {$M$};
\node[draw, circle] (c7) at (2,0) {}; \node[below= 0.1cm of c7] {$M-1$};
\node[draw, circle] (c8) at (3,0) {}; \node[below= 0.1cm of c8] {$M-2$};
\node[draw=none] (c9) at (4,0) {$\cdots$}; \node[below= 0.1cm of c9] {};
\node[draw, circle] (c10) at (5,0)  {}; \node[below= 0.1cm of c10] {$2$};
\node[draw, circle] (c11) at (6,0)  {}; \node[below= 0.1cm of c11] {$1$};
\node[draw, circle] (f1) at (0,1.2)  {$1$}; \node[below= 0.1cm of f1] {};
\draw[draw, solid] (cm1)--(c0)--(c1)--(c2)--(c3)--(c4)--(c6)--(c7)--(c8)--(c9)--(c10)--(c11);
\draw[draw, solid, very thick, blue] (c4)--(f1) node[midway,above, xshift=-1cm] {$p(x+1)-N$} ;
\draw[draw, solid, very thick, red]  (f1)--(c6) node[midway,above, xshift=0.6cm] {$N-px$};
\end{tikzpicture}\\
&\text{+ $\frac{1}{2} (N-p x-1) [p (x+1)-N-1]$ hypermultiplets}~.
}
Since every node in the central line is balanced, the Coulomb branch symmetry of the mirror theory is $SU(N)$.  The Higgs branch symmetry of the mirror theory (excluding the contribution of the free hypermultiplets) is $SU(m_A) \times SU(m_B) \times U(1) = SU(p(x+1)-N) \times SU(N-px) \times U(1)$.

\section{Mirror theories for $D_p^{N-1}(SU(N))$ and $(I_{p-(N-1),N-1},S)$ with $p \geq N-1$} \label{sec:TypeIIp>Nm1}
The method of constructing a mirror theory for the $D_p^{N-1}(SU(N))$ with $p \geq N-1$ is very similar to that discussed in section \ref{sec:DpSUNpgeqN}.  We describe it as follows.  Let us first define the shorthand notations:
\bes{
m= \GCD(p, N-1)~, \qquad  n= \frac{N-1}{m}~, \qquad q= \frac{p}{m}~.
}
Then, the procedure of construction of the mirror theory is the following:
\ben
\item Start with a complete graph with $m$ vertices such that each vertex corresponds to a $U(1)$ gauge group and each edge has multiplicity 
\bes{ n(q-n)=\frac{1}{m^2}(N-1)(p-N+1)~.}  The multiplicity of the edge is equal to the number of bifundamental hypermultiplet between the corresponding pair of gauge groups. 
\item Take another $U(1)$ gauge node and connect it with each $U(1)$ node in the complete graph by the edges, each with equal multiplicity 
\bes{q-n=\frac{1}{m}(p-N+1)~.} \label{step2}
\item Construct the tail $(N-1)-(N-2)-\cdots-2-1$.
\item Connect the $(N-1)$ node of the tail to the complete graph such that the node $U(N-1)$ of the tail is connected to each of the $m$ $U(1)$ nodes in the complete graph by the edges, each with equal multiplicity $n =(N-1)/m$.  
\item Connect the $(N-1)$ node of the  tail to the $U(1)$ node in step \ref{step2} with an edge with multiplicity 1.  Note that this quiver has an overall $U(1)$ that needs to be decoupled.  \label{step3}
\item The number of free hypermultiplets is equal to 
\bes{ \label{freehypTypeIIp>Nm1}
H_{\text{free}} &= \frac{1}{2}m(n-1)(q-n-1)\\
&= \frac{1}{2m} (N-1-m)(p-N+1-m)~. 
} \label{step4}
\item The quiver obtained in Step \ref{step3}, together with the free hypermultiplets in Step \ref{step4}, defines a mirror theory for a circle reduction of the $D^{N-1}_p(SU(N))$ with $p\geq N-1$.
\een

After decoupling the tail $(N-1)-(N-2)-\cdots-2-1$, we obtain the mirror theory for the 3d reduction of the 4d $(I_{p-(N-1),N-1},S)$ theory.  As described below \eref{ccentral}, the latter can be realised as the 6d $\mathcal{N}=(2,0)$ of Type $A_{p-N}$ theory on the sphere with one regular simple puncture (denoted by $S$) and the same irregular puncture of Type I that engineers the $I_{p-(N-1), N-1}= (A_{p-N},A_{N-2})$ theory.  Such a mirror theory consists of the following components:
\ben
\item A complete graph of $m$ $U(1)$ nodes, where each edge has multiplicity $n(q-n)=\frac{1}{m^2}(N-1)(p-N+1)$.
\item Another $U(1)$ gauge node connected each of the $m$ nodes in the complete graph with the edges, each with multiplicity $q-n=\frac{1}{m}(p-N+1)$.
\item $H_{\text{free}}$ free hypermultiplets.
\een

\subsubsection*{The Coulomb branch of the mirror theory}
The Coulomb branch symmetry of the mirror theory for $D^{N-1}_p(SU(N))$ is, in general, $SU(N) \times U(1)^{m}$, where the $SU(N)$ factor comes from the tail and the $U(1)^{m}$ comes from the topological symmetries of the $m+1$ $U(1)$ nodes in the quiver, where there is an overall $U(1)$ that decouples.  Note that this can get enhanced further when certain $U(1)$ gauge nodes are balanced.  In any case, the rank of the Coulomb branch symmetry is $(N-1)+m$.  This agrees with the rank of the flavour symmetry of the $D^{N-1}_p(SU(N))$ theory given by \eref{ranks}.   The Coulomb branch dimension of the mirror theory described above is 
\bes{ \label{dimCmirrTypeIIp>N-1}
\frac{1}{2}N(N-1) +m~.
}

As before, we can compute the difference between the value of $24(c-a)$ for the $D^{N-1}_p(SU(N))$
\bes{\label{24cmaDN1p}
&\text{$24(c-a)[D^{N-1}_p(SU(N))]$} \\
&= \frac{1}{2} (N^2+m -1) - \frac{3(N-1)}{p} \left[\sum_{j=1}^{N-2}  \left \{ \frac{j p}{N-1} \right \}   \left(1-  \left \{  \frac{j p}{N-1} \right \} \right) \right]~.
}
and the dimension \eref{dimCmirrTypeIIp>N-1} of the Coulomb branch of the 3d mirror:
\bes{ 
\delta_H(N,p) &= \eref{24cmaDN1p}- \eref{dimCmirrTypeIIp>N-1}\\
&=  \frac{1}{2} (N-m-1) - \frac{3(N-1)}{p} \left[\sum_{j=1}^{N-2}  \left \{ \frac{j p}{N-1} \right \}   \left(1-  \left \{  \frac{j p}{N-1} \right \} \right) \right]~.  
}
This quantity corresponds to the value of $24(c-a)$ of the non higgsable SCFTs.  We tabulate a few values of $(N,p)$, where $\delta_H(N, p)$ is non-zero, and identify the ``non higgsable'' SCFTs:
\be
\begin{tabular}{|c|c|c|c|}
\hline
$(N,p)$ & $\delta_H$ & non higgsable SCFT & $H_{\text{free}} $\\
\hline
$(3,5)$ & $1/5$ & $I_{2,3} = (A_1, A_2)$ & 1 \\
$(3,7)$ & $2/7$ & $I_{2,5}=(A_1,A_4)$ & 2 \\
$(4,5)$ & $1/5$ &  $I_{2,3} = (A_1, A_2)$ & 1\\
$(4,7)$ & $3/7$ & $I_{3,4} = (A_2, A_3)$ &3\\
$(5,7)$ & $3/7$ & $I_{3,4} = (A_2, A_3)$ &3\\
$(6,7)$ & $2/7$ & $I_{2,5}=(A_1,A_4)$ & 2 \\
\hline
\end{tabular}
\ee 
This is in agreement with the general discussion around \eref{hbflow2}. Note that $H_{\text{free}} $ is equal to the rank of the corresponding non higgsable SCFT.

\subsubsection*{The Higgs branch of the mirror theory}
The Higgs branch dimension of the mirror quiver theory is
\bes{\label{dimHmirrTypeIIp>N-1}
\frac{1}{2} \left[ (N-1)(p-1) -\frac{(N-1)(p-N+1)}{m}+2p-2m  \right]~.
}
After adding to it the contribution $H_{\text{free}}$, this agrees with the rank of the corresponding 4d $D^{N-1}_p(SU(N))$ theory given by \eref{ranks}.
The Higgs branch symmetry (neglecting the contribution of the free hypermultiplets) is, in general, $U(1)^{m+1} \times SU(n)^m \times SU(q-n)^m \times SU(nq-n^2)^{\frac{1}{2}m(m-1)}$ (assuming that $m \geq 2$).\footnote{It should be noted that for $m=1$ we have one $U(1)$ factor; see \eref{mirrTypeIIgcd1}.}  This, again, can be compared with the Coulomb branch symmetry of the linear quiver coming from the 3d reduction of the $D^{N-1}_p(SU(N))$ theory (after dualising and neglecting all possible twisted hypermultiplets)\footnote{It can be checked that the quaternionic Coulomb branch dimension of this linear quiver is equal to \eref{dimHmirrTypeIIp>N-1}; adding to it $H_{\text{free}}$, which is the number of twisted hypermultiplets obtained upon compactification, this is equal to the rank of the $D_p(SU(N))$ theory.  Moreover, the quaternionic Higgs branch dimension of this linear quiver is \eref{dimCmirrTypeIIp>N-1}; adding to it the values of $24(c-a)$ of the non higgsable theories, we obtain the value of $24(c-a)$ of the $D_p(SU(N))$ theory.}:
\be \label{linearquiv3dDpSUnTypeII}
\scalebox{0.8}{$
\begin{split} 
&[N]-(N-1)-(N-2)-\cdots-(N-n+1)-(N-n)^{q-n}-SU(N-n)\\ 
&-(N-n-1)-(N-n-2)-\cdots-(N-2n+1)-(N-2n)^{q-n}-SU(N-2n)- \cdots \\
&-SU(2n+1)-(2n)- \cdots-(n+2)-(n+1)^{q-n}-SU(n+1) - (n) - \cdots -(2)-(1)^{q-n}-[1]
\end{split}$}
\ee
where there are in total $(m-1)$ special unitary gauge groups $SU(N-jn)$ with $1\leq j \leq m-1$ and the notation $(x)^{q-n}$ indicates that the chain $(x)-\cdots-(x)$ where $(x)$ appears $q-n$ times.  The explanation for the Coulomb branch symmetry of \eref{linearquiv3dDpSUnTypeII} is similar to that of \eref{linearquiv3dDpSUn} and is given in Appendix \ref{app:SUenhanced}.  It is instructive to compare the former to the latter: we have additional factors of $U(1)$ and $SU(q-n)^m$ in the former.  The additional $U(1)$ comes from the leftmost overbalanced $U(1)$ gauge node inside the chain $(1)^{q-n}$.  Inside the chain $(1)^{q-n}$ there are also $q-n-1$ balanced $U(1)$ gauge nodes which give rise to a factor of $SU(q-n)$.  Since quiver \eref{linearquiv3dDpSUnTypeII} has $[1]$ at the right end, the collection of balanced nodes in each chain $(N-j n)^{q-n}$, with $1\leq j \leq m$, gives rise to the factor $SU(q-n)^m$.

\subsubsection*{Comparison with SUSY enhancing RG flows}

We can again compare our findings with the prediction from the Maruyoshi--Song (MS) flow. As in the $b=N$ case we can parametrize the trial R-symmetry of the RG flow such that all CB operators $u_{ij}$ have charge $D_{UV}(u_{ij})(1+\epsilon)$. In this case (see \cite{Giacomelli:2017ckh}) the trial $a$ central charge is maximized at $\epsilon=-\frac{p+3N-3}{3p+3N-3}$ and therefore the dimension of CB operators in the IR is 
\be D_{IR}(u_{ij})=D_{UV}(u_{ij})\frac{p}{p+N-1}.\ee 
We therefore conclude that only CB operators whose dimension in the UV is in the range $1<D_{UV}(u_{ij})\leq 1+\frac{N-1}{p}$ can hit the unitarity bound and decouple. Since the dimension in the UV of $u_{ij}$ is $D_{UV}(u_{ij})=N-j-\frac{N-1}{p}i$ and the dimension of $z$ is $\frac{N-1}{p}$, we easily see that for each value of $j$ there is exactly one value of $i$ such that $u_{ij}$ decouples in the infrared, giving in total $N-1$ decoupled operators. 

Again, this implies that in the 3d mirror we have to flip $N-1$ bilinears in the bifindamental hypermultiplets charged under the node $U(N-1)$ and the $m+1$ abelian nodes in the complete graph. As a result, when using the flip-flip duality and remove the $T(SU(N))$ tail to implement the MS flow, we produce $m\frac{n^2-n}{2}$ free hypermultiplets. Along the way, we increase by $n^2$ the number of edges between the abelian nodes in the complete graph and by $n$ the number of bifundamentals between the nodes of the complete graph and the extra $U(1)$ node introduced in step \ref{step2}. 

The interpretation is the following: Starting from $D_p^{N-1}(SU(N))$ and following the MS flow we get in the IR the same SCFT we find by closing the regular puncture of $D_{p+N-1}^{N-1}(SU(N))$ theory. We therefore expect that with the shift $p\rightarrow p+N-1$ the number of free hypermultiplets increases by  $m\frac{n^2-n}{2}$, the number of edges in the complete graph increases by $n^2$ and the number of edges between the $U(1)$ node of step \ref{step2} and the nodes in the complete graph increases by $n$. This is indeed perfectly consistent with our algorithm and provides again a solid consistency check of our claim.

\subsection{Examples}
For $\GCD(N-1,p)=1$, we have $m=1$, $n=N-1$ and $q=p$, the reduction to 3d of $D_p^{N-1}(SU(N))$ gives
\be
\begin{split}
&T^{[p-1,1^N]}_{[2^{N-1},1^{p-N-1}]} (SU(N+p-1)): \\ 
& \qquad [N]-(N-1)-(N-2)-\cdots-2-1-\underbrace{1-\cdots-1}_{p-N}-[1]~,
\end{split}
\ee
together with $\frac{1}{2}(N-2)(p-N)$ twisted hypermultiplets.  The mirror theory of this system can be written as 
\bes{ \label{mirrTypeIIgcd1}
(D^{N-1}_p(SU(N)))^{\GCD(p,N-1)=1}_{\text{3d mirr}}: \quad\,\,
&\scalebox{0.9}{
\begin{tikzpicture}[baseline=0, font=\small]
\tikzstyle{every node}=[minimum size=0.5cm]
\node[draw, circle] (c2) at (-2,0) {$1$};
\node[draw, circle] (c3) at (0,0) {} node[draw=none] at (0,-0.5) {\footnotesize $N-1$};
\node[draw, circle] (c4) at (1,0) {} node[draw=none] at (1,-0.5) {\footnotesize $N-2$}; 
\node[draw=none] (c5) at (2,0)  {$\cdots$}; 
\node[draw, circle] (c6) at (3,0)  {$2$}; 
\node[draw, circle] (c7) at (4,0)  {$1$}; 
\node[draw, circle] (f1) at (-1,1)  {$1$}; 
\draw[draw, solid] (c2)--(c3);
\draw[draw, solid] (c3)--(c4)--(c5)--(c6)--(c7);
\draw[draw, solid] (f1)--(c2);
\draw[very thick, gray] (c2)--(f1) node at (-2.4,0.6)  {\footnotesize $p-N+1$};
\draw[very thick, red] (c3)--(f1) node at (0,0.6)  {\footnotesize $N-1$};
\end{tikzpicture}} \\
& \text{+ $\frac{1}{2}(N-2)(p-N)$ hypers}~.
}
This is in agreement with the result obtained from the above procedure.  The complete graph consists of only one node, which is depicted as the top $U(1)$ node.   The left bottom $U(1)$ gauge node is that mentioned in step \ref{step2}.  The Coulomb branch symmetry of the quiver is $SU(N) \times U(1)$ and the Higgs branch symmetry of the quiver is $SU(p-N+1) \times SU(N-1) \times U(1)$. Upon decoupling the tail, we obtain the mirror theory of $(I_{p-(N-1), \, N-1},S)$ as follows:
\bes{
(I_{p-(N-1), \, N-1},S)^{\GCD(p,N-1)=1}_{\text{3d mirr}}: \qquad  &1-[p-N+1] \\
& \text{+ $\frac{1}{2}(N-2)(p-N)$ hypers}~.
}

Another interesting example is when $\GCD(p,N)=3$. In which case, we have the following mirror theories:
\bes{
(D^{N-1}_p(SU(N)))^{\GCD(p,N-1)=3}_{\text{3d mirr}}:  \quad
&\scalebox{0.8}{
\begin{tikzpicture}[baseline=0]
\def\n{5}
\node[circle,minimum size=3 cm] (b) {};
\foreach\x in{1,...,4}{
  \node[minimum size=0.75cm,draw,circle] (n-\x) at (b.{360/\n*\x}){1};
}
\foreach\x in{5}{
  \node[minimum size=0.75cm,draw,circle] (n-\x) at (b.{360/\n*\x}) {} node[draw=none] at (1.6,-0.5) {\scriptsize $N-1$};
}
\foreach\x in{1,...,3}{
  \foreach\y in{1,...,3}{
    \ifnum\x=\y\relax\else
      \draw (n-\x) edge[very thick, blue] (n-\y);
    \fi
  }
 \foreach\y in{5}{
    \ifnum\x=\y\relax\else
      \draw (n-\x) edge[very thick, red] (n-\y);
    \fi
  } 
 \foreach\y in{4}{
    \ifnum\x=\y\relax\else
      \draw (n-5) edge[thick, black] (n-\y);
      \draw (n-\x) edge[very thick, gray] (n-\y);
    \fi
  }   
\node[draw=none] at (-2,0) {\blue $n(q-n)$};  
\node[draw=none] at (1.2,1) {\red $n$};  
\node[draw=none] at (1.2,-1) {$1$};  
\node[draw=none] at (-0.5,-1.5) {\gray $q-n$};  
\node[draw,circle, minimum size=0.75cm,] (6a) at (3,0) {} node[draw=none] at (3,-0.5) {\scriptsize $N-2$};
\node[draw=none] (4a) at (4.5,0) {$\cdots$};
\node[draw,circle] (2a) at (6,0) {2};
\node[draw,circle] (1a) at (7.5,0) {1};
\draw (n-5)--(6a)--(4a)--(2a)--(1a);
}
\end{tikzpicture}}\\
&\text{+ $\frac{1}{6}(N-4)(p-N-2)$ hypers}
}
The Coulomb branch symmetry of the quiver is $SU(N) \times U(1)^3$, and the Higgs branch symmetry of the quiver is $SU(n(q-n))^3 \times SU(q-n)^3 \times SU(n)^3 \times U(1)^4$.  Upon decoupling the tail, we obtain the mirror theory for the $(I_{p-(N-1), \, N-1},S)$ theory, with ${\GCD(p,N-1)=3}$:
\bes{
(I_{p-(N-1), \, N-1},S)^{\GCD(p,N-1)=3}_{\text{3d mirr}}: \qquad
&\scalebox{0.7}{
\begin{tikzpicture}[baseline=0]
\def\n{4}
\node[draw=none] at (-1.3,1) {\blue $n(q-n)$};
\node[draw=none] at (1.3,1) {\gray $q-n$};
\node[circle,minimum size=3 cm] (b) {};
\foreach\x in{1,...,\n}{
  \node[minimum size=0.75cm,draw,circle] (n-\x) at (b.{360/\n*\x}){1};
}
\foreach\x in{1,...,3}{
  \foreach\y in{1,...,3}{
    \ifnum\x=\y\relax\else
      \draw (n-\x) edge[very thick, blue] (n-\y);
    \fi
  }
 \foreach\y in{4}{
    \ifnum\x=\y\relax\else
      \draw (n-\x) edge[very thick, gray] (n-\y);
    \fi
  }     
} 
\end{tikzpicture}} \\ 
& \text{+ $\frac{1}{6}(N-4)(p-N-2)$ hypers}
}

Finally, let us consider the case of $\GCD(p,N-1)=N-1$, \ie~ $p=(N-1) \fm$ for some integer $\fm \geq 1$.  In this case, $m=N-1$, $n=1$ and $q=\fm$.  The 3d reduction of the corresponding $D^{N-1}_{(N-1) \fm}(SU(N))$ is
\bes{
&[N]-(N-1)^{\fm-1}-SU(N-1)-(N-2)^{\fm-1}-SU(N-2)-\cdots \\
&-(2)^{\fm-1}-SU(2)- (1)^{\fm-1}-[1]
}
where $H_{\text{free}} =0$. This quiver takes the same form as \eref{3dredpmultN}, and indeed we have the identification
\bes{
D^{N-1}_{(N-1) \fm}(SU(N))  = D_{N \fm} (SU(N))~.
}
and so the corresponding mirror theory is the same as described below \eref{3dredpmultN}.  Decoupling the tail $(N-1)-\cdots-(1)$, we obtain the mirror theory for $(I_{(N-1)(\fm-1), \, (N-1) \fm},S) = I_{N (\fm-1), N \fm} $: a complete graph of $N$ $U(1)$ nodes where each edge has multiplicity $\fm-1$.

\section{Mirror theory for $D^{N-1}_p(SU(N))$ with $p \leq N-1$} \label{sec:TypeIIp<Nm1}
The discussion is similar to that in section \ref{sec:DpSUNplN}.  Let us define the following shorthand notations:
\bes{
\begin{array}{lll}
x =\lfloor (N-1)/p \rfloor~,  & \qquad m = \GCD(p,N-1) ~, & \qquad M= N-(x+1)~, \\
n= (N-1)/m ~, & \qquad q= p/m~. &
\end{array}
}
The mirror theory for $D^{N-1}_p(SU(N))$ with $p \leq N-1$ can be constructed as follows:
\ben
\item Start with a complete graph with $m$ $U(1)$ nodes where each edge has multiplicity 
\bes{
m_G = m_A m_B &=(q(1+x)-n) (n-q x) \\
&= \left[ \frac{p(x+1)-N}{m} \right]\left[ \frac{N-1-px}{m} \right]
}
where $m_A = q(1+x)-n$ and $m_B= n-q x$.
\item Construct two types of tails:
\be \nn
\scalebox{1}{$
\begin{split}
&\text{Tail A}: \quad 1-((p-1)+1)-(2(p-1)+1)-\cdots-((p-1)(x-2)+1) \\
& \qquad \qquad \qquad -((p-1)(x-1)+1)-((p-1)x+1) \\
&\text{Tail B}: \quad M-(M-1)-(M-2)-\ldots-2-1
\end{split}
$}
\ee
\item Attach the node $(p-1)x+1$ of tail A to each $U(1)$ node in the complete graph with edges with equal multiplicity 
\bes{
m_A = q(1+x)-n =\frac{p(x+1)-N+1}{m} ~.
}
\item Attach the node $M$ of tail B to each $U(1)$ node in the complete graph with edges with equal multiplicity 
\bes{
m_B = n-q x =\frac{N-1-px}{m} ~.
}
\item Connect the node $(p-1)x$ of tail A to the node $M$ of tail B with an edge with multiplicity $1$.
\item The quiver constructed above has an overall $U(1)$ that needs to be decoupled.
\item There are 
\bes{
H_{\text{free}} &= \frac{1}{2}m(n-qx-1)(q(x+1)-n-1) \\
&= \frac{(N-p x-1-m) (p (x+1)-N+1-m)}{2m} 
}
free hypermultiplets.
\item The above quiver, together with $H_{\text{free}}$ free hypermultiplets, is the mirror theory for the $D^{N-1}_p(SU(N))$ theory with $p \leq N-1$.
\een

\subsubsection*{The Coulomb branch of the mirror theory}
In general, the Coulomb branch symmetry of this mirror theory is $SU(N) \times U(1)^{m}$. The quaternionic dimension of the Coulomb branch of such a theory is
\bes{ \label{dimCTypeIIp<N-1}
\frac{1}{2}N(N-1)+m+ \frac{1}{2} p x (x+1) - (N-1)x~.  
}
As before, we can compute the difference between the value of $24(c-a)$ for the $D^{N-1}_p(SU(N))$ theory and \eref{dimCTypeIIp<N-1}:
\bes{ 
\delta_H(N,p) &= \eref{24cmaDN1p}- \eref{dimCTypeIIp<N-1}\\
&=  \frac{1}{2} (N-m-1) -\frac{1}{2} p x (x+1) + (N-1)x \\
& \qquad  - \frac{3(N-1)}{p} \left[\sum_{j=1}^{N-2}  \left \{ \frac{j p}{N-1} \right \}   \left(1-  \left \{  \frac{j p}{N-1} \right \} \right) \right]~.  
}
This quantity corresponds to the value of $24(c-a)$ of the non higgsable SCFTs.  We tabulate a few values of $(N,p)$, where $\delta_H(N, p)$ is non-zero, and identify the ``non higgsable'' SCFTs:
\be
\begin{tabular}{|c|c|c|c|}
\hline
$(N,p)$ & $\delta_H$ & non higgsable SCFT & $H_{\text{free}} $\\
\hline
$(8,5)$ & $1/5$ & $I_{2,3} = (A_1, A_2)$ & 1 \\
$(9,5)$ & $1/5$ & $I_{2,3} = (A_1, A_2)$ & 1 \\
$(10,7)$ & $2/7$ & $I_{2,5}=(A_1,A_4)$ & 2 \\
$(11,7)$ & $3/7$ & $I_{3,4} = (A_2, A_3)$ & 3 \\
$(12,7)$ & $3/7$ & $I_{3,4} = (A_2, A_3)$ & 3\\
$(12,8)$ & $1/2$ &  $I_{3,5} = (A_2,A_4)$ & 4 \\
$(12,9)$ & $1/3$ & $I_{2,7}=(A_1,A_6)$ & 3\\
\hline
\end{tabular}
\ee 
Note that $H_{\text{free}} $ is equal to the rank of the corresponding non higgsable SCFT.

\subsubsection*{The Higgs branch of the mirror theory}
It can be checked that the quaternionic Higgs branch dimension of the quiver theory constructed using the above prescription is
\bes{ \label{dimHTypeIIp<N-1}
\frac{1}{2} \left[(N-1) (p-1)-\frac{(N-p x-1) (p (x+1)-N+1)}{m}+2p-2 m \right]~.
}
After adding to it $H_{\text{free}}$, we obtain the rank of the 4d $D_{p}(SU(N))$ theory given by \eref{ranks}.    In general the Higgs branch symmetry (neglecting the contribution of the free hypermultiplets) is $U(1)^{m+1} \times SU(m_A)^m\times SU(m_B)^m \times SU(m_G)^{\frac{1}{2}m(m-1)}$, where we emphasise that $m_G= m_A m_B$.  It is interesting to compare with this with the Coulomb branch symmetry of the linear quiver coming from the 3d reduction of the $D_p(SU(N))$ theory (after dualising and neglecting all possible twisted hypermultiplets)\footnote{It can be checked that the quaternionic Coulomb branch dimension of this linear quiver is \eref{dimHTypeIIp<N-1}.  After adding to it $H_{\text{free}}$, which is the number of twisted hypermultiplets obtained upon compactification, is equal to the rank of the $D_p(SU(N))$ theory.  Moreover, the quaternionic Higgs branch dimension of this linear quiver is \eref{dimCTypeIIp<N-1}; after adding to it the values of $24(c-a)$ of the non higgsable theories, this is equal to $24(c-a)$ of the $D_p(SU(N))$ theory.}:
\be \label{linearquiv3dDpSUnp<nII}
\scalebox{0.9}{$
\begin{split} 
&[N]-(N-(x+1))-\cdots -(N-m_B(x+1)) -C[N-n]-SU(N-n)\\ 
&-(N-n-(x+1))-\cdots -(N-n-m_B(x+1)) -C[N-2n]-SU(N-2n)- \cdots \\
&-SU(2n+1)-(2n+1-(x+1))-\cdots -(2n+1-m_B(x+1))-C[n+1]-SU(n+1)\\
&  - (n+1-(x+1))  - \cdots -(n+1-m_B(x+1))-C[1]-[1]
\end{split}$}
\ee
where we use the shorthand notation for the following chain:
\bes{
C[Y] = (Y+(m_A-1) x )- (Y+(m_A-2) x ) -\ldots -(Y+2x)-(Y+x)~.
}
Note that there are in total $(m-1)$ special unitary gauge groups.


\subsection{Examples}
We first consider the case of $D^{10}_4(SU(11))$.  The reduction to 3d gives the following linear quiver
\bes{
[11]-8-SU(6)-3-[1]
}
where this can be viewed as gluing $[11]-8-[6]$ and $[6]-3-[1]$.   In terms of the mirror theories, this can be written as
\bes{
\begin{array}{lllll}
1-2-3-\cdots -7 -&8-&8-7-\cdots-3-2-1 \quad + \quad  1-2-&3-&3-2-1 \\
& \, | & \, | & \, | & \, | \\
&\! [1] & \! [1] & \! [1] & \! [1] 
\end{array}
}
Following the steps discussed above, we have the following mirror theory
\bes{ \label{D4SU11IImirr}
(D^{10}_4(SU(11))_{\text{3d mirr}}: \quad
\scalebox{0.85}{
\begin{tikzpicture}[baseline=0, font=\scriptsize]
\tikzstyle{every node}=[minimum size=0.5cm]
\node[draw, circle] (c2) at (-3,0) {1}; \node[below= 0.1cm of c2] {};
\node[draw, circle] (c3) at (-2,0) {4}; \node[below= 0.1cm of c3] {};
\node[draw, circle] (c4) at (-1,0) {7}; \node[below= 0.13cm of c4] {};
\node[draw, circle] (c6) at (1,0)  {8}; \node[below= 0.1cm of c6] {};
\node[draw, circle] (c7) at (2,0) {7}; \node[below= 0.1cm of c7] {};
\node[draw, circle] (c8) at (3,0) {6}; \node[below= 0.1cm of c8] {};
\node[draw=none] (c9) at (4,0) {$\cdots$}; \node[below= 0.1cm of c9] {};
\node[draw, circle] (c10) at (5,0)  {2}; \node[below= 0.1cm of c10] {};
\node[draw, circle] (c11) at (6,0)  {1}; \node[below= 0.1cm of c11] {};
\node[draw, circle] (f1) at (0,1)  {$1$}; \node[below= 0.1cm of f1] {};
\node[draw, circle] (f2) at (0,-1)  {$1$}; \node[below= 0.1cm of f1] {};
\draw[draw, solid] (c2)--(c3)--(c4)--(c6)--(c7)--(c8)--(c9)--(c10)--(c11);
\draw[draw, solid] (c4)--(f1) node[midway,above, xshift=0.1cm] {} ;
\draw[draw, solid] (c4)--(f2) node[midway,above, xshift=0.1cm] {} ;
\draw[draw, solid]  (f1)--(c6) node[midway,above, xshift=-0.1cm] {};
\draw[draw, solid]  (f2)--(c6) node[midway,above, xshift=-0.1cm] {};
\draw[draw, solid]  (f1)--(f2) node[midway,above, xshift=-0.1cm] {};
\end{tikzpicture}}}
where we have $m_A=m_B=m_G=1$ and $H_{\text{free}}=0$.  The Coulomb branch symmetry is $SU(11) \times U(1)^2$ and the Higgs branch symmetry is $U(1)^3$.

Next we consider $D_{9}(SU(13))$, where we have $m=3$, $n=4$, $q=3$, $x=1$, $m_A=2$, $m_B=1$ and $m_G=2$.  The reduction to 3d gives
\bes{
(D_{9}(SU(13)))_{3d}:  \quad [13]-11-10-SU(9)-7-6-SU(5)-3-2-[1]
}
Following the above procedure, we obtain the mirror theory
\be
\tdplotsetmaincoords{70}{110}
\begin{tikzpicture}[tdplot_main_coords,baseline=0,font=\scriptsize]
\tikzstyle{every node}=[minimum size=0.5cm]
\node[draw, circle] (c3) at (0,-3,0) {1}; \node[below= 0.13cm of c3] {};
\node[draw, circle] (c4) at (0,-1,0) {9}; \node[below= 0.13cm of c4] {};
\node[draw, circle] (c6) at (0,3,0)  {11}; \node[below= 0.1cm of c6] {};
\node[draw, circle] (c7) at (0,4,0) {10}; \node[below= 0.1cm of c7] {};
\node[draw, circle] (c8) at (0,5,0) {9}; \node[below= 0.1cm of c8] {};
\node[draw=none] (c9) at (0,6,0) {$\ldots$}; \node[below= 0.1cm of c9] {};
\node[draw, circle] (c10) at (0,7,0)  {2}; \node[below= 0.1cm of c10] {};
\node[draw, circle] (c11) at (0,8,0)  {1}; \node[below= 0.1cm of c11] {};
\node[draw, circle] (f1) at (0,1,2)  {$1$}; \node[below= 0.1cm of f1] {};
\node[draw, circle] (f2) at (0,1,-2)  {$1$}; \node[below= 0.1cm of f1] {};
\node[draw, circle] (f3) at (-4.5,1,0)  {$1$}; \node[below= 0.1cm of f1] {};
\draw[draw, solid] (c3)--(c4)--(c6)--(c7)--(c8)--(c9)--(c10)--(c11);
\draw[draw, solid, double, thick] (c4)--(f1) node[midway,above, xshift=0.1cm] {} ;
\draw[draw, solid, double, thick] (c4)--(f2) node[midway,above, xshift=0.1cm] {} ;
\draw[draw, solid, double, thick] (c4)--(f3) node[midway,above, xshift=0.1cm] {} ;
\draw[draw, solid]  (f1)--(c6) node[midway,above, xshift=-0.1cm] {};
\draw[draw, solid]  (f2)--(c6) node[midway,above, xshift=-0.1cm] {};
\draw[draw, solid]  (f3)--(c6) node[midway,above, xshift=-0.1cm] {};
\draw[draw, solid,double, thick]  (f1)--(f2)--(f3)--(f1) node[midway,above, xshift=-0.1cm] {};
\end{tikzpicture}
\ee
with $H_{\text{free}} =0$. The Higgs branch symmetry of this quiver is $U(1)^4 \times SU(2)^6$.  The Coulomb branch symmetry of this quiver is $SU(13) \times U(1)^3$.

Let us now consider the case of $N-1=p\fm$ with a positive integer $\fm$.  The $D^{p\fm}_p(SU(p\fm+1))$ theory and the 3d theory upon reduction admit the following quiver description:
\bes{ \label{linearSUII}
[p \fm+1] - SU((p-1)\fm+1) - SU((p-2)\fm+1) -\cdots -SU(\fm+1)-[1]
}
In this case we have $x=\fm$, $m=p$, $M=(p-1)\fm$, $m_A=1$, $m_B=m_G=0$ and $H_{\text{free}} =0$, and so we have the mirror theory:
\bes{
&\begin{tikzpicture}[baseline=0, font=\scriptsize]
\tikzstyle{every node}=[minimum size=0.5cm]
\node[draw,circle] (cm0) at (-7,0) {}; \node[below= 0.1cm of cm0] {\!\!\!\!\!\! $1$};
\node[draw,circle] (cm1) at (-6,0) {}; \node[below= 0.1cm of cm1] {\!\!\!\!\!\! $\substack{(p-1)\\+1}$};
\node[draw,circle] (c0) at (-5,0) {}; \node[below= 0.1cm of c0] {\,\,\, $\substack{2(p-1)\\+1}$};
\node[draw=none] (c1) at (-4,0) {$\cdots$}; \node[below= 0.1cm of c1] {};
\node[draw, circle] (c2) at (-3,0) {}; \node[below= 0.1cm of c2] {\!\!\!\!\!\! $\substack{(p-1)\\ (\fm-2) \\ +1}$};
\node[draw, circle] (c3) at (-2,0) {}; \node[below= 0.1cm of c3] {$\substack{(p-1) \\(\fm-1) \\ +1}$};
\node[draw, circle] (c4) at (-1,0) {}; \node[below= 0.13cm of c4] {\,\,\,\, $\substack{(p-1)\fm \\ +1}$};
\node[draw, circle] (c6) at (1,0)  {}; \node[below= 0.1cm of c6] {$\substack{(p-1)\fm}$};
\node[draw, circle] (c7) at (2,0) {}; \node[below= 0.1cm of c7] {$\substack{(p-1)\fm \\-1}$};
\node[draw, circle] (c8) at (3,0) {}; \node[below= 0.1cm of c8] {$\substack{(p-1)\fm \\-2}$};
\node[draw=none] (c9) at (4,0) {$\cdots$}; \node[below= 0.1cm of c9] {};
\node[draw, circle] (c10) at (5,0)  {}; \node[below= 0.1cm of c10] {$2$};
\node[draw, circle] (c11) at (6,0)  {}; \node[below= 0.1cm of c11] {$1$};
\node[draw,circle] (b1) at (-2,1) {$1$}; \node[above= 0.1cm of b1] {};
\node[draw,circle] (b2) at (-1,1) {$1$}; \node[above= 0.1cm of b2] {};
\node[draw=none] (b3) at (0,1) {$\ldots$}; \node[above= 0.1cm of b3] {};
\node[draw,circle] (b4) at (1,1) {$1$}; \node[above= 0.1cm of b3] {};
\draw[draw, solid] (cm0)--(cm1)--(c0)--(c1)--(c2)--(c3)--(c4)--(c6)--(c7)--(c8)--(c9)--(c10)--(c11);
\draw[draw, solid] (c4)--(b1);
\draw[draw, solid] (c4)--(b2);
\draw[draw, solid] (c4)--(b4);
\draw [decorate,decoration={brace,amplitude=10pt}] (-2.5,1.3) -- (1.5,1.3) node [black,midway, yshift=0.6cm] {\footnotesize $p$};
\end{tikzpicture}
}
This is as expected for the mirror theory for \eref{linearSUII} (see \eg~ \cite{Hanany:2018vph, Bourget:2019rtl, talkZhong}).  For $p=2$, this is simply the mirror theory of the $SU(\fm+1)$ gauge theory with $2\fm+2$ flavours \cite{Hanany:1996ie}.

Finally, let us consider the case of $\GCD(p,N-1)=1$.  We have $m=1$, $n=N-1$ and $q=p$.  The 3d reduction of the corresponding $D^{N-1}_{p\leq N-1}(SU(N))$ is
\be
\scalebox{0.9}{$
\begin{split}
&[N] - (N-(x+1))- (N-2(x+1)) - \cdots- (N-(N-p x-1)(x+1))- \\
&- (N-(N-p x-1)(x+1)-x) - (N-(N-p x-1)(x+1)-2x)- \cdots \\
&- (2x+1)-(x+1)-[1]~,
\end{split}$}
\ee
together with $\frac{1}{2} (N-p x-2) [p (x+1)-N]$ twisted hypermultiplets.
Note that all gauge nodes are balanced, except $(N-(N-p x-1)(x+1))$, which is overbalanced.
The corresponding mirror theory is 
\bes{ \label{TypeIIp<Nm1gcd1}
&\begin{tikzpicture}[baseline=0, font=\scriptsize]
\tikzstyle{every node}=[minimum size=0.5cm]
\node[draw,circle] (cm0) at (-7,0) {}; \node[below= 0.1cm of cm0] {\!\!\!\!\!\! $1$};
\node[draw,circle] (cm1) at (-6,0) {}; \node[below= 0.1cm of cm1] {\!\!\!\!\!\! $\substack{(p-1) \\ +1}$};
\node[draw,circle] (c0) at (-5,0) {}; \node[below= 0.1cm of c0] {\,\,\, $\substack{2(p-1) \\ +1}$};
\node[draw=none] (c1) at (-4,0) {$\cdots$}; \node[below= 0.1cm of c1] {};
\node[draw, circle] (c2) at (-3,0) {}; \node[above= 0cm of c2, yshift=-0.1cm] {\!\!\!\!\!\! \!\!\! \!\!\! \!\!\!  $\substack{(p-1)(x-2) \\ +1 }$};
\node[draw, circle] (c3) at (-2,0) {}; \node[below= 0.1cm of c3] {\!\!\! \!\!\! \!\!\! \!\!\! \!\!\! \!\!\!  $\substack{(p-1)(x-1) \\ +1 }$};
\node[draw, circle] (c4) at (-1,0) {}; \node[below= 0.13cm of c4] {$ \substack{(p-1)x\\ +1}$};
\node[draw, circle] (c6) at (1,0)  {}; \node[below= 0.1cm of c6] {$\substack{N \\-(x+1)}$};
\node[draw, circle] (c7) at (2,0) {}; \node[below= 0.1cm of c7] {$\substack{N \\-(x+1)\\-1}$};
\node[draw, circle] (c8) at (3,0) {}; \node[below= 0.1cm of c8] {$\substack{N \\-(x+1)\\-2}$};
\node[draw=none] (c9) at (4,0) {$\cdots$}; \node[below= 0.1cm of c9] {};
\node[draw, circle] (c10) at (5,0)  {}; \node[below= 0.1cm of c10] {$2$};
\node[draw, circle] (c11) at (6,0)  {}; \node[below= 0.1cm of c11] {$1$};
\node[draw, circle] (f1) at (0,1.2)  {$1$}; \node[below= 0.1cm of f1] {};
\draw[draw, solid] (cm0)--(cm1)--(c0)--(c1)--(c2)--(c3)--(c4)--(c6)--(c7)--(c8)--(c9)--(c10)--(c11);
\draw[draw, solid, very thick, blue] (c4)--(f1) node[midway,above, xshift=-1cm] {$p(1+x)-N+1$} ;
\draw[draw, solid, very thick, red]  (f1)--(c6) node[midway,above, xshift=0.6cm] {$N-1-px$};
\end{tikzpicture}\\
&\text{+ $\frac{1}{2} (N-p x-2) [p (x+1)-N]$ hypermultiplets}~.
}
Since every node in the central line, except the leftmost $U(1)$, is balanced, the Coulomb branch symmetry of the mirror theory is $SU(N)\times U(1)$.  The Higgs branch symmetry of the mirror theory (excluding the contribution of the free hypermultiplets) is $SU(N-1-px)\times SU(p(1+x)-N+1) \times U(1)$.

\acknowledgments
We are indebted to Cyril Closset and Wolfger Peelaers for many useful discussions. The work of S.G. is supported by the ERC Consolidator Grant 682608 Higgs bundles: Supersymmetric Gauge Theories and Geometry (HIGGSBNDL). M.S. is partially supported by the ERC-STG grant 637844-HBQFTNCER, by the University of Milano-Bicocca grant 2016-ATESP0586, by the MIUR-PRIN contract 2017CC72MK003, and by the INFN.

\appendix 
\section{Aharony duality for quivers}
\label{app:aharony}

In this appendix we review the effect of applying Aharony duality \cite{Aharony:1997gp} locally on a node of a linear quiver gauge theory. We will focus in particular on the case of quivers with unitary groups only \cite{Pasquetti:2019uop,Pasquetti:2019tix} which is relevant to us, but the same results have been recently generalized to $SU$, $USp$ and $SO$ groups in \cite{Benvenuti:2020wpc}. Similar manipulations and variants involving the application of dualities with monopole superpotentials \cite{Benini:2017dud} have been used for example in \cite{Benvenuti:2017kud,Giacomelli:2017vgk,Pasquetti:2019uop,Pasquetti:2019tix,Garozzo:2019xzi,Hwang:2020wpd, Benvenuti:2020wpc,Benvenuti:2020gvy}.

Let's consider a generic linear quiver of unitary gauge groups with fundamental, bifundamental and adjoint matter:
\be\label{quiver1}
\begin{tikzpicture}[thick, scale=0.4]
\node[](L1) at (-8,0) {$\dots$};
\node[](L3) at (-2.5,2.5) {$\Phi_{a}$};
\node[] (L2) at (-5,0.7) {$\widetilde{B}_a,B_a$};
\node[] at (-2.5,-1.7) {$\widetilde{q_a}\; q_a$};
\node[circle, draw, inner sep=2.5](L4) at (-2.5,0){$N_{a}$};
\node[rectangle, draw, inner sep=1.7,minimum height=.6cm,minimum width=.6cm](L5) at (-2.5,-3){$k_a$};
\path[every node/.style={font=\sffamily\small,
  		fill=white,inner sep=1pt}]
(L4) edge [loop, out=55, in=125, looseness=4] (L4);
\draw[-, thick] (L4) -- (L5);
\draw[-, thick] (L1) -- (L4);

\node[](M3) at (3,2.5) {$\Phi_{b}$};
\node[] (M2) at (0.5,0.7) {$\widetilde{B}_b,B_b$};
\node[] at (3,-1.7) {$\widetilde{q_b}\; q_b$};
\node[circle, draw, inner sep=2.5](M4) at (3,0){$N_{b}$};
\node[rectangle, draw, inner sep=1.7,minimum height=.6cm,minimum width=.6cm](M5) at (3,-3){$k_b$};
\path[every node/.style={font=\sffamily\small,
  		fill=white,inner sep=1pt}]
(M4) edge [loop, out=55, in=125, looseness=4] (M4);
\draw[-, thick] (M4) -- (M5);
\draw[-, thick] (L4) -- (M4);

\node[] (N2) at (6,0.7) {$\widetilde{B}_c,B_c$};
\node[] at (8.5,-1.7) {$\widetilde{q_c}\; q_c$};
\node[circle, draw, inner sep=2.5, color=red](N4) at (8.5,0){$N_{c}$};
\node[rectangle, draw, inner sep=1.7,minimum height=.6cm,minimum width=.6cm](N5) at (8.5,-3){$k_c$};

\draw[-, thick] (N4) -- (N5);
\draw[-, thick] (M4) -- (N4);

\node[](P3) at (14,2.5) {$\Phi_{d}$};
\node[] (P2) at (11.5,0.7) {$\widetilde{B}_d,B_d$};
\node[] at (14,-1.7) {$\widetilde{q_d}\; q_d$};
\node[circle, draw, inner sep=2.5](P4) at (14,0){$N_{d}$};
\node[rectangle, draw, inner sep=1.7,minimum height=.6cm,minimum width=.6cm](P5) at (14,-3){$k_d$};
\path[every node/.style={font=\sffamily\small,
  		fill=white,inner sep=1pt}]
(P4) edge [loop, out=55, in=125, looseness=4] (P4);
\draw[-, thick] (P4) -- (P5);
\draw[-, thick] (N4) -- (P4);

\node[](Q3) at (19.5,2.5) {$\Phi_{e}$};
\node[] (Q2) at (17,0.7) {$\widetilde{B}_e,B_e$};
\node[] at (19.5,-1.7) {$\widetilde{q_e}\; q_e$};
\node[circle, draw, inner sep=2.5](Q4) at (19.5,0){$N_{e}$};
\node[rectangle, draw, inner sep=1.7,minimum height=.6cm,minimum width=.6cm](Q5) at (19.5,-3){$k_e$};
\path[every node/.style={font=\sffamily\small,
  		fill=white,inner sep=1pt}]
(Q4) edge [loop, out=55, in=125, looseness=4] (Q4);
\draw[-, thick] (Q4) -- (Q5);
\draw[-, thick] (Q4) -- (P4);

\node[] (R2) at (22.5,0.7) {$\widetilde{B}_f,B_f$};
\node[](R4) at (25,0){$\dots$};
\draw[-, thick] (Q4) -- (R4);

\end{tikzpicture}
\ee
We do not include an adjoint for $U(N_c)$ because we want to perform Aharony duality \cite{Aharony:1997gp} there. We do not turn on any superpotential so fundamentals and antifundamentals have the same R-charge, which we parametrize as follows: 
\be\label{Rcharge} R(\widetilde{B}_i)=R(B_i)\equiv r_B^i;\quad R(\widetilde{q}_i)=R(q_i)\equiv r_q^i; \quad R(\Phi_i)\equiv r_{\Phi}^i.\ee 
All the constants $r_q$, $r_B$, $r_{\Phi}$ can be determined using Z-extremization \cite{Jafferis:2010un} but we leave them generic. The case of quivers without adjoints can be recovered simply setting $r_{\Phi}^i=1$. Let us now focus on the R-charge of monopole operators, which can be computed using the methods of \cite{Borokhov:2002cg, Gaiotto:2008ak}. For simplicity we will consider only operators with magnetic flux $(\pm1,0,\dots,0)$ (or trivial) under the gauge groups $U(N_b)$, $U(N_c)$ and $U(N_d)$. We will denote for example a monopole with charge 0 under $U(N_b)$, 1 under $U(N_c)$ and $-1$ under $U(N_d)$ as $\mathfrak{M}_{0+-}$. The charge under all the other groups are not specified, since this will not be essential for our argument. The R-charges are as follows: 
\be\label{moncharge} \begin{array}{|c|c|}
\hline
\text{Operator} & \text{R-charge} \\
\hline
\mathfrak{M}_{0\pm0} &  k_c(1-r_q^c)+N_b(1-r_B^c)+N_d(1-r_B^d)+1-N_c+r_{\mathfrak{M}}\\ 
\hline
\mathfrak{M}_{\pm00} &  k_b(1-r_q^b)+ N_a(1-r_B^b)+(N_b-1)(1-r_{\Phi}^b)+N_c(1-r_B^c)+1-N_b+r'_{\mathfrak{M}}\\
\hline 
\mathfrak{M}_{00\pm} &  k_d(1-r_q^d)+ N_e(1-r_B^e)+(N_d-1)(1-r_{\Phi}^d)+N_c(1-r_B^d)+1-N_d+r''_{\mathfrak{M}}\\
\hline 
\mathfrak{M}_{\pm\pm0} & \begin{array}{l} k_b(1-r_q^b)+ N_a(1-r_B^b)+(N_b-1)(1-r_{\Phi}^b)+ k_c(1-r_q^c)+N_d(1-r_B^d)\\ -r_B^c(N_b+N_c-2)+r'_{\mathfrak{M}}\end{array}\\
\hline 
\mathfrak{M}_{0\pm\pm} & \begin{array}{l} k_d(1-r_q^d)+ N_e(1-r_B^e)+(N_d-1)(1-r_{\Phi}^d)+ k_c(1-r_q^c)+N_b(1-r_B^c)\\ -r_B^d(N_d+N_c-2)+r''_{\mathfrak{M}}\end{array}\\
\hline
\mathfrak{M}_{\pm\pm\pm} & \begin{array}{l}  k_b(1-r_q^b)+ k_c(1-r_q^c)+ k_d(1-r_q^d)+N_a(1-r_B^b)+ N_e(1-r_B^e)+\\
(N_b-1)(1-r_{\Phi}^b)+(N_d-1)(1-r_{\Phi}^d)-r_B^c(N_b+N_c-2)+N_c-1\\
 -r_B^d(N_d+N_c-2)+r_{\mathfrak{M}}'''\\
\end{array}\\
\hline
\end{array}\ee 
where the terms $r_{\mathfrak{M}}$, $r'_{\mathfrak{M}}$ and $r''_{\mathfrak{M}}$ denote the further contributions due to magnetic charges under gauge groups on the left of $U(N_b)$ and on the right of $U(N_d)$. More precisely, if we separate these two contributions $r_{\mathfrak{M}}=r^L_{\mathfrak{M}}+r^R_{\mathfrak{M}}$, we have 
\be\label{restcharge} r'_{\mathfrak{M}}=r'^L_{\mathfrak{M}}+r^R_{\mathfrak{M}};\quad  r''_{\mathfrak{M}}=r^L_{\mathfrak{M}}+r''^R_{\mathfrak{M}};\quad r'''_{\mathfrak{M}}=r'^L_{\mathfrak{M}}+r''^R_{\mathfrak{M}}.\ee 
With this notation we find 
\be R(\mathfrak{M}_{\pm0\pm})=R(\mathfrak{M}_{\pm00})+R(\mathfrak{M}_{00\pm})-r^L_{\mathfrak{M}}-r^R_{\mathfrak{M}}+r'^L_{\mathfrak{M}}+r''^R_{\mathfrak{M}}.\ee   
Let us now compare with the theory we obtain by performing Aharony duality on the node $U(N_c)$. The resulting quiver is 
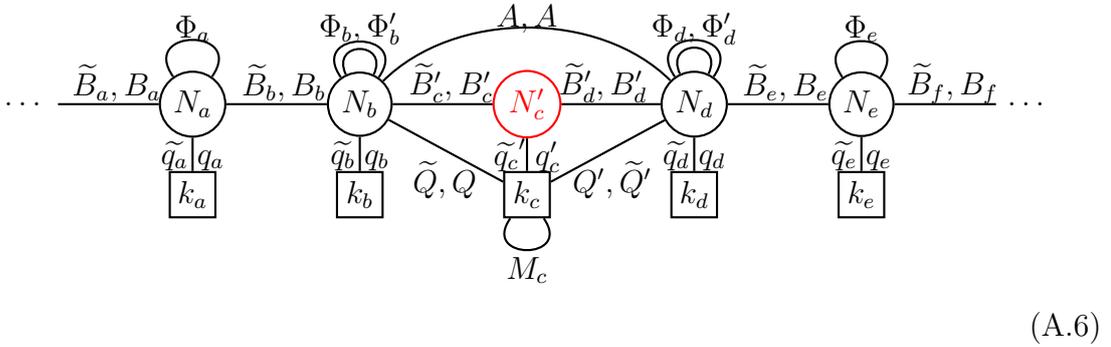
\be\label{quiver2}
\begin{tikzpicture}[thick, scale=0.4]
\node[](L1) at (-8,0) {$\dots$};
\node[](L3) at (-2.5,2.5) {$\Phi_{a}$};
\node[] (L2) at (-5,0.7) {$\widetilde{B}_a,B_a$};
\node[] at (-2.5,-1.7) {$\widetilde{q_a}\; q_a$};
\node[circle, draw, inner sep=2.5](L4) at (-2.5,0){$N_{a}$};
\node[rectangle, draw, inner sep=1.7,minimum height=.6cm,minimum width=.6cm](L5) at (-2.5,-3){$k_a$};
\path[every node/.style={font=\sffamily\small,
  		fill=white,inner sep=1pt}]
(L4) edge [loop, out=55, in=125, looseness=4] (L4);
\draw[-, thick] (L4) -- (L5);
\draw[-, thick] (L1) -- (L4);

\node[](M3) at (3,2.5) {$\Phi_{b},\Phi'_b$};
\node[] (M2) at (0.5,0.7) {$\widetilde{B}_b,B_b$};
\node[] at (3,-1.7) {$\widetilde{q_b}\; q_b$};
\node[circle, draw, inner sep=2.5](M4) at (3,0){$N_{b}$};
\node[rectangle, draw, inner sep=1.7,minimum height=.6cm,minimum width=.6cm](M5) at (3,-3){$k_b$};
\path[every node/.style={font=\sffamily\small,
  		fill=white,inner sep=1pt}]
(M4) edge [loop, out=55, in=125, looseness=4] (M4);
\path[every node/.style={font=\sffamily\small,
  		fill=white,inner sep=1pt}]
(M4) edge [loop, out=65, in=115, looseness=3.5] (M4);
\draw[-, thick] (M4) -- (M5);
\draw[-, thick] (L4) -- (M4);

\node[](N8) at (8.5,-5.5) {$M_c$};
\node[](N9) at (8.5,3) {$A,\widetilde{A}$};
\node[] (N2) at (6,0.7) {$\widetilde{B}'_c,B'_c$};
\node[] at (8.5,-1.7) {$\widetilde{q_c}'\; q'_c$};
\node[circle, draw, inner sep=2.5, color=red](N4) at (8.5,0){$N_{c}'$};
\node[rectangle, draw, inner sep=1.7,minimum height=.6cm,minimum width=.6cm](N5) at (8.5,-3){$k_c$};
\node[] (N6) at (5.8,-2.5) {$\widetilde{Q},Q$};
\node[] (N7) at (11.3,-2.5) {$Q',\widetilde{Q}'$};
\path[every node/.style={font=\sffamily\small,
  		fill=white,inner sep=1pt}]
(N5) edge [loop, out=305, in=235, looseness=4] (N5);

\node[](P3) at (14,2.5) {$\Phi_{d},\Phi'_d$};
\node[] (P2) at (11,0.7) {$\widetilde{B}'_d,B'_d$};
\node[] at (14,-1.7) {$\widetilde{q_d}\; q_d$};
\node[circle, draw, inner sep=2.5](P4) at (14,0){$N_{d}$};
\node[rectangle, draw, inner sep=1.7,minimum height=.6cm,minimum width=.6cm](P5) at (14,-3){$k_d$};
\path[every node/.style={font=\sffamily\small,
  		fill=white,inner sep=1pt}]
(P4) edge [loop, out=55, in=125, looseness=4] (P4);
\path[every node/.style={font=\sffamily\small,
  		fill=white,inner sep=1pt}]
(P4) edge [loop, out=65, in=115, looseness=3.5] (P4);
\draw[-, thick] (P4) -- (P5);
\draw[-, thick] (N4) -- (P4);
\path[every node/.style={font=\sffamily\small,
  		fill=white,inner sep=1pt}]
(M4) edge [loop, out=45, in=135, looseness=.9] (P4);
\draw[-, thick] (N4) -- (N5);
\draw[-, thick] (M4) -- (N4);
\draw[-, thick] (M4) -- (N5);
\draw[-, thick] (P4) -- (N5);

\node[](Q3) at (19.5,2.5) {$\Phi_{e}$};
\node[] (Q2) at (17,0.7) {$\widetilde{B}_e,B_e$};
\node[] at (19.5,-1.7) {$\widetilde{q_e}\; q_e$};
\node[circle, draw, inner sep=2.5](Q4) at (19.5,0){$N_{e}$};
\node[rectangle, draw, inner sep=1.7,minimum height=.6cm,minimum width=.6cm](Q5) at (19.5,-3){$k_e$};
\path[every node/.style={font=\sffamily\small,
  		fill=white,inner sep=1pt}]
(Q4) edge [loop, out=55, in=125, looseness=4] (Q4);
\draw[-, thick] (Q4) -- (Q5);
\draw[-, thick] (Q4) -- (P4);

\node[] (R2) at (22.5,0.7) {$\widetilde{B}_f,B_f$};
\node[](R4) at (25,0){$\dots$};
\draw[-, thick] (Q4) -- (R4);

\end{tikzpicture}
\ee
where $N_c'=k_c+N_b+N_d-N_c$. From the duality we know we have the following assignment of R-charges: 
\be\label{rduals}\begin{array}{l}
R(A)=R(\widetilde{A})=r_B^c+r_B^d;\quad R(Q)=R(\widetilde{Q})=r_B^c+r_q^c;\quad R(Q')=R(\widetilde{Q}')=r_B^d+r_q^c;\\ R(M_c)=2r_q^c;\quad R(\Phi'_b)=2r_B^c;\quad R(\Phi'_d)=2r_B^d.\end{array}\ee 
We also have two more gauge singlets $a^{\pm}$ and the following superpotential 
\be\label{suppot} \begin{array}{ll} \mathcal{W}= &
\widetilde{B}'_c\widetilde{B}'_d\widetilde{A}+B'_dB'_cA+\widetilde{B}'_c\widetilde{q_c}'\widetilde{Q}+q_c'B_c'Q+q_c'\widetilde{B}_d'Q'+B_d'\widetilde{q_c}'\widetilde{Q}'+\widetilde{B}'_c\Phi'_bB'_c+B'_d\Phi'_d\widetilde{B}'_d \\ 
& +M_c\widetilde{q}'_cq'_c+a^+\mathfrak{M}'_{0+0}+a^-\mathfrak{M}'_{0-0}.\end{array}\ee
From (\ref{rduals}) and (\ref{suppot}) we also find 
\be\label{rdual2} R(\widetilde{q_c}')=R(q_c')=1-r_q^c;\quad R(\widetilde{B}'_c)=R(B'_c)=1-r_B^c;\quad R(\widetilde{B}'_d)=R(B'_d)=1-r_B^d.\ee 
Notice first from \eqref{rduals} that when $r_B^c=r_B^d=\frac{1}{2}$ we have $R(A)=R(\tilde{A})=1$. This suggests that a superpotential term of the form $A\tilde{A}$ is generated, provided that all the other quantum numbers are correct. In other words, the fields $A$ and $\tilde{A}$ are actually massive and no link between the $U(N_b)$ and the $U(N_d)$ nodes are created. We used this fact extensively in Subsection \ref{subsec:DpSUNplN:Examples} for the derivation of the 3d mirror of the circle reduction of $D_4(SU(6))$ and we will use it again in the next appendices. Moreover,
we can now compute the R-charge of $\mathfrak{M}'_{0\pm0}$ and from (\ref{suppot}) we find as expected that the singlets $a^{\pm}$ have the same R-charge as the monopoles of the original theory charged under $U(N_c)$ only. They are indeed dual operators. All other fields appearing in the dual quiver retain the same R-charge as the corresponding fields in the original quiver. In particular, we conclude from this that the scaling dimension of monopole operators with trivial magnetic charge under $U(N_b)$, $U(N_d)$ and $U(N_c')$ is not affected by the duality. Said differently, the duality acts trivially on them. By the same token, the contributions $r_{\mathfrak{M}}$, $r'_{\mathfrak{M}}$ and $r''_{\mathfrak{M}}$ appearing in (\ref{moncharge}) are unchanged.

We now  focus on the R-charge of the other monopole operators in the dual theory (\ref{quiver2}). We find 
\be\label{moncharge2} \begin{array}{|c|c|}
\hline
\text{Operator} & \text{R-charge} \\
\hline
\mathfrak{M}'_{\pm00} & \begin{array}{l} k_b(1-r_q^b)+ N_a(1-r_B^b)+(N_b-1)(1-r_{\Phi}^b)+ k_c(1-r_q^c)+N_d(1-r_B^d)\\ -r_B^c(N_b+N_c-2)+r'_{\mathfrak{M}}\end{array}\\
\hline 
\mathfrak{M}'_{00\pm} &  \begin{array}{l} k_d(1-r_q^d)+ N_e(1-r_B^e)+(N_d-1)(1-r_{\Phi}^d)+ k_c(1-r_q^c)+N_b(1-r_B^c)\\ -r_B^d(N_d+N_c-2)+r''_{\mathfrak{M}}\end{array}\\
\hline 
\mathfrak{M}'_{\pm\pm0} &  k_b(1-r_q^b)+ N_a(1-r_B^b)+(N_b-1)(1-r_{\Phi}^b)+N_c(1-r_B^c)+1-N_b+r'_{\mathfrak{M}}\\
\hline 
\mathfrak{M}'_{0\pm\pm} & k_d(1-r_q^d)+ N_e(1-r_B^e)+(N_d-1)(1-r_{\Phi}^d)+N_c(1-r_B^d)+1-N_d+r''_{\mathfrak{M}}\\
\hline
\mathfrak{M}'_{\pm\pm\pm} & \begin{array}{l}  k_b(1-r_q^b)+ k_c(1-r_q^c)+ k_d(1-r_q^d)+N_a(1-r_B^b)+ N_e(1-r_B^e)+\\
(N_b-1)(1-r_{\Phi}^b)+(N_d-1)(1-r_{\Phi}^d)-r_B^c(N_b+N_c-2)+N_c-1\\
 -r_B^d(N_d+N_c-2)+r_{\mathfrak{M}}'''\\
\end{array}\\
\hline
\end{array}\ee 
By comparing (\ref{moncharge}) and (\ref{moncharge2}) we find that the duality acts in the following way on the monopole operators of the theory: 
\be\label{monmap} \mathfrak{M}_{\pm00}\leftrightarrow\mathfrak{M}'_{\pm\pm0}; \; \mathfrak{M}_{\pm\pm0}\leftrightarrow\mathfrak{M}'_{\pm00}; \; \mathfrak{M}_{00\pm}\leftrightarrow\mathfrak{M}'_{0\pm\pm}; \; \mathfrak{M}_{0\pm\pm}\leftrightarrow\mathfrak{M}'_{00\pm}; \; \mathfrak{M}_{\pm\pm\pm}\leftrightarrow\mathfrak{M}'_{\pm\pm\pm}.\ee
We stress that the above relation means that the magnetic charge under all the gauge groups other than $U(N_b)$, $U(N_c')$ and $U(N_d)$ (if any) are not affected by the duality. For example, the monopole charged under $U(N_a)$ and $U(N_b)$ in the original theory (\ref{quiver1}) is mapped in the dual quiver to the monopole charged under $U(N_a)$, $U(N_b)$ and $U(N_c')$.

The same conclusion can be derived from a slightly different perspective that was adopted in \cite{Pasquetti:2019uop,Pasquetti:2019tix}. Aharony duality indeed holds up to mixed contact terms between the topological symmetry and the flavor symmetry\footnote{Such contact terms can be detected, for example, with the $S^3_b$ partition function, see the first exponential factor in the second line of eq.~(2.12) of \cite{Pasquetti:2019uop}.}. When we apply the duality on a node of a quiver, the flavor symmetry is identified with the gauge symmetries of the adjacent nodes and the contact term becomes a BF coupling between the topological symmetry at the dualized node and the gauge symmetry of the adjacent nodes. As a consequence, after the dualization the topological symmetries of the adjacent nodes get mixed with the one of the dualized node. Explicitly, denoting with $U(1)_{T_b}$, $U(1)_{T_c}$, $U(1)_{T_d}$ the topological symmetries at the $U(N_b)$, $U(N_c)$, $U(N_d)$ nodes of the quiver \eqref{quiver1} respectively, after applying Aharony duality to the node $U(N_c)$ we have
\be
U(1)_{T_b}\to U(1)_{T_b} - U(1)_{T_c},\qquad U(1)_{T_d}\to U(1)_{T_d} - U(1)_{T_c}\,.
\ee
The map \eqref{monmap} then immediately follows. For example, the monopole $\mathfrak{M}'_{\pm\pm0}$ carries charge $\pm1$ under $U(1)_{T_b}$ and $\pm(1-1)=0$ under $U(1)_{T_c}$ and is thus mapped to the monopole $\mathfrak{M}_{\pm00}$. Moreover, since the BF coupling is produced only for the adjacent nodes of the quiver, we recover the fact that all of the monopoles with no magnetic flux for $U(N_b)$, $U(N_c)$ and $U(N_d)$ are not affected by the dualization and map straightforwardly to themselves.

\section{The flip-flip duality for $T(SU(N))$}
\label{app:flipflip}

The observation of the previous appendix can be used to provide a field theoretic derivation \cite{Hwang:2020wpd} of a recently discovered duality for $T(SU(N))$ \cite{Aprile:2018oau}. The theory consists of a $\mathcal{N}=4$ linear quiver of unitary gauge groups \cite{Gaiotto:2008ak}: 
\be\label{tsun} U(1)-U(2)-\dots -U(N-1)-\boxed{N}\ee 
and has $SU(N)^2$ global symmetry. One $SU(N)$ acts on the $N$ flavors on the right, whereas the other arises from the enhancement of the $U(1)^{N-1}$ topological symmetry due to the presence of monopole operators of scaling dimension one. The corresponding moment map can be described explicitly and for e.g. $T(SU(4))$ is 
\be\label{monmatrix}\left(\begin{array}{cccc}
 &  \mathfrak{M}_{+00} &  \mathfrak{M}_{++0} &  \mathfrak{M}_{+++} \\
 \mathfrak{M}_{-00} & &  \mathfrak{M}_{0+0} &  \mathfrak{M}_{0++} \\
 \mathfrak{M}_{--0} &  \mathfrak{M}_{0-0} & &  \mathfrak{M}_{00+}\\ 
 \mathfrak{M}_{---} &  \mathfrak{M}_{0--} &  \mathfrak{M}_{00-} & \\
\end{array}\right)\ee
where $ \mathfrak{M}_{abc}$ denotes the monopole operator with magnetic charge $a$ under $U(1)$, $(b,0)$ under $U(2)$ and $(c,0,0)$ under $U(3)$. The Cartan components are given by the trace part of the adjoint chirals in the $\mathcal{N}=4$ vector multiplets. More in general, the off-diagonal components are given by monopole operators with minimal magnetic charge under gauge groups forming a connected subset of the quiver (\ref{tsun}). 

It was recently observed in \cite{Aprile:2018oau} that $T(SU(N))$ is dual to the same theory with the moment maps for the two $SU(N)$ symmetries flipped (flip-flip duality), meaning that we add by hand a chiral multiplet $M$ in the adjoint of $SU(N)$ and we turn on the superpotential term $\Tr(\mu M)$, where $\mu$ denotes the $SU(N)$ moment map. In \cite{Hwang:2020wpd} it was observed that flip-flip duality can be derived by sequentially applying Aharony duality locally in the quiver, similarly to what we have seen in the previous appendix. We are now going to review this derivation, as it is also instructive to understand similar manipulations that we perfomed in Subsection \ref{subsec:DpSUNplN:Examples} and that we are going to perfom in the next appendix. 

For $T(SU(2))$ flip-flip duality is just a neat application of Aharony duality \cite{Aprile:2018oau}: this is SQED with 2 flavors $q_{1,2}$, $\tilde{q}_{1,2}$ and the superpotential is 
\be\mathcal{W}=\phi\tilde{q}_iq^i.\ee 
By performing Aharony duality we find the same theory with superpotential 
\be\label{dualsu2}\mathcal{W}=M_{i}^j\tilde{q}_jq^i+\phi\Tr M+a^+\mathfrak{M}_++a^-\mathfrak{M}_-.\ee
In the above formula the trace part of $M$ couples to the trace of the meson and plays the role of the singlet sitting in the $\mathcal{N}=4$ abelian vector multiplet. Therefore the traceless part of $M$ is identified with the flipping field for the $SU(2)$ symmetry acting on the flavors and the matrix 
\be\left(\begin{array}{cc}
\phi/2 & a^+ \\
a^- & -\phi/2 \\
\end{array}\right)\ee 
is identified with the flipping field for the topological $SU(2)$ symmetry. 

In order to understand the generalization to arbitrary $N$, let us focus on $T(SU(3))$. The superpotential is 
\be\mathcal{W}=\phi_1\tilde{q}q+\Tr[\phi_2(p\tilde{p}-q\tilde{q})],\ee 
where $\phi_1$ and $\phi_2$ are the $U(1)$ and $U(2)$ adjoints respectively. $\tilde{q}$ and $q$ are the $U(1)\times U(2)$ bifundamentals and $\tilde{p}$, $p$ denote the 3 $U(2)$ fundamentals. If we now dualize at the abelian node, we make $\phi_2$ massive (as can be seen from (\ref{dualsu2})) and therefore we can now apply Aharony duality to the $U(2)$ node. This generates the flipping field for the $SU(3)$ acting on the flavors $\tilde{p}$, $p$. Notice that the gauge group is still $U(1)\times U(2)$. In principle we also generate new $U(1)\times SU(3)$ bifundamentals, but these turn out to be massive. Finally, with a second dualization at the abelian node we reintroduce the $U(2)$ adjoint chiral. 

Let us now look at the action of this chain of dualities on monopole operators: The first dualization introduces flipping terms in the superpotential for $\mathfrak{M}_{\pm0}$. Then when we dualize at the $U(2)$ node we generate flipping terms for $\mathfrak{M}_{0\pm}$ as well. The crucial point now is that, due to (\ref{monmap}), the operators $\mathfrak{M}_{\pm0}$ are mapped to $\mathfrak{M}_{\pm\pm}$ after the second duality. This in particular implies that the third and last dualization does not remove the flipping terms generated at the first step, which is what would have happened if we had simply dualized twice at the abelian node without dualizing the $U(2)$ gauge group. In conclusion, after the three duality steps we end up with flipping terms for the monopole operators $\mathfrak{M}_{\pm0}$, $\mathfrak{M}_{\pm\pm}$ and $\mathfrak{M}_{0\pm}$, i.e. all the off-diagonal components of the moment map for the $SU(3)$ topological symmetry, in agreement with the flip-flip duality. 

We can now proceed with the analysis of the general case by induction: we know the duality follows from Aharony duality for $T(SU(2))$ and we assume it holds for $T(SU(N-1))$. We then proceed by dualizing at all the gauge groups in (\ref{tsun}) starting from the abelian node. Once we have dualized the node $U(N-1)$ we end up with the quiver theory
\be\label{tsunind}
\begin{tikzpicture}[thick, scale=0.4]
\node[](N1) at (0,0) {$T(SU(N-1))$};
\node[](N2) at (8,0) {$U(N-1)$};
\node[] at (11.5,0.7) {$\widetilde{P}, P$};
\node[] at (4.5,0.7) {$\widetilde{Q}, Q$};
\node[] at (16,0) {$M$};
\node[rectangle, draw, inner sep=1.7,minimum height=.6cm,minimum width=.6cm](N3) at (13.5,0){$N$};
\path[every node/.style={font=\sffamily\small,
  		fill=white,inner sep=1pt}]
(N3) edge [loop, out=35, in=325, looseness=4] (N3);
\draw[-, thick] (N1) -- (N2);
\draw[-, thick] (N2) -- (N3);
\end{tikzpicture}
\ee
and superpotential 
\be\label{suptah}\mathcal{W}=\mathcal{W}^{\mathcal{N}=4}_{T(SU(N-1))}-Q\widetilde{Q}P\widetilde{P}+M\widetilde{P}P+\mathcal{W}_{monopole}\ee 
where $\mathcal{W}_{monopole}$ is the part involving monopole operators and from (\ref{monmap}) we can easily see that it includes the $2N-2$ flipping terms for monopoles of dimension one and magnetic charge $(\pm,0,\dots,0)$ under $U(N-1)$. These are the entries in the last row and last column of the moment map (\ref{monmatrix}). We now proceed with the inductive step and replace $T(SU(N-1))$ with its flipped-flipped version. Assuming this is equivalent to performing a sequence of Aharony dualities along the nodes of $T(SU(N-1))$, we conclude that under the flip-flip duality all the monopoles appearing in $\mathcal{W}_{monopole}$ are mapped to themselves and therefore the flipping terms are not removed by the dualization. Overall, (\ref{suptah}) becomes 
\be\label{suptah2}\mathcal{W}=\mathcal{W}^{\mathcal{N}=4}_{T(SU(N-1))}+NQ\widetilde{Q}-NP\widetilde{P}+M\widetilde{P}P+\mathcal{W}'_{monopole},\ee 
where $N$ is the flipping field for the $SU(N-1)$ moment map and now plays the role of the $U(N-1)$ adjoint chiral. We therefore recover the matter content and superpotential of $T(SU(N))$, but with the trace part of all the adjoints in the $U(n_i)$ vector multiplets flipped and integrated out (notice that the matrix $N$ is traceless). The monopole superpotential now includes the $2N-2$ flipping terms generated along the initial sequence of $N-1$ dualizations and also all the flipping terms produced by the dualization of 
$T(SU(N-1))$. Overall, all the monopole operators appearing in the moment map (\ref{monmatrix}) are now flipped. We therefore recover the flipped-flipped dual of $T(SU(N))$. 

In summary, we have to perform the following chain of Aharony dualities: we first dualize all the nodes starting from $U(1)$ up to $U(N-1)$, then we start over from $U(1)$ and dualize all the nodes up to $U(N-2)$, then from $U(1)$ to $U(N-3)$ and so on. Overall we perform $N(N-1)/2$ Aharony dualities which provide the expected $N(N-1)$ flipping terms for monopole operators. At the end of this procedure we find the flipped-flipped $T(SU(N))$.

\section{Derivation of the 3d mirror of the circle reduction of $D_4(SU(10))$}
\label{app:deriv3dmirrD4SU10}

In this appendix we give more details on the derivation of the 3d mirror \eqref{D4SU10mirr} of the circle reduction of $D_4(SU(10))$.
Remember that the 3d linear quiver \eqref{D4SU10linear} coming from the reduction was
\bes{
(D_4(SU(10))_{3d}: \quad [10]-7-SU(5)-2
}
This can be realised as gluing $[10]-7-[5]$ and $[5]-2$ via an $SU(5)$ gauge group.  In terms of the mirror theories, this can be written as
\bes{ \label{D4SU10gluingapp}
\begin{tikzpicture}[baseline=0, font=\scriptsize]
\tikzstyle{every node}=[minimum size=0.5cm]
\node[draw, circle] (a1) at (0,0) {1};
\node[draw, circle] (a2) at (1,0) {2};
\node[draw=none] (a3) at (2,0) {$\cdots$};
\node[draw, circle] (a6) at (3,0) {6};
\node[draw, circle] (a7) at (4,0) {7};
\node[draw, circle] (a8) at (5,0) {7};
\node[draw, circle] (a9) at (6,0) {6};
\node[draw=none] (a12) at (7,0) {$\cdots$};
\node[draw, circle] (a13) at (8,0) {2};
\node[draw, circle] (a14) at (9,0) {1};
\node[draw, circle] (b) at (4.5,-1) {1};
\draw[draw, solid] (a1)--(a2);
\draw[draw, solid] (a2)--(a3);
\draw[draw, solid] (a3)--(a6);
\draw[draw, solid] (a6)--(a7);
\draw[draw, solid] (a7)--(a8);
\draw[draw, solid] (a8)--(a9);
\draw[draw, solid] (a9)--(a12);
\draw[draw, solid] (a12)--(a13);
\draw[draw, solid] (a13)--(a14);
\draw[draw, solid] (a7)--(b);
\draw[draw, solid] (a8)--(b);
\end{tikzpicture}
\quad + \quad
\begin{tikzpicture}[baseline=0, font=\scriptsize]
\tikzstyle{every node}=[minimum size=0.5cm]
\node[draw, circle] (a1) at (1,0) {1};
\node[draw, circle] (a2) at (2,0) {2};
\node[draw, circle] (a3) at (3,0) {2};
\node[draw, circle] (a4) at (4,0) {1};
\node[draw, circle] (b) at (2.5,-1) {1};
\draw[draw, solid] (a1)--(a2);
\draw[draw, solid] (a2)--(a3);
\draw[draw, solid] (a3)--(a4);
\draw[draw, solid] (a2)--(b);
\draw[draw, solid] (a3)--(b);
\end{tikzpicture}
}
In order to understand the result of this gluing, we proceed as we did in Subsection \ref{subsec:DpSUNplN:Examples} for $D_4(SU(6))$. The left quiver in \eqref{D4SU10gluingapp} can be understood as $T_{[3,2]}(SU(5))$. Hence, we can first perfom the gluing replacing that quiver with the one of $T(SU(5))$ and then turn on the vev $[3,2]$ afterwards. The gluing with $T(SU(5))$ is indeed standard
\bes{ 
\begin{tikzpicture}[baseline=0, font=\scriptsize]
\tikzstyle{every node}=[minimum size=0.5cm]
\node[draw, circle] (a1) at (0,0) {1};
\node[draw, circle] (a2) at (1,0) {2};
\node[draw=none] (a3) at (2,0) {$\cdots$};
\node[draw, circle] (a6) at (3,0) {6};
\node[draw, circle] (a7) at (4,0) {7};
\node[draw, circle] (a8) at (5,0) {7};
\node[draw, circle] (a9) at (6,0) {6};
\node[draw=none] (a12) at (7,0) {$\cdots$};
\node[draw, circle] (a13) at (8,0) {2};
\node[draw, circle] (a14) at (9,0) {1};
\node[draw, circle] (b) at (4.5,-1) {1};
\draw[draw, solid] (a1)--(a2);
\draw[draw, solid] (a2)--(a3);
\draw[draw, solid] (a3)--(a6);
\draw[draw, solid] (a6)--(a7);
\draw[draw, solid] (a7)--(a8);
\draw[draw, solid] (a8)--(a9);
\draw[draw, solid] (a9)--(a12);
\draw[draw, solid] (a12)--(a13);
\draw[draw, solid] (a13)--(a14);
\draw[draw, solid] (a7)--(b);
\draw[draw, solid] (a8)--(b);
\end{tikzpicture}
\quad + \quad
\begin{tikzpicture}[baseline=0, font=\scriptsize]
\tikzstyle{every node}=[minimum size=0.5cm]
\node[draw, circle] (a1) at (1,0) {1};
\node[draw, circle] (a2) at (2,0) {2};
\node[draw, circle] (a3) at (3,0) {3};
\node[draw, circle] (a4) at (4,0) {4};
\node[draw, circle] (a5) at (5,0) {5};
\draw[draw, solid] (a1)--(a2);
\draw[draw, solid] (a2)--(a3);
\draw[draw, solid] (a3)--(a4);
\draw[very thick, blue] (a4)--(a5) node at (4.5,0.3)  {\footnotesize $5$};
\end{tikzpicture}
}
and yields the following quiver theory:
\bes{
\begin{tikzpicture}[baseline=0, font=\scriptsize]
\tikzstyle{every node}=[minimum size=0.5cm]
\node[draw, circle] (a1) at (0,0) {1};
\node[draw, circle] (a2) at (1,0) {2};
\node[draw=none] (a3) at (2,0) {$\cdots$};
\node[draw, circle] (a6) at (3,0) {6};
\node[draw, circle] (a7) at (4,0) {7};
\node[draw, circle] (a8) at (5,0) {7};
\node[draw, circle] (a9) at (6,0) {6};
\node[draw, circle] (a10) at (7,0) {1};
\node[draw, circle] (b) at (4.5,-1) {1};
\draw[draw, solid] (a1)--(a2);
\draw[draw, solid] (a2)--(a3);
\draw[draw, solid] (a3)--(a6);
\draw[draw, solid] (a6)--(a7);
\draw[draw, solid] (a7)--(a8);
\draw[draw, solid] (a8)--(a9);
\draw[draw, solid] (a7)--(b);
\draw[draw, solid] (a8)--(b);
\draw[very thick, blue] (a9)--(a10) node at (6.5,0.3)  {\footnotesize $5$};
\end{tikzpicture}
}
Now we give the nilpotent vev $[3,2]$, which makes $T(SU(5))$ reduce to $T_{[3,2]}(SU(5))$, to the five flavors at the right end of the tail depicted in blue. The effect of this vev can be more easily understood in the flip-flip dual frame, where it is mapped to a massive deformation for some of those flavors. After integrating out the massive fields we get
\bes{
\begin{tikzpicture}[baseline=0, font=\scriptsize]
\tikzstyle{every node}=[minimum size=0.5cm]
\node[draw, circle] (a1) at (0,0) {1};
\node[draw, circle] (a2) at (1,0) {2};
\node[draw=none] (a3) at (2,0) {$\cdots$};
\node[draw, circle] (a6) at (3,0) {6};
\node[draw, circle] (a7) at (4,0) {7};
\node[draw, circle] (a8) at (5,0) {7};
\node[draw, circle] (a9) at (6,0) {6};
\node[draw, circle] (a10) at (7,0) {1};
\node[draw, circle] (b) at (4.5,-1) {1};
\draw[draw, solid] (a1)--(a2);
\draw[draw, solid] (a2)--(a3);
\draw[draw, solid] (a3)--(a6);
\draw[draw, solid] (a6)--(a7);
\draw[draw, solid] (a7)--(a8);
\draw[draw, solid] (a8)--(a9);
\draw[draw, solid] (a7)--(b);
\draw[draw, solid] (a8)--(b);
\draw[red,solid] (a9) edge [out=30,in=150,loop,looseness=1]  (a10);\draw[brown,solid] (a9) edge [out=-30,in=-150,loop,looseness=1]  (a10);
\end{tikzpicture}
}
We stress the fact that this representation of the theory is schematic at this stage. The different colors for the links connecting the $U(6)$ and $U(1)$ nodes on the right stand for the fact that the corresponding pairs of chirals interact differently with the adjoint chiral at the $U(6)$ gauge node. Specifically, denoting by $\Gp$ such adjoint chiral, by $Q$, $\tilde{Q}$ the chirals in red and by $P$, $\tilde{P}$ the chirals in brown, the interactions are respectively $\tilde{Q}\Gp^2Q$ and $\tilde{P}\Gp^3P$. Moreover, we avoid specifying additional gauge singlet chiral fields that are present at this stage. Since the vev deformation we are considering is $\mathcal{N}=4$ peserving, these should all disappear at the end of the manipulations that we are going to perform.

In order to obtain the desired 3d mirror from this theory, we have to sequentially apply Aharony duality along the quiver starting from the $U(1)$ node where the adjoint chiral is just a singlet, similarly to what we did in Subsection \ref{subsec:DpSUNplN:Examples} and in Appendix \ref{app:flipflip}. We first apply the duality up to the rightmost $U(6)$ gauge node and obtain the theory (again we stress that we are neglecting all singlets, including those coming from applying Aharony duality)
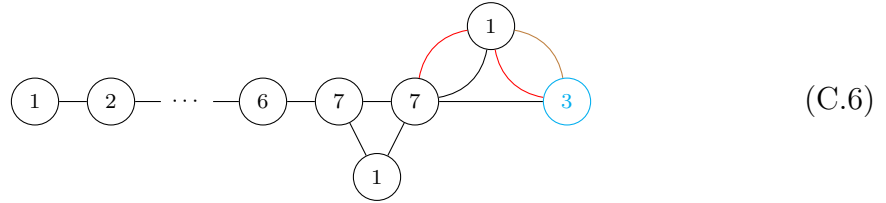
\bes{
\begin{tikzpicture}[baseline=0, font=\scriptsize]
\tikzstyle{every node}=[minimum size=0.5cm]
\node[draw, circle] (a1) at (0,0) {1};
\node[draw, circle] (a2) at (1,0) {2};
\node[draw=none] (a3) at (2,0) {$\cdots$};
\node[draw, circle] (a6) at (3,0) {6};
\node[draw, circle] (a7) at (4,0) {7};
\node[draw, circle] (a8) at (5,0) {7};
\node[draw, circle, cyan] (a9) at (7,0) {3};
\node[draw, circle] (u) at (6,1) {1};
\node[draw, circle] (b) at (4.5,-1) {1};
\draw[draw, solid] (a1)--(a2);
\draw[draw, solid] (a2)--(a3);
\draw[draw, solid] (a3)--(a6);
\draw[draw, solid] (a6)--(a7);
\draw[draw, solid] (a7)--(a8);
\draw[draw, solid] (a8)--(a9);
\draw[draw, solid] (a7)--(b);
\draw[draw, solid] (a8)--(b);
\draw[red,solid] (a9) edge [out=170,in=280,loop,looseness=1]  (u);\draw[brown,solid] (a9) edge [out=100,in=350,loop,looseness=1]  (u);
\draw[black,solid] (a8) edge [out=10,in=260,loop,looseness=1]  (u);\draw[red,solid] (a8) edge [out=80,in=190,loop,looseness=1]  (u);
\end{tikzpicture}
}
Observe that the rank of the rightmost $U(6)$ node was decreased to $7+2-6=3$ after the last dualization. Moreover, we now lack the adjoint chiral at such $U(3)$ gauge node (to highlight this fact we coloured the node without the adjoint chiral in light blue). To restore it, we need to apply again Aharony duality along the quiver starting from the left $U(1)$ node, but this time stopping at the rightmost $U(7)$ node. This gives the theory
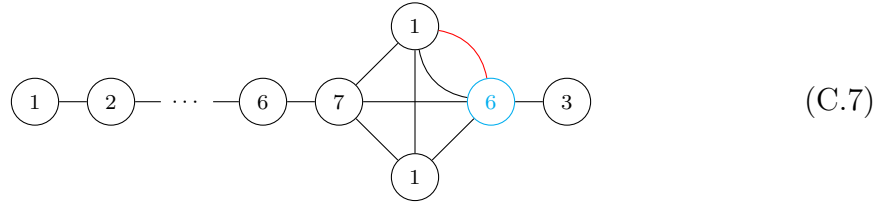
\bes{
\begin{tikzpicture}[baseline=0, font=\scriptsize]
\tikzstyle{every node}=[minimum size=0.5cm]
\node[draw, circle] (a1) at (0,0) {1};
\node[draw, circle] (a2) at (1,0) {2};
\node[draw=none] (a3) at (2,0) {$\cdots$};
\node[draw, circle] (a6) at (3,0) {6};
\node[draw, circle] (a7) at (4,0) {7};
\node[draw, circle, cyan] (a8) at (6,0) {6};
\node[draw, circle] (a9) at (7,0) {3};
\node[draw, circle] (u) at (5,1) {1};
\node[draw, circle] (b) at (5,-1) {1};
\draw[draw, solid] (a1)--(a2);
\draw[draw, solid] (a2)--(a3);
\draw[draw, solid] (a3)--(a6);
\draw[draw, solid] (a6)--(a7);
\draw[draw, solid] (a7)--(a8);
\draw[draw, solid] (a8)--(a9);
\draw[draw, solid] (a7)--(b);
\draw[draw, solid] (a8)--(b);
\draw[draw, solid] (a7)--(u);
\draw[draw, solid] (u)--(b);
\draw[black,solid] (a8) edge [out=170,in=280,loop,looseness=1]  (u);\draw[red,solid] (a8) edge [out=100,in=350,loop,looseness=1]  (u);
\end{tikzpicture}
}
Observe that the rank of the rightmost $U(7)$ node was decreased to $7+3+2+1-7=6$ after the last dualization. Moreover, this time the adjoint chiral that we are missing is the one at such $U(6)$ node. Hence, we need to iterate again the procedure of applying Aharony duality along the quiver from left to right, but this time we have to stop at the rightmost $U(7)$ node. The result is now
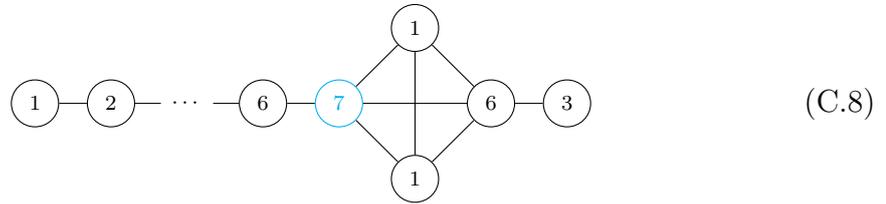
\bes{
\begin{tikzpicture}[baseline=0, font=\scriptsize]
\tikzstyle{every node}=[minimum size=0.5cm]
\node[draw, circle] (a1) at (0,0) {1};
\node[draw, circle] (a2) at (1,0) {2};
\node[draw=none] (a3) at (2,0) {$\cdots$};
\node[draw, circle] (a6) at (3,0) {6};
\node[draw, circle, cyan] (a7) at (4,0) {7};
\node[draw, circle] (a8) at (6,0) {6};
\node[draw, circle] (a9) at (7,0) {3};
\node[draw, circle] (u) at (5,1) {1};
\node[draw, circle] (b) at (5,-1) {1};
\draw[draw, solid] (a1)--(a2);
\draw[draw, solid] (a2)--(a3);
\draw[draw, solid] (a3)--(a6);
\draw[draw, solid] (a6)--(a7);
\draw[draw, solid] (a7)--(a8);
\draw[draw, solid] (a8)--(a9);
\draw[draw, solid] (a7)--(b);
\draw[draw, solid] (a8)--(b);
\draw[draw, solid] (a8)--(u);
\draw[draw, solid] (a7)--(u);
\draw[draw, solid] (u)--(b);
\end{tikzpicture}
}
This is almost the 3d mirror \eqref{D4SU10mirr} we are looking for, but we have to remember that the $U(7)$ adjoint chiral is now missing and that there are still some singlets around. This can be solved by iterating the previous procedure until we only apply Aharony duality to the $U(1)$ node on the left, which is equivalent to applying flip-flip duality to the left $T(SU(7))$ tail of the quiver. This eventually gives a manifestly 3d $\mathcal{N}=4$ theory which coincides with the claimed 3d mirror \eqref{D4SU10mirr}.

\section{Symmetry enhancement in linear quivers with special unitary gauge groups} \label{app:SUenhanced}
In this appendix, we discuss symmetry enhancement in linear quivers with special unitary gauge groups\footnote{See also \cite{Beratto:2020wmn} for other examples of symmetry enhancements in quiver theories with mixed types of gauge groups.}.  We shall not give a general here but provide some observations that are relevant to the material in the main text.

Let us first consider the 3d $\CN=4$ $SU(N)$ gauge theory with $N_f \geq 2N-1$ flavours.  For $N_f > 2N-1$, this theory has no Coulomb branch symmetry.  However, for $N_f= 2N-1$, the Coulomb branch symmetry is $U(1)$.  The latter is an emergent symmetry in the IR that corresponds to the monopole operator with flux $(1,0,\ldots,0, -1)$ that carry $R$-charge $1$.  When an $SU(N)$ gauge group has $2N-1$ flavours of hypermultiplets transforming under the fundamental representation, it is said to be balanced.  In the following discussion, we assume that this is the case.  

It is worth comparing this with a unitary gauge group $U(N)$ with $N_f$ flavours. If $N_f > 2N$, the $U(N)$ gauge group is said to be overbalanced and, if $N_f=2N$, it is said to be balanced \cite{Gaiotto:2008ak}.  If a linear quiver consists of precisely $m$ consecutive balanced unitary gauge groups of the form $(N_1)-(N_2)-\cdots-(N_m)$, then the Coulomb branch symmetry contains a factor $SU(m+1)$ \cite{Gaiotto:2008ak}.  On the other hand, each overbalanced unitary gauge group gives rise to a $U(1)$ factor.

Now let us consider the quiver of the following form: 
\be [F_1]-(N_1)-SU(M)-(N_2)-[F_2]~, 
\ee
with $N_2+N_3 = 2M-1$ and for $F_1\neq0$, $N_1\neq 1$\footnote{The theories for $F_1=0$, $N_1=1$ should be regarded as particular cases of \eqref{oneSUwTSUtail}, whose enhancement of the Coulomb branch symmetry we discuss momentarily.}.  The Coulomb branch symmetry is as follows:
\bi
\item If both $U(N_1)$ and $U(N_2)$ are overbalanced, then the Coulomb branch symmetry is $U(1)^3$, where two factors of $U(1)$ come from the overbalanced $U(N_1)$ and $U(N_2)$ nodes and the other $U(1)$ comes from $SU(M)$.   
\item If precisely one of $U(N_1)$ and $U(N_2)$ is overbalanced and the other is balanced, the Coulomb branch symmetry is $SU(2)^2 \times U(1)^2$, where the $U(1)^2$ factor comes from $SU(M)$ and the overbalanced gauge node, whereas an $SU(2)$ factor comes from the balanced node but this symmetry gets doubled, similarly to the discussion of the symplectic gauge group in \cite[section 5.3]{Gaiotto:2008ak}. 
\item If both $U(N_1)$ and $U(N_2)$ are balanced, then the Coulomb branch symmetry is $SU(2)^2 \times SU(4) \times U(1)$, where each $SU(2)$ comes from the balanced $U(N_1)$ and $U(N_2)$ nodes, the $U(1)$ factor comes from $SU(M)$, and the parameter $4$ in $SU(4)$ comes from the product $2 \times 2$ such that each $2$ comes from the aforementioned $SU(2)^2$ factor.  
\ei
The latter can be seen, for example, from the Coulomb branch Hilbert series for the quiver $[4]-4-SU(4)-3-[2]$, whose computation is described in \cite{Cremonesi:2013lqa}:
\bes{
1 + \left(z_1 z_2+\frac{z_1}{z_2}+\frac{z_2}{z_1}+\frac{1}{z_1 z_2}+3 z_1+\frac{3}{z_1} +3 z_2+ \frac{3}{z_2}+6 \right) t +\cdots~.
}
where $z_1$ and $z_2$ are the topological fugacities for $U(4)$ and $U(3)$ gauge groups respectively and the coefficient of $t$ corresponds to the Coulomb branch operators with $R$-charge $1$.  The characters $z_1+z_1^{-1} +1$ and $z_2+z_2^{-1}+1$ corresponds to the adjoint representation of each factor of $SU(2)$ in $SU(2)^2$. The character $z_1 z_2+\frac{z_1}{z_2}+\frac{z_2}{z_1}+\frac{1}{z_1 z_2}+2 z_1+\frac{2}{z_1} +2 z_2+ \frac{2}{z_2}+3$ corresponds to the adjoint representation of $SU(4)$, where it should be observed that not only two out of three Cartan variables of $SU(4)$ appear in the character; this is due to the fact that $SU(4)$ does not have a $U(1)$ topological symmetry.  The remaining term $1$ corresponds to the $U(1)$ factor.

This result can be generalised to a quiver of the form 
\be [F]-(N_1)-\cdots (N_{L-1}) -SU(M)-(N'_{R-1})-\cdots-(N'_1)-[F']~, 
\ee
with $N_{L-1}+N'_{R-1}=2M-1$.  If all $U(N_i)$ and $U(N'_j)$ nodes are balanced, the Coulomb branch symmetry is $SU(L) \times SU(R) \times SU(LR) \times U(1)$.   For example, the coefficient of $t$ in the Coulomb branch Hilbert series of 
\be \label{444321}
[4]-4-SU(4)-3-2-[1] 
\ee
is 
\bes{
&9+ \Big[ 4 z_1+3
   z_2+3
   z_3 +\frac{z_1}{z_2} +\frac{z_1}{z_3}+\frac{z_1}{z_2 z_3}+z_1 z_2 +z_1 z_3+3 z_2 z_3+z_1 z_2 z_3 \\
 &\qquad + (z_j \, \rightarrow \,z_j^{-1}) \Big]
}
where $z_1$, $z_2$ and $z_3$ are topological fugacities for $U(4)$, $U(3)$ and $U(2)$ respectively.  This indicates the Coulomb branch symmetry of \eref{444321} is $SU(2) \times SU(3) \times SU(6) \times U(1)$.  Suppose that there is one node, say $U(N'_{\hat{R}})$ (with $1 \leq \hat{R} \leq R-2$), which is overbalanced and all of the other nodes are balanced.  The Coulomb branch symmetry is $SU(L) \times U(1)_M \times SU(R-\hat{R}) \times SU(LR-L\hat{R}) \times U(1)_{\hat{R}} \times SU(\hat{R}) $, where $U(1)_M$ indicates the factor arising from the gauge node $SU(M)$ and $U(1)_{\hat{R}}$ indicates the factor arising from the gauge node $U(N'_{\hat{R}-1})$.  For example, the Coulomb branch symmetry of $[4]-4-SU(4)-3-2-2-[2]$ is $SU(2) \times U(1) \times SU(2) \times SU(4) \times U(1) \times SU(2)$.  This can be easily generalised to the cases in which there are more than one overbalanced nodes.

Let us now consider the quiver of the following form
\bes{
\label{oneSUwTSUtail}
[F]-(N_1)-\cdots- (N_{L-1}) -SU(M)-(M-1)-(M-2)- \cdots-2-1
}
with $N_{L-1}+(M-1)=2M$.  If all $U(N_j)$ nodes are balanced, then the Coulomb branch symmetry is $SU(M) \times SU(LM)$.  As an example, we can consider the quiver
\be \label{444321g}
[4]-4-SU(4)-3-2-1 
\ee
This comes from gauging $[1]$ in \eref{444321}.  Classically, we gain a $U(1)$ factor in the Coulomb branch symmetry.  However, the Coulomb branch symmetry of \eref{444321} after such gauging, namely $SU(2) \times SU(3) \times SU(6) \times U(1) \times U(1)$, gets enhanced to $SU(4) \times SU(8)$ in the IR.  This can be checked, for example, using the Coulomb branch Hilbert series \cite{Cremonesi:2013lqa}.  Observe that we do not have a $U(1)$ factor in the Coulomb branch symmetry due to the $SU(M)$ gauge node.  Suppose that one node, say $U(N_{\hat{L}})$ (with $1 \leq \hat{L} \leq L-2$), which is overbalanced and all of the other nodes are balanced.  Then the Coulomb branch symmetry is 
\be \label{CsymoneSUoneOB}
SU(\hat{L}) \times U(1)_{\hat{L}} \times SU(M) \times SU(ML-M\hat{L})~.  
\ee 
We have discussed an example for this case in the context of the $D_{14}(SU(8))$ theory in the main text.  The relevant linear quiver is 
\bes{
[8]-7-6-5-{\blue 4}-4-4-SU(4)-3-2-1
}
where the overbalanced node is highlighted in blue.
The Coulomb branch symmetry of this quiver is 
\be
SU(4)_{7-6-5} \times SU(4)_{3-2-1} \times SU(12)_{4-4,\, 3-2-1} \times U(1)_{{\blue 4}}~, 
\ee
where the subscripts indicate the part of the quiver that gives rise to each factor of the symmetry.  Note that this in agreement with the Higgs branch symmetry of the proposed mirror theory \eref{mirrD14SU8}.
Again, this result can be easily generalised to the cases in which there are more than one overbalanced nodes.

We now discuss the quiver with more than one special unitary gauge groups.  Let us consider the following quiver
\bes{
&[F]-(N_1)-(N_2)- \cdots- (N_{L-2})- (N_{L-1}) - (SU(M_1)) \\
&- (N'_{R-1})- (N'_{R-2})- \cdots - {\blue (N'_{\hat{R}})} -\cdots - (N'_2)-(N'_1)-(SU(M_2)) \\
&-(M_2-1)-(M_2-2)- \cdots -1
}
where we assume that every gauge node is balanced, except $U(N'_{\hat{R}})$, which is highlighted in blue.  One of a very interesting phenomena here is that there is also a factor $SU(LM_2)$ appears in the Coulomb branch symmetry even though the chains $(M_2-1) -(M_2-2) - \cdots - 1$ and $(N_1)-(N_2)-\cdots-(N_{L-1})$ are disconnected.  In fact, the full Coulomb branch symmetry of the above quiver is
\bes{
&SU(M_2) \times SU(R-\hat{R}) \times SU(M_2 \hat{R}) \times SU(L(R-\hat{R})) \times SU(LM_2)\\
& \times U(1)_{\hat{R}} \times U(1)_{M_1}~.
}
where the subscripts indicate the part of the quiver that gives rise to each factor of the symmetry.  As an example, the quiver
\bes{
[4]-4-4-SU(4)-3-{\blue 2}-2-SU(2)-1
}
has the Coulomb branch symmetry
\bes{
&SU(2)_{1} \times SU(2)_3 \times SU(4)_{2,1} \times SU(6)_{4-4, \, 1} \times SU(6)_{4-4,\, 3} \\
&\times U(1)_{\blue 2} \times U(1)_{SU(4)}~.
}

Suppose that there is an additional overbalanced node, for example, 
\bes{ \label{fulllinear}
&[F]-(N_1)-(N_2) -\ldots-{\blue (N_{\hat{L}})}-\cdots - (N_{L-2})- (N_{L-1}) - (SU(M_1)) \\
&- (N'_{R-1})- (N'_{R-2})- \cdots - {\blue (N'_{\hat{R}})} -\cdots - (N'_2)-(N'_1)-(SU(M_2)) \\
&-(M_2-1)-(M_2-2)- \cdots -1
}
where we assume that every gauge node is balanced, except $U(N_{\hat{L}})$ and $U(N'_{\hat{R}})$, which are highlighted in blue.  In this case, we can consider two subquivers resulting from cutting the above quiver at ${\blue (N_{\hat{L}})}$ and apply the above rules.  The Coulomb branch symmetry of \eref{fulllinear} is
\be \label{symCfulllinear}
\scalebox{0.9}{$
\begin{split}
&SU(M_2) \times SU(R-\hat{R}) \times SU(M_2 \hat{R}) \times SU((L-\hat{L})(R-\hat{R})) \times SU((L-\hat{L})M_2) \\
& \times SU(L) \times U(1)_{\hat{L}} \times U(1)_{\hat{R}} \times U(1)_{M_1} ~.
\end{split}$}
\ee
Observe that the number of $U(1)$ factors is equal to the number of special unitary gauge groups plus one in this case.

We are now ready to consider the Coulomb branch symmetry of \eref{linearquiv3dDpSUn}, where we have $c-1$ balanced special unitary gauge groups.  We can make use of \eref{symCfulllinear} and implement the following changes:
\bes{
\scalebox{0.8}{$
\begin{array}{ll}
SU(M_2) \quad &\longrightarrow  \quad SU(a) \\
SU(R-\hat{R})  \quad &\longrightarrow  \quad SU(a)^{c-1} \\
SU(M_2 \hat{R}) \times SU((L-\hat{L})(R-\hat{R})) \times SU((L-\hat{L})M_2)  \quad &\longrightarrow  \quad SU(a(b-a))^{\frac{1}{2}(c-1)[(c-1)+1]} \\
SU(L)  \quad &\longrightarrow  \quad SU(a)\\
U(1)_{\hat{L}} \times U(1)_{\hat{R}} \times U(1)_{M_1} \quad &\longrightarrow  \quad U(1)^c 
\end{array}$}
}
In summary, we deduce that the Coulomb branch symmetry of \eref{linearquiv3dDpSUn} is $U(1)^c \times SU(a)^c \times SU(ab-a^2)^{\frac{1}{c}(c-1)}$.  In the special case of $c=2$, we have precisely one special unitary gauge group, and so there is one $U(1)$ factor as discussed in \eref{CsymoneSUoneOB}.  For $c=1$, we do not have any special unitary gauge group in \eref{linearquiv3dDpSUn} and as discussed around \eref{gcd1} there is no $U(1)$ factor.

\bibliographystyle{ytphys}
\bibliography{ref}

\providecommand{\href}[2]{#2}\begingroup\raggedright\begin{thebibliography}{10}

\bibitem{Seiberg:1994rs}
N.~Seiberg and E.~Witten, ``{Electric - magnetic duality, monopole
  condensation, and confinement in N=2 supersymmetric Yang-Mills theory},''
  \href{http://dx.doi.org/10.1016/0550-3213(94)90124-4}{{\em Nucl. Phys. B}
  {\bfseries 426} (1994) 19--52},
  \href{http://arxiv.org/abs/hep-th/9407087}{{\ttfamily arXiv:hep-th/9407087}}.
  [Erratum: Nucl.Phys.B 430, 485--486 (1994)].

\bibitem{Seiberg:1994aj}
N.~Seiberg and E.~Witten, ``{Monopoles, duality and chiral symmetry breaking in
  N=2 supersymmetric QCD},''
  \href{http://dx.doi.org/10.1016/0550-3213(94)90214-3}{{\em Nucl. Phys.}
  {\bfseries B431} (1994) 484--550},
\href{http://arxiv.org/abs/hep-th/9408099}{{\ttfamily arXiv:hep-th/9408099
  [hep-th]}}.

\bibitem{Argyres:1995jj}
P.~C. Argyres and M.~R. Douglas, ``{New phenomena in SU(3) supersymmetric gauge
  theory},'' \href{http://dx.doi.org/10.1016/0550-3213(95)00281-V}{{\em Nucl.
  Phys. B} {\bfseries 448} (1995) 93--126},
  \href{http://arxiv.org/abs/hep-th/9505062}{{\ttfamily arXiv:hep-th/9505062}}.

\bibitem{Argyres:1995xn}
P.~C. Argyres, M.~Plesser, N.~Seiberg, and E.~Witten, ``{New N=2 superconformal
  field theories in four-dimensions},''
  \href{http://dx.doi.org/10.1016/0550-3213(95)00671-0}{{\em Nucl. Phys. B}
  {\bfseries 461} (1996) 71--84},
  \href{http://arxiv.org/abs/hep-th/9511154}{{\ttfamily arXiv:hep-th/9511154}}.

\bibitem{Eguchi:1996vu}
T.~Eguchi, K.~Hori, K.~Ito, and S.-K. Yang, ``{Study of N=2 superconformal
  field theories in four-dimensions},''
  \href{http://dx.doi.org/10.1016/0550-3213(96)00188-5}{{\em Nucl. Phys. B}
  {\bfseries 471} (1996) 430--444},
  \href{http://arxiv.org/abs/hep-th/9603002}{{\ttfamily arXiv:hep-th/9603002}}.

\bibitem{Eguchi:1996ds}
T.~Eguchi and K.~Hori, ``{N=2 superconformal field theories in four-dimensions
  and A-D-E classification},'' in {\em {Conference on the Mathematical Beauty
  of Physics (In Memory of C. Itzykson)}}, pp.~67--82.
\newblock 7, 1996.
\newblock \href{http://arxiv.org/abs/hep-th/9607125}{{\ttfamily
  arXiv:hep-th/9607125}}.

\bibitem{Shapere:1999xr}
A.~D. Shapere and C.~Vafa, ``{BPS structure of Argyres-Douglas superconformal
  theories},'' \href{http://arxiv.org/abs/hep-th/9910182}{{\ttfamily
  arXiv:hep-th/9910182}}.

\bibitem{Cecotti:2010fi}
S.~Cecotti, A.~Neitzke, and C.~Vafa, ``{R-Twisting and 4d/2d
  Correspondences},'' \href{http://arxiv.org/abs/1006.3435}{{\ttfamily
  arXiv:1006.3435 [hep-th]}}.

\bibitem{Bonelli:2011aa}
G.~Bonelli, K.~Maruyoshi, and A.~Tanzini, ``{Wild Quiver Gauge Theories},''
  \href{http://dx.doi.org/10.1007/JHEP02(2012)031}{{\em JHEP} {\bfseries 02}
  (2012) 031}, \href{http://arxiv.org/abs/1112.1691}{{\ttfamily arXiv:1112.1691
  [hep-th]}}.

\bibitem{Xie:2012hs}
D.~Xie, ``{General Argyres-Douglas Theory},''
  \href{http://dx.doi.org/10.1007/JHEP01(2013)100}{{\em JHEP} {\bfseries 01}
  (2013) 100}, \href{http://arxiv.org/abs/1204.2270}{{\ttfamily arXiv:1204.2270
  [hep-th]}}.

\bibitem{Wang:2015mra}
Y.~Wang and D.~Xie, ``{Classification of Argyres-Douglas theories from M5
  branes},'' \href{http://dx.doi.org/10.1103/PhysRevD.94.065012}{{\em Phys.
  Rev. D} {\bfseries 94} no.~6, (2016) 065012},
  \href{http://arxiv.org/abs/1509.00847}{{\ttfamily arXiv:1509.00847
  [hep-th]}}.

\bibitem{Gaiotto:2009hg}
D.~Gaiotto, G.~W. Moore, and A.~Neitzke, ``{Wall-crossing, Hitchin Systems, and
  the WKB Approximation},'' \href{http://arxiv.org/abs/0907.3987}{{\ttfamily
  arXiv:0907.3987 [hep-th]}}.

\bibitem{Intriligator:1996ex}
K.~A. Intriligator and N.~Seiberg, ``{Mirror symmetry in three-dimensional
  gauge theories},'' \href{http://dx.doi.org/10.1016/0370-2693(96)01088-X}{{\em
  Phys. Lett.} {\bfseries B387} (1996) 513--519},
\href{http://arxiv.org/abs/hep-th/9607207}{{\ttfamily arXiv:hep-th/9607207
  [hep-th]}}.

\bibitem{Ferlito:2017xdq}
G.~Ferlito, A.~Hanany, N.~Mekareeya, and G.~Zafrir, ``{3d Coulomb branch and 5d
  Higgs branch at infinite coupling},''
  \href{http://dx.doi.org/10.1007/JHEP07(2018)061}{{\em JHEP} {\bfseries 07}
  (2018) 061}, \href{http://arxiv.org/abs/1712.06604}{{\ttfamily
  arXiv:1712.06604 [hep-th]}}.

\bibitem{Cabrera:2018jxt}
S.~Cabrera, A.~Hanany, and F.~Yagi, ``{Tropical Geometry and Five Dimensional
  Higgs Branches at Infinite Coupling},''
  \href{http://dx.doi.org/10.1007/JHEP01(2019)068}{{\em JHEP} {\bfseries 01}
  (2019) 068}, \href{http://arxiv.org/abs/1810.01379}{{\ttfamily
  arXiv:1810.01379 [hep-th]}}.

\bibitem{Cabrera:2019izd}
S.~Cabrera, A.~Hanany, and M.~Sperling, ``{Magnetic quivers, Higgs branches,
  and 6d $N$=(1,0) theories},''
  \href{http://dx.doi.org/10.1007/JHEP06(2019)071}{{\em JHEP} {\bfseries 06}
  (2019) 071}, \href{http://arxiv.org/abs/1904.12293}{{\ttfamily
  arXiv:1904.12293 [hep-th]}}. [Erratum: JHEP 07, 137 (2019)].

\bibitem{Bourget:2019rtl}
A.~Bourget, S.~Cabrera, J.~F. Grimminger, A.~Hanany, and Z.~Zhong, ``{Brane
  Webs and Magnetic Quivers for SQCD},''
  \href{http://dx.doi.org/10.1007/JHEP03(2020)176}{{\em JHEP} {\bfseries 03}
  (2020) 176}, \href{http://arxiv.org/abs/1909.00667}{{\ttfamily
  arXiv:1909.00667 [hep-th]}}.

\bibitem{Xie:2013jc}
D.~Xie and P.~Zhao, ``{Central charges and RG flow of strongly-coupled N=2
  theory},'' \href{http://dx.doi.org/10.1007/JHEP03(2013)006}{{\em JHEP}
  {\bfseries 03} (2013) 006}, \href{http://arxiv.org/abs/1301.0210}{{\ttfamily
  arXiv:1301.0210 [hep-th]}}.

\bibitem{boalch2008irregular}
P.~Boalch, ``{Irregular connections and Kac-Moody root systems},''
  \href{http://arxiv.org/abs/0806.1050}{{\ttfamily arXiv:0806.1050 [math.DG]}}.

\bibitem{DelZotto:2014kka}
M.~Del~Zotto and A.~Hanany, ``{Complete Graphs, Hilbert Series, and the Higgs
  branch of the 4d $\mathcal{N} =$ 2 $(A_n,A_m)$ SCFTs},''
  \href{http://dx.doi.org/10.1016/j.nuclphysb.2015.03.017}{{\em Nucl. Phys. B}
  {\bfseries 894} (2015) 439--455},
  \href{http://arxiv.org/abs/1403.6523}{{\ttfamily arXiv:1403.6523 [hep-th]}}.

\bibitem{Xie:2017vaf}
D.~Xie and S.-T. Yau, ``{Argyres-Douglas matter and N=2 dualities},''
  \href{http://arxiv.org/abs/1701.01123}{{\ttfamily arXiv:1701.01123
  [hep-th]}}.

\bibitem{Dey:2020hfe}
A.~Dey, ``{Three Dimensional Mirror Symmetry beyond $ADE$ quivers and
  Argyres-Douglas theories},''
  \href{http://arxiv.org/abs/2004.09738}{{\ttfamily arXiv:2004.09738
  [hep-th]}}.

\bibitem{Maruyoshi:2016tqk}
K.~Maruyoshi and J.~Song, ``{Enhancement of Supersymmetry via Renormalization
  Group Flow and the Superconformal Index},''
  \href{http://dx.doi.org/10.1103/PhysRevLett.118.151602}{{\em Phys. Rev.
  Lett.} {\bfseries 118} no.~15, (2017) 151602},
\href{http://arxiv.org/abs/1606.05632}{{\ttfamily arXiv:1606.05632 [hep-th]}}.

\bibitem{Maruyoshi:2016aim}
K.~Maruyoshi and J.~Song, ``{$ \mathcal{N}=1 $ deformations and RG flows of $
  \mathcal{N}=2 $ SCFTs},''
  \href{http://dx.doi.org/10.1007/JHEP02(2017)075}{{\em JHEP} {\bfseries 02}
  (2017) 075},
\href{http://arxiv.org/abs/1607.04281}{{\ttfamily arXiv:1607.04281 [hep-th]}}.

\bibitem{Agarwal:2016pjo}
P.~Agarwal, K.~Maruyoshi, and J.~Song, ``{$ \mathcal{N} $ =1 Deformations and
  RG flows of $ \mathcal{N} $ =2 SCFTs, part II: non-principal deformations},''
  \href{http://dx.doi.org/10.1007/JHEP12(2016)103,
  10.1007/JHEP04(2017)113}{{\em JHEP} {\bfseries 12} (2016) 103},
  \href{http://arxiv.org/abs/1610.05311}{{\ttfamily arXiv:1610.05311
  [hep-th]}}.
[Addendum: JHEP04,113(2017)].

\bibitem{Benvenuti:2017lle}
S.~Benvenuti and S.~Giacomelli, ``{Supersymmetric gauge theories with decoupled
  operators and chiral ring stability},''
  \href{http://dx.doi.org/10.1103/PhysRevLett.119.251601}{{\em Phys. Rev.
  Lett.} {\bfseries 119} no.~25, (2017) 251601},
  \href{http://arxiv.org/abs/1706.02225}{{\ttfamily arXiv:1706.02225
  [hep-th]}}.

\bibitem{Benvenuti:2017kud}
S.~Benvenuti and S.~Giacomelli, ``{Abelianization and sequential confinement in
  $2+1$ dimensions},'' \href{http://dx.doi.org/10.1007/JHEP10(2017)173}{{\em
  JHEP} {\bfseries 10} (2017) 173},
  \href{http://arxiv.org/abs/1706.04949}{{\ttfamily arXiv:1706.04949
  [hep-th]}}.

\bibitem{Agarwal:2017roi}
P.~Agarwal, A.~Sciarappa, and J.~Song, ``{$ \mathcal{N} $ =1 Lagrangians for
  generalized Argyres-Douglas theories},''
  \href{http://dx.doi.org/10.1007/JHEP10(2017)211}{{\em JHEP} {\bfseries 10}
  (2017) 211}, \href{http://arxiv.org/abs/1707.04751}{{\ttfamily
  arXiv:1707.04751 [hep-th]}}.

\bibitem{Benvenuti:2017bpg}
S.~Benvenuti and S.~Giacomelli, ``{Lagrangians for generalized Argyres-Douglas
  theories},'' \href{http://dx.doi.org/10.1007/JHEP10(2017)106}{{\em JHEP}
  {\bfseries 10} (2017) 106}, \href{http://arxiv.org/abs/1707.05113}{{\ttfamily
  arXiv:1707.05113 [hep-th]}}.

\bibitem{Buican:2015ina}
M.~Buican and T.~Nishinaka, ``{On the superconformal index of
  Argyres\textendash{}Douglas theories},''
  \href{http://dx.doi.org/10.1088/1751-8113/49/1/015401}{{\em J. Phys. A}
  {\bfseries 49} no.~1, (2016) 015401},
  \href{http://arxiv.org/abs/1505.05884}{{\ttfamily arXiv:1505.05884
  [hep-th]}}.

\bibitem{Buican:2015hsa}
M.~Buican and T.~Nishinaka, ``{Argyres\textendash{}Douglas theories, S$^1$
  reductions, and topological symmetries},''
  \href{http://dx.doi.org/10.1088/1751-8113/49/4/045401}{{\em J. Phys. A}
  {\bfseries 49} no.~4, (2016) 045401},
  \href{http://arxiv.org/abs/1505.06205}{{\ttfamily arXiv:1505.06205
  [hep-th]}}.

\bibitem{Buican:2015tda}
M.~Buican and T.~Nishinaka, ``{Argyres-Douglas Theories, the Macdonald Index,
  and an RG Inequality},''
  \href{http://dx.doi.org/10.1007/JHEP02(2016)159}{{\em JHEP} {\bfseries 02}
  (2016) 159}, \href{http://arxiv.org/abs/1509.05402}{{\ttfamily
  arXiv:1509.05402 [hep-th]}}.

\bibitem{Dedushenko:2019mnd}
M.~Dedushenko and Y.~Wang, ``{4d/2d $\rightarrow $ 3d/1d: A song of protected
  operator algebras},'' \href{http://arxiv.org/abs/1912.01006}{{\ttfamily
  arXiv:1912.01006 [hep-th]}}.

\bibitem{Simone:2020}
C.~Closset, S.~Giacomelli, S.~Schafer-Nameki, and Y.-N. Wang, ``{5d and 4d
  SCFTs: Canonical Singularities, Trinions and S-Dualities},''
  \href{http://arxiv.org/abs/2012.12827}{{\ttfamily arXiv:2012.12827
  [hep-th]}}.

\bibitem{Nanopoulos:2010bv}
D.~Nanopoulos and D.~Xie, ``{More Three Dimensional Mirror Pairs},''
  \href{http://dx.doi.org/10.1007/JHEP05(2011)071}{{\em JHEP} {\bfseries 05}
  (2011) 071}, \href{http://arxiv.org/abs/1011.1911}{{\ttfamily arXiv:1011.1911
  [hep-th]}}.

\bibitem{Gang:2021hrd}
D.~Gang, S.~Kim, K.~Lee, M.~Shim, and M.~Yamazaki, ``{Non-unitary TQFTs from 3D
  $\mathcal{N}=4$ rank 0 SCFTs},''
  \href{http://arxiv.org/abs/2103.09283}{{\ttfamily arXiv:2103.09283
  [hep-th]}}.

\bibitem{Aharony:1997gp}
O.~Aharony, ``{IR duality in d = 3 N=2 supersymmetric USp(2N(c)) and U(N(c))
  gauge theories},''
  \href{http://dx.doi.org/10.1016/S0370-2693(97)00530-3}{{\em Phys. Lett.}
  {\bfseries B404} (1997) 71--76},
\href{http://arxiv.org/abs/hep-th/9703215}{{\ttfamily arXiv:hep-th/9703215
  [hep-th]}}.

\bibitem{Aprile:2018oau}
F.~Aprile, S.~Pasquetti, and Y.~Zenkevich, ``{Flipping the head of $T[SU(N)]$:
  mirror symmetry, spectral duality and monopoles},''
  \href{http://dx.doi.org/10.1007/JHEP04(2019)138}{{\em JHEP} {\bfseries 04}
  (2019) 138}, \href{http://arxiv.org/abs/1812.08142}{{\ttfamily
  arXiv:1812.08142 [hep-th]}}.

\bibitem{Gaiotto:2008ak}
D.~Gaiotto and E.~Witten, ``{S-Duality of Boundary Conditions In N=4 Super
  Yang-Mills Theory},''
  \href{http://dx.doi.org/10.4310/ATMP.2009.v13.n3.a5}{{\em Adv. Theor. Math.
  Phys.} {\bfseries 13} no.~3, (2009) 721--896},
\href{http://arxiv.org/abs/0807.3720}{{\ttfamily arXiv:0807.3720 [hep-th]}}.

\bibitem{Giacomelli:2017ckh}
S.~Giacomelli, ``{RG flows with supersymmetry enhancement and geometric
  engineering},'' \href{http://dx.doi.org/10.1007/JHEP06(2018)156}{{\em JHEP}
  {\bfseries 06} (2018) 156},
\href{http://arxiv.org/abs/1710.06469}{{\ttfamily arXiv:1710.06469 [hep-th]}}.

\bibitem{Shapere:2008zf}
A.~D. Shapere and Y.~Tachikawa, ``{Central charges of N=2 superconformal field
  theories in four dimensions},''
  \href{http://dx.doi.org/10.1088/1126-6708/2008/09/109}{{\em JHEP} {\bfseries
  09} (2008) 109}, \href{http://arxiv.org/abs/0804.1957}{{\ttfamily
  arXiv:0804.1957 [hep-th]}}.

\bibitem{Xie:2016evu}
D.~Xie, W.~Yan, and S.-T. Yau, ``{Chiral algebra of Argyres-Douglas theory from
  M5 brane},'' \href{http://arxiv.org/abs/1604.02155}{{\ttfamily
  arXiv:1604.02155 [hep-th]}}.

\bibitem{Cecotti:2012jx}
S.~Cecotti and M.~Del~Zotto, ``{Infinitely many N=2 SCFT with ADE flavor
  symmetry},'' \href{http://dx.doi.org/10.1007/JHEP01(2013)191}{{\em JHEP}
  {\bfseries 01} (2013) 191}, \href{http://arxiv.org/abs/1210.2886}{{\ttfamily
  arXiv:1210.2886 [hep-th]}}.

\bibitem{Cecotti:2013lda}
S.~Cecotti, M.~Del~Zotto, and S.~Giacomelli, ``{More on the N=2 superconformal
  systems of type $D_p(G)$},''
  \href{http://dx.doi.org/10.1007/JHEP04(2013)153}{{\em JHEP} {\bfseries 04}
  (2013) 153}, \href{http://arxiv.org/abs/1303.3149}{{\ttfamily arXiv:1303.3149
  [hep-th]}}.

\bibitem{Chacaltana:2012zy}
O.~Chacaltana, J.~Distler, and Y.~Tachikawa, ``{Nilpotent orbits and
  codimension-two defects of 6d N=(2,0) theories},''
  \href{http://dx.doi.org/10.1142/S0217751X1340006X}{{\em Int. J. Mod. Phys.}
  {\bfseries A28} (2013) 1340006},
\href{http://arxiv.org/abs/1203.2930}{{\ttfamily arXiv:1203.2930 [hep-th]}}.

\bibitem{Tachikawa:2015bga}
Y.~Tachikawa, ``{A review of the $T_N$ theory and its cousins},''
  \href{http://dx.doi.org/10.1093/ptep/ptv098}{{\em PTEP} {\bfseries 2015}
  no.~11, (2015) 11B102}, \href{http://arxiv.org/abs/1504.01481}{{\ttfamily
  arXiv:1504.01481 [hep-th]}}.

\bibitem{Buican:2016arp}
M.~Buican and T.~Nishinaka, ``{Conformal Manifolds in Four Dimensions and
  Chiral Algebras},''
  \href{http://dx.doi.org/10.1088/1751-8113/49/46/465401}{{\em J. Phys. A}
  {\bfseries 49} no.~46, (2016) 465401},
  \href{http://arxiv.org/abs/1603.00887}{{\ttfamily arXiv:1603.00887
  [hep-th]}}.

\bibitem{Tachikawa:2011yr}
Y.~Tachikawa and S.~Terashima, ``{Seiberg-Witten Geometries Revisited},''
  \href{http://dx.doi.org/10.1007/JHEP09(2011)010}{{\em JHEP} {\bfseries 09}
  (2011) 010}, \href{http://arxiv.org/abs/1108.2315}{{\ttfamily arXiv:1108.2315
  [hep-th]}}.

\bibitem{Chacaltana:2010ks}
O.~Chacaltana and J.~Distler, ``{Tinkertoys for Gaiotto Duality},''
  \href{http://dx.doi.org/10.1007/JHEP11(2010)099}{{\em JHEP} {\bfseries 11}
  (2010) 099},
\href{http://arxiv.org/abs/1008.5203}{{\ttfamily arXiv:1008.5203 [hep-th]}}.

\bibitem{Xie:2016uqq}
D.~Xie and S.-T. Yau, ``{New N = 2 dualities},''
  \href{http://arxiv.org/abs/1602.03529}{{\ttfamily arXiv:1602.03529
  [hep-th]}}.

\bibitem{Hwang:2020wpd}
C.~Hwang, S.~Pasquetti, and M.~Sacchi, ``{4d mirror-like dualities},''
  \href{http://dx.doi.org/10.1007/JHEP09(2020)047}{{\em JHEP} {\bfseries 09}
  (2020) 047}, \href{http://arxiv.org/abs/2002.12897}{{\ttfamily
  arXiv:2002.12897 [hep-th]}}.

\bibitem{Giacomelli:2018ziv}
S.~Giacomelli, ``{Infrared enhancement of supersymmetry in four dimensions},''
  \href{http://dx.doi.org/10.1007/JHEP10(2018)041}{{\em JHEP} {\bfseries 10}
  (2018) 041},
\href{http://arxiv.org/abs/1808.00592}{{\ttfamily arXiv:1808.00592 [hep-th]}}.

\bibitem{Benvenuti:2011ga}
S.~Benvenuti and S.~Pasquetti, ``{3D-partition functions on the sphere: exact
  evaluation and mirror symmetry},''
  \href{http://dx.doi.org/10.1007/JHEP05(2012)099}{{\em JHEP} {\bfseries 05}
  (2012) 099},
\href{http://arxiv.org/abs/1105.2551}{{\ttfamily arXiv:1105.2551 [hep-th]}}.

\bibitem{Hanany:2018vph}
A.~Hanany and G.~Zafrir, ``{Discrete Gauging in Six Dimensions},''
  \href{http://dx.doi.org/10.1007/JHEP07(2018)168}{{\em JHEP} {\bfseries 07}
  (2018) 168},
\href{http://arxiv.org/abs/1804.08857}{{\ttfamily arXiv:1804.08857 [hep-th]}}.

\bibitem{talkZhong}
Z.~Zhong, {\em {Quivers with both unitary and special unitary gauge groups}}.
\newblock The quiver meeting (Imperial College London), July 31, 2020.
\newblock
  \url{https://us02web.zoom.us/rec/play/OOSzAbwcuDicSPfecguLKESXTTjxilRZ9IonqbXtNuc6D1HVYWgfEy3niuyTCdAkb-bRpVnqshrf0XTr.j98KG6wIs5LWNjSJ?continueMode=true&_x_zm_rtaid=inKqbRQkTfe3gkvu3xGb4Q.1607710216024.1d8f8f3e1380bc6b05419ec185efef00&_x_zm_rhtaid=147}.

\bibitem{Hanany:1996ie}
A.~Hanany and E.~Witten, ``{Type IIB superstrings, BPS monopoles, and
  three-dimensional gauge dynamics},''
  \href{http://dx.doi.org/10.1016/S0550-3213(97)00157-0}{{\em Nucl.Phys.}
  {\bfseries B492} (1997) 152--190},
\href{http://arxiv.org/abs/hep-th/9611230}{{\ttfamily arXiv:hep-th/9611230
  [hep-th]}}.

\bibitem{Pasquetti:2019uop}
S.~Pasquetti and M.~Sacchi, ``{From 3$d$ dualities to 2$d$ free field
  correlators and back},''
  \href{http://dx.doi.org/10.1007/JHEP11(2019)081}{{\em JHEP} {\bfseries 11}
  (2019) 081},
\href{http://arxiv.org/abs/1903.10817}{{\ttfamily arXiv:1903.10817 [hep-th]}}.

\bibitem{Pasquetti:2019tix}
S.~Pasquetti and M.~Sacchi, ``{3d dualities from 2d free field correlators:
  recombination and rank stabilization},''
  \href{http://dx.doi.org/10.1007/JHEP01(2020)061}{{\em JHEP} {\bfseries 01}
  (2020) 061},
\href{http://arxiv.org/abs/1905.05807}{{\ttfamily arXiv:1905.05807 [hep-th]}}.

\bibitem{Benvenuti:2020wpc}
S.~Benvenuti, I.~Garozzo, and G.~L. Monaco, ``{Monopoles and dualities in $3d$
  $\mathcal{N}=2$ quivers},'' \href{http://arxiv.org/abs/2012.08556}{{\ttfamily
  arXiv:2012.08556 [hep-th]}}.

\bibitem{Benini:2017dud}
F.~Benini, S.~Benvenuti, and S.~Pasquetti, ``{SUSY monopole potentials in 2+1
  dimensions},'' \href{http://dx.doi.org/10.1007/JHEP08(2017)086}{{\em JHEP}
  {\bfseries 08} (2017) 086},
\href{http://arxiv.org/abs/1703.08460}{{\ttfamily arXiv:1703.08460 [hep-th]}}.

\bibitem{Giacomelli:2017vgk}
S.~Giacomelli and N.~Mekareeya, ``{Mirror theories of 3d $ \mathcal{N} $ = 2
  SQCD},'' \href{http://dx.doi.org/10.1007/JHEP03(2018)126}{{\em JHEP}
  {\bfseries 03} (2018) 126},
\href{http://arxiv.org/abs/1711.11525}{{\ttfamily arXiv:1711.11525 [hep-th]}}.

\bibitem{Garozzo:2019xzi}
I.~Garozzo, N.~Mekareeya, and M.~Sacchi, ``{Duality walls in the 4d $
  \mathcal{N} $ = 2 SU(N) gauge theory with $2N$ flavours},''
  \href{http://dx.doi.org/10.1007/JHEP11(2019)053}{{\em JHEP} {\bfseries 11}
  (2019) 053},
\href{http://arxiv.org/abs/1909.02832}{{\ttfamily arXiv:1909.02832 [hep-th]}}.

\bibitem{Benvenuti:2020gvy}
S.~Benvenuti, I.~Garozzo, and G.~L. Monaco, ``{Sequential deconfinement in $3d$
  $\mathcal{N}\!=\!2$ gauge theories},''
  \href{http://arxiv.org/abs/2012.09773}{{\ttfamily arXiv:2012.09773
  [hep-th]}}.

\bibitem{Jafferis:2010un}
D.~L. Jafferis, ``{The Exact Superconformal R-Symmetry Extremizes Z},''
  \href{http://dx.doi.org/10.1007/JHEP05(2012)159}{{\em JHEP} {\bfseries 05}
  (2012) 159},
\href{http://arxiv.org/abs/1012.3210}{{\ttfamily arXiv:1012.3210 [hep-th]}}.

\bibitem{Borokhov:2002cg}
V.~Borokhov, A.~Kapustin, and X.-k. Wu, ``{Monopole operators and mirror
  symmetry in three-dimensions},''
  \href{http://dx.doi.org/10.1088/1126-6708/2002/12/044}{{\em JHEP} {\bfseries
  12} (2002) 044}, \href{http://arxiv.org/abs/hep-th/0207074}{{\ttfamily
  arXiv:hep-th/0207074}}.

\bibitem{Beratto:2020wmn}
E.~Beratto, S.~Giacomelli, N.~Mekareeya, and M.~Sacchi, ``{3d mirrors of the
  circle reduction of twisted A$_{2N}$ theories of class S},''
  \href{http://dx.doi.org/10.1007/JHEP09(2020)161}{{\em JHEP} {\bfseries 09}
  (2020) 161}, \href{http://arxiv.org/abs/2007.05019}{{\ttfamily
  arXiv:2007.05019 [hep-th]}}.

\bibitem{Cremonesi:2013lqa}
S.~Cremonesi, A.~Hanany, and A.~Zaffaroni, ``{Monopole operators and Hilbert
  series of Coulomb branches of $3d$ $\mathcal{N} = 4$ gauge theories},''
  \href{http://dx.doi.org/10.1007/JHEP01(2014)005}{{\em JHEP} {\bfseries 01}
  (2014) 005},
\href{http://arxiv.org/abs/1309.2657}{{\ttfamily arXiv:1309.2657 [hep-th]}}.

\end{thebibliography}\endgroup
\end{document}